\documentclass[mnsc,sglanonrev]{informs4}
\usepackage{eqndefns-left} 
\RequirePackage{tgtermes}
\RequirePackage{newtxtext}
\RequirePackage{newtxmath}
\RequirePackage{bm}
\RequirePackage{endnotes}

\OneAndAHalfSpacedXII 

\usepackage{algorithm}
\usepackage{algpseudocode}
\usepackage{tikz}
\usepackage{multirow}
\usepackage{fix-cm}
\usepackage{subcaption}
\usepackage{arydshln}
\usepackage{float}
\usepackage{booktabs}

\usepackage{natbib}
 \bibpunct[, ]{(}{)}{,}{a}{}{,}%
 \def\bibfont{\small}%
\EquationsNumberedThrough    

\TheoremsNumberedThrough     
\ECRepeatTheorems  %

\MANUSCRIPTNO{MNSC-0001-2024.00}

\begin{document}


\RUNAUTHOR{Tan and Kok}

\RUNTITLE{Explainable AI for Comprehensive Risk Assessment for Financial Reports}

\TITLE{Explainable AI for Comprehensive Risk Assessment for Financial Reports: A Lightweight Hierarchical Transformer Network Approach}

\ARTICLEAUTHORS{%
\AUTHOR{Xue Wen Tan}
\AFF{Asian Institute of Digital Finance,
National University of Singapore, \EMAIL{xuewen@u.nus.edu}}

\AUTHOR{Stanley Kok}
\AFF{Department of Information Systems and Analytics,
National University of Singapore, \EMAIL{skok@comp.nus.edu.sg}}
} 

\ABSTRACT{%
Every publicly traded U.S. company files an annual 10-K report containing critical insights into financial health and risk. We propose Tiny eXplainable Risk Assessor (TinyXRA), a lightweight and explainable transformer-based model that automatically assesses company risk from these reports. Unlike prior work that relies solely on the standard deviation of excess returns (adjusted for the Fama-French model), which indiscriminately penalizes both upside and downside risk, TinyXRA incorporates skewness, kurtosis, and the Sortino ratio for more comprehensive risk assessment. We leverage TinyBERT as our encoder to efficiently process lengthy financial documents, coupled with a novel dynamic, attention-based word cloud mechanism that provides intuitive risk visualization while filtering irrelevant terms. This lightweight design ensures scalable deployment across diverse computing environments with real-time processing capabilities for thousands of financial documents which is essential for production systems with constrained computational resources. We employ triplet loss for risk quartile classification, improving over pairwise loss approaches in existing literature by capturing both the direction and magnitude of risk differences. Our TinyXRA achieves state-of-the-art predictive accuracy across seven test years on a dataset spanning 2013–2024, while providing transparent and interpretable risk assessments. We conduct comprehensive ablation studies to evaluate our contributions and assess model explanations both quantitatively by systematically removing highly attended words and sentences, and qualitatively by examining explanation coherence. The paper concludes with findings, practical implications, limitations, and future research directions.
}%





\maketitle


\section{Introduction}
\label{sec:Intro}

The financial industry has traditionally depended on conventional risk assessment methods, such as credit score evaluation, debt-to-income ratio analysis, and financial statement scrutiny \citep{https://doi.org/10.1111/j.1540-6261.1968.tb00843.x}. While these approaches have been effective in certain risk-related applications, like assessing the riskiness of borrowers, investments, and financial instruments, they are inherently limited in scope and necessitate time-consuming human intervention. To address these challenges, research over the past few decades has increasingly focused on applying statistical methodologies to financial risk analysis \citep{FAMA19933, RePEc:chb:bcchsb:v15c12pp371-410, TOMA2014712}. For instance, \citet{FAMA19933} identified firm size and book-to-market ratios as primary indicators of financial risk using statistical methods, yet they failed to detect additional subtle risk drivers. This gap highlights the inherent constraints of traditional statistical approaches in capturing nuanced relationships and complex interactions, particularly within the vast and complex datasets prevalent in today's modern financial environment.

Machine learning (ML) has emerged as a transformative technology for financial risk classification, addressing many limitations inherent in traditional and statistical methodologies. ML techniques improves risk assessment by effectively utilizing large datasets and advanced algorithms, producing models that are adept at predicting potential risk factors with precision and uncovering complex relationships between variables \citep{shi2022machine}. This technological advancement has substantially enhanced risk assessment capabilities, allowing financial organizations to make well-informed decisions in a fraction of the time required by traditional methods or basic statistical techniques.

A notable evolution within the broader machine learning field is the rise of natural language processing (NLP), which has profoundly reshaped financial analysis practices. NLP facilitates the extraction of valuable insights from unstructured textual data, significantly improving risk assessment and enhancing investment decision-making capabilities. The effectiveness of NLP techniques has prompted researchers to explore novel textual data sources, such as financial news, analyst reviews, and corporate reports, to uncover previously unrecognized risk factors. Concurrently, various text-mining and analytic methods have been evaluated to optimize insights extraction from textual data \citep{Ding2015DeepLF, 10.5555/1620754.1620794, nopp-hanbury-2015-detecting, rekabsaz-etal-2017-volatility, TSAI2017243}. Among these approaches, deep learning stands out due to its ability to generate semantically meaningful representations without extensive manual feature engineering. Deep learning architectures, including Convolutional Neural Networks (CNN; \citealp{726791}), Gated Recurrent Units (GRU; \citealp{69e088c8129341ac89810907fe6b1bfe}), and Bidirectional Encoder Representations from Transformers (BERT; \citealp{DBLP:conf/naacl/DevlinCLT19}), have demonstrated significant performance improvements in NLP tasks such as document classification and sentiment analysis \citep{akhtar-etal-2017-multilayer, Santos2014DeepCN}.

Despite these advantages, deep learning models frequently face criticism due to their opaque decision-making processes, often described as a black-box issue. The inherent lack of interpretability in these models raises concerns among regulators, stakeholders, and financial institutions \citep{federal2018big}. The obscured nature of deep learning decisions impedes users' ability to comprehend and trust the rationale behind predictions, particularly in scenarios involving consequential decisions, impacting individuals or organizations \citep{ribeiro-etal-2016-trust}. As a result, institutions may be hesitant to adopt such technologies, even when their predictive accuracy is demonstrated \citep{10.2307/25148718, 10.2307/23015508}. Moreover, limited interpretability poses compliance challenges, as financial organizations are often required to justify and document their risk assessment methodologies explicitly \citep{bussmann2021explainable}.

Because of these concerns, there is a growing need in financial risk classification for artificial intelligence (AI) models that can provide clear and understandable reasons for their decisions, commonly referred to as explainable AI. Explainable AI merges the analytical strengths of machine learning models with transparent decision-making processes to create a more transparent and accountable system, which is essential for maintaining trust and compliance in the financial sector \citep{8400040}. Integrating explainable AI into risk assessment facilitates deeper insight into influential risk factors, empowering financial institutions with clearer rationale for credit evaluations, investment decisions, and overall risk management strategies \citep{bussmann2021explainable}. Additionally, XAI methodologies can uncover previously undetected risk factors and mitigate potential biases inherent in machine learning models, leading to more accurate, fair, and reliable risk assessments \citep{doi:10.1080/10580530.2020.1849465, e23010018}.

The concept of explainable AI approaches in natural language processing (NLP) is not entirely new. Explainable AI approaches for NLP applications have previously been explored, predominantly offering interpretability at the word level \citep{sun-lu-2020-understanding, Zhao2020SHAPVF, 10356_154295} and occasionally extending to sentence-level explanations \citep{10.1007/978-3-030-86514-6_16, mathews2019explainable}. However, these techniques often focus solely on individual documents, missing the opportunity to leverage their explanations further. We believe that high-level explanations are crucial for promoting effective communication between AI-driven models and their users. It enables users to grasp the broader context of risk factors, ensuring they do not miss the forest for the trees. By identifying overarching themes and trends in the entire corpus, users can recognize wide-ranging patterns that might be overlooked when focusing solely on word-level or sentence-level explanations. Financial professionals, stakeholders, and regulators often need to interpret and explain complex risk assessments to others, and these high-level explanations allow them to present a coherent and comprehensive narrative that connects all significant risk factors \citep{fritz2022financial}.

Additionally, existing literature on financial risk measurement has largely concentrated on volatility metrics \citep{du2024financial, ge2022neural, mashrur2020machine}, typically quantified as the standard deviation of excess returns, while other forms of risk measurement receive comparatively less attention. However, critics have argued that volatility alone inadequately captures risk complexities, particularly under conditions involving asymmetric returns and extreme market conditions \citep{eftekhari2000volatility, christoffersen2000relevant}.

To further advance the field of explainable AI in the finance sector, we introduce Tiny eXplainable Risk Assessor (TinyXRA), a lightweight and explainable transformer-based model designed to automatically assess company risk from financial reports. Unlike prior approaches that rely solely on the standard deviation of excess returns (adjusted for the Fama-French model), which indiscriminately penalizes both upside and downside risk, TinyXRA integrates skewness, kurtosis, and the Sortino ratio for a more nuanced risk assessment which we will discuss further in Section \ref{sec:Data}. Additionally, we introduce a word cloud based on the attention mechanism that filters out irrelevant words and sentences to provide a bird’s-eye view of risk factors for the target year. As opposed to prior explainable AI approaches that depended primarily on bag-of-words methods \citep{hajek2018combining} or static embeddings like GloVe \citep{10.1007/978-3-030-86514-6_16}, TinyXRA utilizes contextual transformer-based embeddings. We also recognize the computational challenges posed by lengthy financial documents; therefore, we have adopted TinyBert \citep{jiao2020tinybert} as our encoder, which helps to significantly reduce computational requirements while maintaining prediction accuracy comparable to BERT.

Moreover, the financial industry faces distinctive computational constraints that critically demand lightweight model architectures. Unlike technology companies with massive resources, financial institutions operate under strict regulatory frameworks and limited computational budgets, particularly smaller firms, hedge funds, and asset management companies lacking enterprise-grade infrastructure. Real-time risk assessment during volatile market conditions requires models that can process lengthy financial documents within seconds where delays can result in significant financial losses, making computational efficiency a business necessity rather than preference. Most financial institutions cannot deploy traditional large language models that require hundreds of gigabytes of memory, and stringent data privacy regulations further restrict options by prohibiting cloud-based solutions, leaving institutions dependent on limited on-premise hardware. We therefore define a lightweight model as one capable of operating within consumer-grade GPUs with 11GB VRAM or less, such as the NVIDIA GeForce RTX 2080 Ti used in our experiments. This approach ensures sophisticated financial NLP capabilities remain accessible to practitioners who cannot justify high-end computational infrastructure while delivering performance required for time-sensitive financial decision-making. In summary, our work contributes to the field in several key areas\footnote{The dataset and code will be made publicly available upon acceptance of the paper.}:

\begin{enumerate}
    \item We extend traditional volatility-based risk measures by incorporating skewness, kurtosis, and the Sortino ratio, providing a more comprehensive characterization of financial risk that distinguishes between upside and downside volatility.
    \item We present TinyXRA, an interpretable model that demonstrates superior performance in terms of F1, Kendall's Tau and Spearman's Rho scores compared to existing strong baselines. 
    \item We develop attention-driven word clouds that leverage TinyXRA's inherent interpretability to filter irrelevant textual content and provide intuitive, high-level explanations that align with predicted risk classifications.
    \item We conduct rigorous evaluation of model explanations through both quantitative assessment by systematically removing top-k percentages of highly-attended words and sentences to measure impact on predictive performance, and qualitative analysis of explanation coherence and domain relevance.
    \item We demonstrate empirically that interpretability and predictive accuracy can be mutually reinforcing rather than competing objectives, with our transparent model consistently outperforming existing and more recent benchmarks across all evaluation metrics, thereby challenging the traditional interpretability-performance trade-off paradigm.
\end{enumerate}

\section{Literature Review}\label{sec:Litrev}

\subsection{Text Analytics in the Financial Domain}
\cite{bellstam2021text} employed text analytics to assess the innovativeness of companies based on analyst reports for S\&P 500 firms. Their text-based approach successfully forecasted firm performance and growth opportunities for up to four years. Similarly, \cite{das2016mining} provided an overview of systems and methods for tracking ongoing events from sources such as corporate filings, financial articles, analyst reports, press releases, customer feedback, and news articles. Both studies utilize Latent Dirichlet Allocation (LDA) for classification tasks. LDA is a widely used statistical model in natural language processing (NLP) for topic modeling. It identifies underlying topics within a collection of text documents and estimates their relative frequency. LDA assumes that each document consists of a mixture of topics, with each topic represented by a distribution of words. By analyzing word co-occurrence in documents, LDA uncovers hidden thematic structures and reveals prevalent topics. However, LDA has a notable limitation: it disregards the order of words within the text, potentially leading to the loss of vital information. This issue arises from LDA's reliance on the bag-of-words representation, which only considers word presence and frequency in a document, but not their sequential arrangement. Consequently, LDA might struggle to capture contextual and syntactical nuances within texts, which are critical for understanding the true meaning and significance of the content, especially in complex domains like financial document analysis. Despite these limitations, LDA remains a popular choice in finance-specific applications due to its simplicity, interpretability, and ease of understanding, which are essential in the finance domain. In contrast, transformer models like FinBERT \citep{huang2023finbert}, despite their superior performance, have yet to gain widespread adoption due to their “black box” nature.

The empirical effectiveness of bag-of-words techniques, such as LDA, has been surpassed by the deep learning approach of word embeddings. Word embeddings represent words as numerical vectors that encapsulate both syntactic and semantic associations between words. They are learned from large collections of text data and can be used for various NLP tasks, such as text classification and sentiment analysis. \cite{yang2022analyzing} built on recent progress in representation learning and proposed a novel word embedding method that incorporates external knowledge from a finance-domain lexicon \citep{loughran2011liability}. This approach learns semantic relationships among words in firm reports for better stock volatility prediction. Its empirical results demonstrate that domain-lexicon-enhanced text representation learning significantly outperforms bag-of-words models and generic word embeddings for stock volatility prediction.

The recent eXplainable Risk Ranking (XRR) model \citep{10.1007/978-3-030-86514-6_16} is currently the state-of-the-art system for risk classification in financial reports. It features a bi-level system that provides explanations at the word and sentence levels. XRR trains domain-specific word embeddings using 10-K financial documents by finetuning pre-trained static word embeddings such as GLoVe \citep{pennington2014glove}. Similar to \cite{yang2022analyzing}, their word embeddings are also static, meaning that a word will have the same vector regardless of the context in which it is placed. Consequently, such embeddings cannot adequately represent polysemous words whose semantics depend on their contexts. In contrast, our TinyXRA model utilizes context-dependent word embeddings, allowing it to outperform XRR in predictive accuracy. Moreover, rather than solely focusing on word-level and sentence-level explanations, we enhance interpretability by generating an attention-based word cloud. This approach provides a comprehensive overview across multiple documents, offering a broader perspective on the data.

\subsection{Explainable AI Techniques}

Two popular explainable AI frameworks that provide model-agnostic explanations for machine learning models are Local Interpretable Model-agnostic Explanations (LIME; \citealp{ribeiro-etal-2016-trust}) and SHapley Additive exPlanations (SHAP; \citealp{lundberg2017unified}). LIME generates locally linear explanations for individual predictions, while SHAP uses cooperative game theory to determine the contribution of each feature to a prediction.

In the financial sector, these techniques have been applied to areas such as credit risk classification, in which machine learning models are used to predict the likelihood of borrowers defaulting on loans \citep{gramegna2021shap}. LIME and SHAP assist financial institutions in understanding the factors influencing the models' predictions to promote transparency and enable better decision-making.

However, these techniques exhibit limitations. LIME's explanations may not represent the model's global behavior and can be sensitive to the choice of neighborhood and local model complexities. Although SHAP provides more consistent explanations, it can be computationally expensive, particularly for high dimensional datasets such as text data. Additionally, both techniques face challenges when applied to text inputs, as they rely on perturbing input features to estimate local models or feature attributions. Since the meaning of text inputs is encoded in their sequential structure, this perturbative approach often results in the generation of non-representative, nonsensical text inputs. Despite these limitations, LIME and SHAP continue to offer valuable insights for non-text-based inputs.

Another powerful tool for providing explanations is the attention mechanism of a deep learning system. By assigning varying importance weights to different input elements, the attention mechanism allows a deep-learning model to focus on the most relevant parts of the input when making predictions \citep{vaswani2017attention}. These attention weights can be interpreted as feature importance scores, offering insights into which parts of the input contribute the most to a model's prediction. This interpretability aspect of the attention mechanism makes it an attractive choice for generating explanations \citep{wiegreffe2019attention}. Compared to LIME and SHAP, the attention mechanism offers several advantages. It inherently captures the relationships between the input elements in a sequential structure, which is particularly important for textual data. Additionally, attention-based explanations are directly derived from the model's internal parameters, eliminating the need for additional computation or approximation (like those needed by LIME and SHAP). This results in more efficient and accurate explanations that are better aligned with the model's predictions. For this reason, our TinyXRA model innovates by enhancing attention mechanisms. Specifically, we enhance the existing Hierarchical Attention Networks \citep{yang2016hierarchical} by replacing the word-level attention mechanism with the transformer's multi-head attention mechanism. This modification enables the model to attend to a greater number of important words within a sentence, ensuring that key information is effectively captured. Additionally, instead of using static word embeddings, we incorporate contextualized word embeddings, further improving predictive performance.

In the field of Information Systems (IS), interpretability in AI is a pressing concern, as reflected by recent research. \cite{someh2022building} emphasize the need for businesses to develop AI explanation capabilities, highlighting four critical areas such as decision tracing, bias remediation, boundary setting, and value formulation, to address challenges such as opacity and model drift. Our TinyXRA model aligns with ``decision tracing" by offering a clear pathway to understand how predictions are made. \cite{zhang2020addressing} examine the use of AI in legal practice, illustrating its potential to transform businesses while also shedding light on the unique challenges of developing machine-learning AI systems, especially in knowledge-intensive work. Additionally, \cite{asatiani2020challenges} present a case study at the Danish Business Authority, offering a framework and recommendations to deal with the complexities of explaining the behavior of black-box AI systems and to avoid any potential legal or ethical issues. Similarly, our paper extends the application of explainable AI into the finance domain. Together, these studies reveal different facets of a shared problem: making AI transparent and comprehensible. In line with these works, our research contributes to the field by introducing an explainable AI model that not only classifies financial risk but also provides clear and understandable explanations for the prediction.
\section{Datasets}
\label{sec:Data}
\subsection{10k Financial Reports: Management's Discussion and Analysis}
The U.S. Securities and Exchange Commission (SEC) requires a publicly traded company to provide a comprehensive summary of its financial performance in the form of an annual document called the 10-K financial report. This report encompasses a broad range of information, including financial statements, management's discussion and analysis (MD\&A), business descriptions, risk factors, corporate governance, and executive compensation. By providing this detailed and audited insight into a company's financial health, business strategy, and potential risks, the 10-K report serves as an invaluable resource for investors, analysts, and regulators. 

Each section of the report provides its own unique and valuable insights. For instance, through the report's presentation of balance sheets, income statements, and cash flow statements, analysts can assess a company's profitability, liquidity, and overall financial stability. Furthermore, the MD\&A section, which presents management's viewpoint on the company's performance and future prospects, enables investors to comprehend the strategic direction and management's plans for tackling challenges and seizing opportunities. In addition, the risk factors section, which details potential risks and uncertainties that the company faces, allows investors to gain insights into the possible challenges that could impact a company's performance.

Given the wealth of information contained in the 10-K reports, researchers have employed natural language processing (NLP) techniques to automatically extract valuable insights from the vast amounts of unstructured textual data found in these financial reports, uncovering patterns and connections that might not be easily discernible through manual analysis. The information and insights extracted can be used to facilitate numerous tasks such as sentiment analysis~\citep{ren2013effective, https://doi.org/10.1111/j.1475-679X.2010.00382.x}, topic modeling~\citep{DYER2017221}, and text classification~\citep{BALAKRISHNAN2010789}. For instance, in sentiment analysis, NLP techniques can be applied to the MD\&A section to evaluate the sentiment polarity conveyed by management regarding a company's future prospects. In topic modeling, NLP techniques like Latent Dirichlet Allocation (LDA) can be used to identify the prevalent themes and topics within a report, helping analysts understand the key areas of focus and concern for the company. In text classification, NLP algorithms can be employed to automatically categorize sections of a report or classify the entire document based on specific criteria, such as industry sector or financial performance.

In this study, our objective is to \textbf{harness the Management Discussion and Analysis (MD\&A) sections of 10-K financial document as input to predict the financial risk associated to the company}. Research has highlighted the valuable insights contained within the MD\&A section of financial reports. \cite{li2010information} has identified a correlation between the tone of the MD\&A and future earnings, while \cite{muslu2015forward} have observed that firms whose stock prices do not accurately reflect future earnings are more likely to provide forward-looking information in their MD\&A. Additionally, \cite{davis2012managers} have found that managers tend to convey more pessimistic information in the MD\&A compared to earnings press releases. Collectively, these findings highlight the MD\&A as a key resource for understanding a company's financial status. By offering insights into both current conditions and future prospects, it serves as a valuable asset in assessing and potentially predicting the company's financial risk.

\subsection{Risk Measurements}
\label{subsect:RiskMeasure}
In this study, we aim to measure the risk associated with each company by analyzing post-event excess returns, specifically following the filing of 10-K financial reports. Traditional approaches \citep{ito1998there} often consider daily stock returns over a specified period; however, this method may not adequately capture the effects of specific events on stock prices. To more accurately assess such impacts, we employ the Fama-French Three-Factor Model \citep{FAMA19933}, which accounts for systematic risk factors including market risk, company size, and value characteristics. 

The residuals from this model represent stock-specific, unsystematic risk, portions not explained by the market risk premium, size effect, or value effect as modeled by the Fama-French factors. We calculate excess returns from the $1^{st}$ day to the $252^{nd}$ day after the event and also ensuring a minimum of 60 daily observations to maintain statistical robustness. We first calculate the excess return as:
\begin{equation}
\label{eq:excess_ret}
    R_{\text{excess}, t} = R_t - R_{f,t}, 
\end{equation}
\noindent where \( R_t \) is the stock return at time \( t \) and \( R_{f,t} \) is the risk-free rate at time \( t \). We then estimate with the Fama-French Three-Factor Model, specified in Equation \ref{eq:FF3M}, to regress the excess returns on three macro-economics variables as:
\begin{equation}
\label{eq:FF3M}
\begin{aligned}
    R_{\text{excess},t} &= \alpha + \beta_1 \cdot MKT_{RF,t} \\
    &\quad + \beta_2 \cdot SMB_t + \beta_3 \cdot HML_t + \epsilon_t,
\end{aligned}
\end{equation}
where \( \alpha \) is the intercept (Jensen's Alpha), \( \beta_1, \beta_2, \beta_3 \) are the factor loadings (sensitivities to market, size, and value factors), \( MKT_{RF,t} \) is the market excess return, \( SMB_t \) is the size factor (small-minus-big), \( HML_t \) is the value factor (high-minus-low), \( \epsilon_t \) is the residual (unexplained component of return). After calculating the residuals ($\epsilon_t$) from the Fama-French Three-Factor Model, we proceed to compute various risk metrics to gain a comprehensive understanding of each company's risk profile. These metrics include \textbf{1. Standard Deviation} \citep{liu2022short}, \textbf{2. Skewness}, \textbf{3. Kurtosis} \citep{theodossiou2016skewness}, and the \textbf{4. Sortino Ratio} \citep{makrani2014ranking}.

\textbf{1. Standard Deviation (Volatility)} typically quantified as the standard deviation of excess returns, is a widely used metric for assessing financial risk. 
The standard deviation of residuals \( \sigma_{\epsilon} \) is given by:
\begin{equation}
    \sigma_{\epsilon} = \sqrt{\frac{1}{n} \sum_{t=1}^{n} \epsilon_t^2}.
\end{equation}
However, the effectiveness of volatility as a sole risk measure has been increasingly questioned due to several critical limitations. First, standard deviation penalizes both upward and downward price movements equally, failing to distinguish between beneficial positive returns and detrimental negative returns. This symmetric treatment contradicts typical investor preferences for upside potential over downside risk. Second, volatility assumes that asset returns follow a normal distribution which is an assumption frequently violated in real-world markets characterized by skewed returns and fat tails. This can lead to systematic underestimation of extreme events' likelihood and impact, resulting in misleading risk assessments.

Given these limitations, relying solely on volatility provides an incomplete view of investment risk. A more comprehensive approach incorporates additional metrics: skewness to capture return distribution asymmetry, kurtosis to assess the propensity for extreme values, and the Sortino ratio to distinguish harmful downside volatility from beneficial upside volatility. Together, these metrics offer a more nuanced and accurate characterization of financial risk.

\textbf{2. Skewness} is a statistical measure that quantifies the asymmetry of a distribution around its mean. In the context of financial returns, skewness provides insights into the likelihood and magnitude of deviations from the average return, offering a deeper understanding of potential investment risks and rewards. Mathematically, Skewness \( S \) of the residuals is computed as:
\begin{equation}
S = \frac{\frac{1}{n} \sum_{t=1}^{n} (\epsilon_t - \bar{\epsilon})^3}{\left(\frac{1}{n} \sum_{t=1}^{n} (\epsilon_t - \bar{\epsilon})^2 \right)^{3/2}},
\end{equation}
\noindent where \( \bar{\epsilon} \) is the mean of residuals from the Fama-French Three-Factor Model. A skewness value of zero indicates a perfectly symmetrical distribution, akin to a normal distribution. Positive skewness (right-skewed $>$ 0) suggests that the distribution has a longer tail on the right side, implying a higher probability of large positive returns. Conversely, negative skewness (left-skewed $<$ 0) indicates a longer left tail, signifying a higher probability of substantial negative returns. 

In financial analysis, understanding skewness is important. Investors typically prefer assets with positive skewness \citep{brunnermeier2007optimal}, as these offer the potential for occasional significant gains, even if accompanied by higher volatility. On the other hand, assets with negative skewness but low volatility, such as high-yield or ``junk" bonds, may provide stable returns but carry a higher risk of extreme negative outcomes. These bonds often exhibit negatively skewed returns due to the higher risk of default. While the majority of returns might be relatively stable, occasional defaults can result in significant losses, creating a left-skewed distribution \citep{chiang2016skewness}. Note that in our implementation, the bias is set to False, ensuring that the calculations are adjusted to correct for statistical bias\footnote{https://docs.scipy.org/doc/scipy/reference/generated/scipy.stats.skew.html}.

\textbf{3. Kurtosis} is a statistical measure that describes the ``tailedness" or the propensity of a distribution to produce extreme values, offering insight into the likelihood of outliers in a dataset. Kurtosis \( K \) of the residuals is given by:
\begin{equation}
K = \frac{\frac{1}{n} \sum_{t=1}^{n} (\epsilon_t - \bar{\epsilon})^4}{\left(\frac{1}{n} \sum_{t=1}^{n} (\epsilon_t - \bar{\epsilon})^2 \right)^{2}},
\end{equation}
\noindent where \( \bar{\epsilon} \) is the mean of residuals from the Fama-French Three-Factor Model. This measures the fat tails or extreme movements in residuals. In financial contexts, understanding kurtosis is essential for evaluating the likelihood of significant fluctuations in asset returns. A kurtosis value equal to 3 characterizes a \textit{mesokurtic distribution}, akin to the normal distribution, indicating moderate tails and a standard bell-curve shape. Values exceeding 3 denote a \textit{leptokurtic distribution}, which exhibits fat tails, signaling a higher propensity for extreme events, both gains and losses. Such distributions are often observed in financial markets, especially during periods of heightened volatility or crises, implying elevated risk due to large, unexpected movements \citep{de2023modeling, gabrisch2011extreme}. Conversely, a kurtosis value less than 3 describes a \textit{platykurtic distribution}, characterized by thin tails and fewer extreme events, typically associated with more stable investments like government bonds or blue-chip stocks \citep{Mondello2023, samunderu2021return}. For investors, recognizing the kurtosis of return distributions is essential, as high kurtosis suggests markets are susceptible to extreme events, necessitating robust risk management strategies to mitigate potential adverse impacts. Therefore, investors often prefer low kurtosis, as it implies a more predictable distribution with fewer extreme movements in portfolio returns, aligning with a more stable investment profile. Note that we are using the Fisher definition of kurtosis, meaning that 3.0 is subtracted from the calculated value so that a normal distribution has a kurtosis of 0. Additionally, the bias is set to False, ensuring that kurtosis is computed using k-statistics to eliminate any bias arising from moment estimators\footnote{https://docs.scipy.org/doc/scipy/reference/generated/scipy.stats.kurtosis.html}.

\textbf{4. Sortino Ratio} is a risk-adjusted performance metric that evaluates an investment's return relative to its downside risk, focusing solely on negative deviations from a specified target or required rate of return. It is calculated as:
\begin{equation}
    \text{Sortino Ratio} = \frac{\bar{\epsilon} - \text{MAR}}{\sigma_{\text{down}}},
\end{equation}
\begin{equation}
    \sigma_{\text{down}} = \sqrt{\frac{1}{n} \sum_{t=1}^{n} \min(\epsilon_t, 0)^2},
\end{equation}
\noindent where \( \bar{\epsilon} \) is the mean of residuals from the Fama-French Three-Factor Model, \( \sigma_{\text{down}} \) is the downside standard deviation. This formula considers only negative residuals, thereby focusing on downside risk. In this context, the minimum acceptable return (MAR) is 0, as the risk-free rate has already been subtracted when calculating the residuals.

In contrast, the Sharpe ratio assesses risk-adjusted return by considering total volatility, encompassing both upward and downward fluctuations. It is calculated by dividing the excess return (expected return minus risk-free rate) by the standard deviation of the investment's returns. The Sortino ratio refines this approach by isolating downside risk, aligning more closely with investors' primary concern of potential losses. A higher Sortino ratio indicates better risk-adjusted returns, as it signifies higher returns per unit of downside risk.

\subsection{Dataset Statistics}
\label{subsect:Data}
We collected and scraped SEC filings from the EDGAR database, specifically extracting the Management’s Discussion and Analysis (MD\&A) section using the open-source library edgar-crawler\footnote{https://github.com/lefterisloukas/edgar-crawler}. Since SEC filings identify companies using their Central Index Key (CIK) rather than traditional stock tickers, we must convert CIKs to ticker symbols. To achieve this, we utilized the official mapping provided by the SEC, available in JSON format\footnote{https://www.sec.gov/files/company\_tickers.json}. Once the ticker symbols were retrieved, we used them to obtain historical stock price data from Yahoo Finance via the yfinance API\footnote{https://github.com/ranaroussi/yfinance}. Using the extracted stock prices, we computed the risk measures as outlined in subsection \ref{subsect:RiskMeasure}, which provides the methodological framework for assessing financial risk based on stock price movements. 
\begin{table}[htb!]
\TABLE
{Yearly distribution of datapoints segmented into three percentile bins: 0-30\%, 30-70\%, and 70-100\%.\label{tab:data_count1}}
{\begin{tabular}{p{2cm} p{2cm} p{2cm} p{2cm} p{2cm}}
\hline\up
Year & 0-30\% & 30-70\% & 70-100\% & Total \\ \hline\up
        2024 & 1371 & 1828 & 1371 & 4570 \\
        2023 & 1336 & 1780 & 1336 & 4452 \\
        2022 & 1283 & 1711 & 1283 & 4277 \\
        2021 & 1169 & 1557 & 1169 & 3895 \\
        2020 & 1049 & 1398 & 1049 & 3496 \\
        2019 & 970  & 1294 & 970  & 3234 \\
        2018 & 921  & 1228 & 921  & 3070 \\
        2017 & 858  & 1144 & 858  & 2860 \\
        2016 & 821  & 1095 & 821  & 2737 \\
        2015 & 782  & 1043 & 782  & 2607 \\
        2014 & 734  & 979  & 734  & 2447 \\
        2013 & 690  & 920  & 690  & 2300 \down\\ \hline
\end{tabular}}{}
\end{table}

The input for our analysis is the MD\&A section of the company’s filings, while the labels correspond to one of four risk measures derived from the excess stock returns: standard deviation, skewness, kurtosis, and Sortino ratio. However, instead of directly predicting the raw risk values, we categorize the data into three bins based on percentile rankings of the calculated risk values: 0-30th percentile, 30-70th percentile, and 70-100th percentile. This binning approach addresses the inherent volatility and complexity of financial data, enhancing model performance and robustness. By focusing on relative rankings, we align our assessment with common practices in credit risk modeling \citep{Zeng2014ANC}, where categorization helps manage outliers and noisy data, leading to more reliable predictions. Consequently, we have reframed the problem from regression to ranking, emphasizing a company's performance relative to its peers, which is more insightful than analyzing raw risk scores. Table \ref{tab:data_count1} presents the distribution of data points across various bins for each year from 2013 to 2024.
\begin{table}[htb!]
\TABLE
{Summary of Training and Testing Dataset Splits Over Different Years.\label{tab:data_count2}}
{\begin{tabular}{c c c c}
\hline\up
Training Years & Testing Year & Number of Training Samples & Number of Testing Samples \\
\hline\up
2013-2017 & 2018 & 12951 & 3070 \\
2014-2018 & 2019 & 13721 & 3234 \\
2015-2019 & 2020 & 14508 & 3496 \\
2016-2020 & 2021 & 15397 & 3895 \\
2017-2021 & 2022 & 16555 & 4277 \\
2018-2022 & 2023 & 17972 & 4452 \\
2019-2023 & 2024 & 19354 & 4570 \down\\ \hline
\end{tabular}}{}
\end{table}

To evaluate TinyXRA's performance, we employ a rolling-origin evaluation strategy \citep{tashman2000out} that partitions the dataset such that each testing year is preceded by a five-year training period. This five-year window ensures models are trained on sufficient recent data while maintaining temporal relevance to the prediction year. The rolling-origin approach prevents data leakage by strictly maintaining chronological order where models can only use historical information to predict future outcomes, mimicking realistic deployment scenarios. Table \ref{tab:data_count2} illustrates this partitioning scheme, detailing the training and testing sample counts for each evaluation year from 2018 to 2024. This methodology enables robust assessment of model generalization across seven distinct time periods, each representing different market conditions and economic environments.

Moreover, we have included Figures \ref{fig:std_dev} to \ref{fig:sortino} in Appendix \ref{appendix:datasets} to illustrate the trends of each financial risk measure from 2013 to 2024 at the 30th, 50th, and 70th percentiles. This provides a clearer understanding of the overall dataset and its associated risk distribution. Overall, the risk distribution remains relatively stable, except for the Sortino ratio risk measurement setting in 2022, which shows a sharp increase. This suggests that most stocks experienced favorable returns during that year, likely driven by a post-COVID market recovery characterized by improved investor sentiment and economic rebound.
\section{Our TinyXRA Architecture}\label{sec:Method}

\begin{figure}[htb!]
     \FIGURE
     {\includegraphics[width=0.9\textwidth]{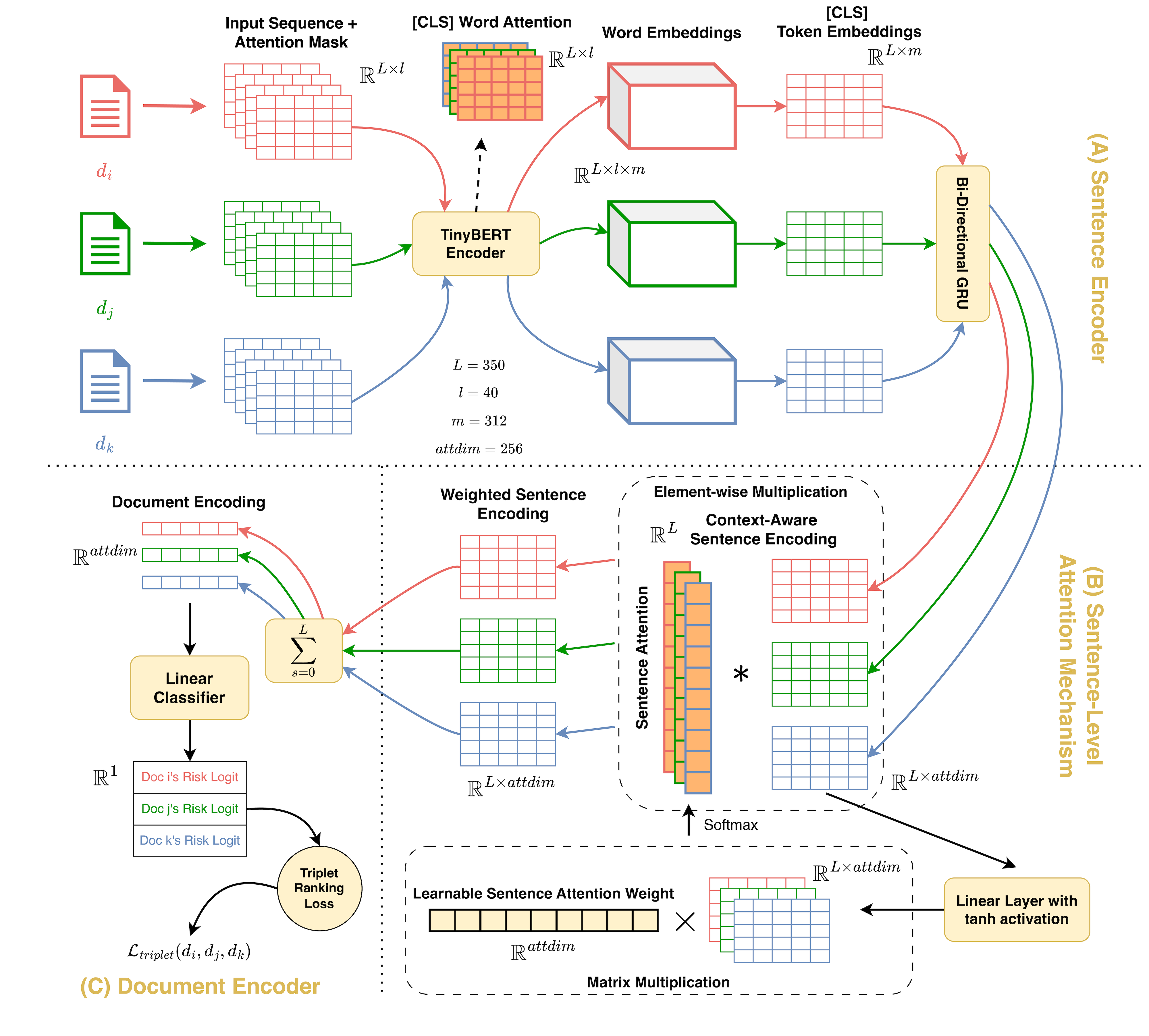}}
     {Our TinyXRA Architecture. \label{fig:architecture}}
     {}
\end{figure}

Figure \ref{fig:architecture} illustrates the overall structure of our TinyXRA model, which is designed for risk assessing in the financial domain while offering explainability for its prediction. We have divided the model into three main components: Sentence Encoder, Sentence-Level Attention Mechanism, Document Encoder.

\subsection{Sentence Encoder}
\subsubsection{Text Preprocessing}
As depicted in Figure~\ref{fig:architecture}(A), we begin with text preprocessing to prepare textual data for input into the TinyBERT module. This preprocessing involves several steps. Initially, raw text is \textit{tokenized} using the BERT tokenizer, converting it into tokens by breaking down words into subwords or characters and assigning unique numerical identifiers to each token. Subsequently, we introduce \textit{special tokens} such as \texttt{[CLS]} and \texttt{[SEP]} to structure the input appropriately. Specifically, the \texttt{[CLS]} token is placed at the beginning of each input sequence, and the \texttt{[SEP]} token marks the separation between different segments, typically appearing at the end of sequences or between sentence pairs. To maintain consistent sequence lengths, we employ \textit{padding and truncation} techniques; sequences shorter than the defined length are padded with the \texttt{[PAD]} token, while longer sequences are truncated to adhere to the model's maximum input constraint. In this study, we set the maximum sentence length at $l=40$ tokens and limit each document to a maximum of $L=350$ sentences. Documents exceeding this limit are truncated accordingly, whereas shorter documents are padded. Lastly, an \textit{attention mask} is created as a binary tensor of identical length to the input sequences, where tokens corresponding to actual input data are marked with ``1", and padding tokens are indicated by ``0".

\subsubsection{Sentence Embeddings}
TinyBERT \citep{jiao2020tinybert} is a compact and efficient variant of the well-known Bidirectional Encoder Representations from Transformers (BERT; \citealp{DBLP:conf/naacl/DevlinCLT19}) model. Developed through knowledge distillation, TinyBERT retains much of BERT’s natural language understanding capabilities while being significantly smaller and faster. Despite being approximately 7.5 times smaller and 9.4 times faster at inference than BERT, it achieves comparable performance across various natural language processing tasks. Notably, TinyBERT can run efficiently on GPUs with as little as 11 GB of VRAM (tested on an Nvidia GeForce RTX 2080 Ti) while handling lengthy Management Discussion and Analysis (MD\&A) content. This makes it particularly suitable for resource-constrained environments and tasks that require processing long-form inputs without sacrificing effectiveness.

As depicted in Figure \ref{fig:architecture}(A), we input the prepared data into TinyBERT to generate both word embeddings for the text and attention for the words. Word embeddings are continuous vector representations of words that effectively capture their semantic and syntactic properties. These representations enable the model to understand relationships and similarities between words in a given context. On the other hand, attention is a mechanism that assigns importance scores to different words in the text, allowing the model to selectively focus on specific words and their relationships with others. One key aspect of TinyBERT is the \texttt{[CLS]} token, a special symbol added at the beginning of each input sequence. This token is designed to capture the overall meaning of the input and is often used to represent the entire sequence in tasks like sentiment analysis and classification, hence it is the ideal candidate to represent the embeddings for each sentence. However, to ensure the best performance in our financial risk assessment task, we need to fine-tune TinyBERT on our dataset. Fine-tuning a pre-trained language model like TinyBERT involves training the model on a smaller, domain-specific dataset while employing a lower learning rate. This approach ensures that the language model (in this case TinyBERT)'s internal parameters are only slightly adjusted, allowing it to learn task-specific patterns and relationships without drastically altering the general linguistic knowledge acquired during pre-training on a large corpus.

The \texttt{[CLS]} embeddings, which represent the sentence embeddings, are subsequently passed through a bidirectional Gated Recurrent Unit ($\text{GRU}$; \citealp{69e088c8129341ac89810907fe6b1bfe}) to capture dependencies and information flow between sentences. This process results in context-aware sentence encoding. At the end of the process, we obtain the context-aware sentence encoding, represented as $h$:
\begin{equation}
h = \overrightarrow{\text{GRU}(s)} \oplus \overleftarrow{\text{GRU}(s)},
\end{equation}
\noindent where $s \in \mathbb{R}^{L \times m}$ is the sentence embeddings from the \texttt{[CLS]} token, $h \in \mathbb{R}^{L \times attdim}$ is the context-aware sentence encoding, and $m=312$ is the standard embedding size from TinyBERT. $L=350$ is the maximum number of sentences in a document and \(\textit{attdim} = 256\) as the dimensionality used for learning the sentence-level attention weights. $\oplus$ is the concatenation operator. The right arrow ($\rightarrow$) and left arrow ($\leftarrow$) on top of the operators and variables in the equation represent the forward and backward direction of the \text{GRU} respectively. The inclusion of the bidirectional \text{GRU} allows the model to better understand the relationships between sentences in a given document by capturing both forward and backward dependencies. This context-awareness is important as the meaning of a sentence is often influenced by the surrounding context. By combining both directions, the model can create a more comprehensive understanding of the relationships between sentences in the document, leading to improved performance in tasks like financial risk assessment.

\subsubsection{Word Level Attention Explanation}

\begin{figure}[htb!]
     \FIGURE
     {\includegraphics[width=0.8\textwidth]{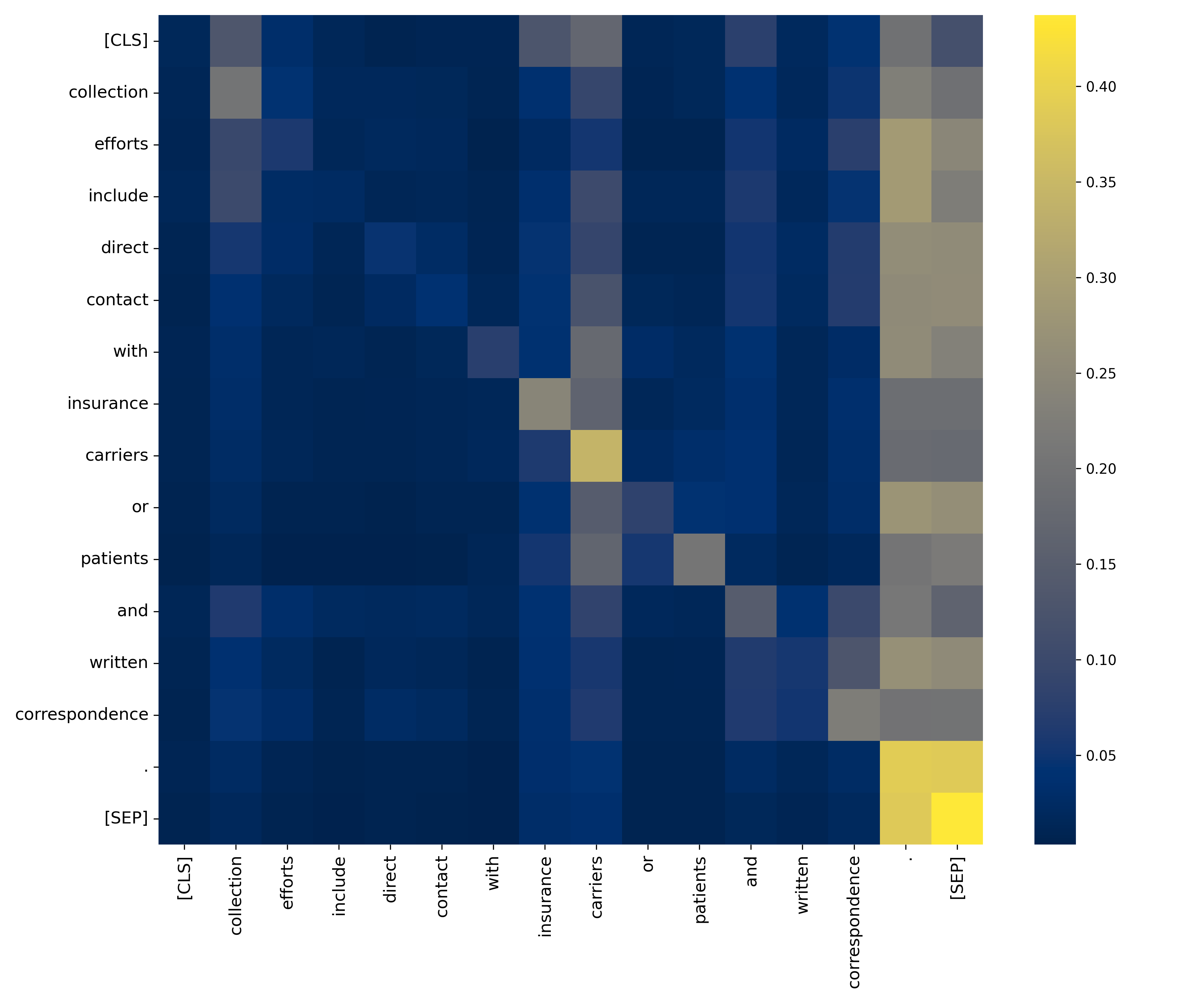}}
     {Raw attention heatmap of TinyBERT’s final layer \\ (Showing for all tokens in a sentence averaged across 12 heads). \label{fig:TinyBertAttn}}
     {}
\end{figure}

The attention output generated by TinyBERT captures pairwise attention values between words in the input text as shown in Figure~\ref{fig:TinyBertAttn}. Since we use the \texttt{[CLS]} token to represent the sentence embedding, it is logical to focus on the attention values associated with the \texttt{[CLS]} token when determining the importance of individual words. By analyzing which tokens receive the highest attention from the \texttt{[CLS]} token, we can effectively identify the most influential words in the sentence. 

Given that TinyBERT has multiple attention layers, we select the final layer for analysis, as it captures the most refined and contextually relevant information. Additionally, since TinyBERT employs multiple attention heads (12 in total), we compute the mean attention scores across these heads to obtain a more generalized measure of word importance. The presence of multiple attention heads allows the model to capture different aspects of word relationships, leading to a richer and more nuanced understanding of contextual dependencies.

This approach provides a structured method for deriving word-level explanations, offering insights into how TinyBERT prioritizes specific words when constructing its overall sentence representation. In the context of financial risk assessment, identifying the most influential words enhances interpretability, enabling users to better understand the model’s decision-making process. By highlighting key terms that contribute most significantly to the model's predictions, this method improves transparency and empowers users to make more informed and confident judgments based on the model’s outputs. To formalize this, we define the word-level attention score as follows:
\begin{equation}
\label{eq:attn_word}
\alpha_{w} = \frac{1}{H} \sum_{h=1}^{H} \alpha_{\texttt{CLS}}^{(h)} \ ,
\end{equation}
\noindent where $\alpha_{w} \in \mathbb{R}^{l}$ represents the mean attention score of tokens in a sentence with respect to the \texttt{[CLS]} token, and $H$ denotes the number of attention heads (12 in TinyBERT). The term $\alpha_{\texttt{CLS}}^{(h)}$ represents the attention score from the \texttt{[CLS]} token to each token in the sentence within the (h)-th attention head of the final layer. To enhance the quality of word-level explanations, we filter out special tokens, stopwords, and punctuation marks during the visualization step.

\subsection{Sentence-Level Attention Mechanism}
As seen in Figure \ref{fig:architecture}(B), the first step in computing attention scores involves applying a linear transformation to the context-aware sentence encoding $h \in \mathbb{R}^{L \times attdim}$. This transformation uses a weight matrix $W \in \mathbb{R}^{L \times attdim \times attdim}$ and a bias vector $b \in \mathbb{R}^{L \times attdim}$. The transformed sentence encoding is then processed by a non-linear activation function, such as the hyperbolic tangent (tanh), ensuring that the value $u$ remains within a specific range and introducing non-linearity into the model, as shown in equation \ref{eq:us}:
\begin{equation}
\label{eq:us}
u = \tanh( W \cdot h +b),
\end{equation}
\begin{equation}
\label{eq:tanh}
\tanh(x) = \frac{e^x - e^{-x}}{e^x + e^{-x}}.
\end{equation}
After obtaining the transformed sentence encoding $u \in \mathbb{R}^{L \times attdim}$, the next step is to calculate the attention scores. This involves taking the dot product between $u$ and a learnable sentence attention weight vector $U \in \mathbb{R}^{attdim}$ as seen in equation \ref{eq:attnscore}. The dot product measures the similarity between the transformed sentence encoding and the weight vector, with higher values indicating greater similarity and hence higher importance of the sentence in the document.
\begin{equation}
\label{eq:attnscore}
\alpha_{s} =\text{softmax}(u^{T} \cdot U).
\end{equation}
The attention scores are then normalized using the \text{softmax} function, converting the scores into a probability distribution. This normalization ensures that the attention scores sum up to 1, allowing them to be used as weights when combining sentence embeddings. The normalized attention score $\alpha_{s} \in \mathbb{R}^{L}$ also serves as the basis for the sentence-level explanation for the document. The \text{softmax} function is defined in Equation \ref{eq:softmax}.
\begin{equation}
\label{eq:softmax}
\text{softmax}(x_i) = \frac{\exp({x_i})}{\sum_{j} \exp({x_j})}.
\end{equation}
\subsection{Document Encoding}
As shown in Figure \ref{fig:architecture}(C), the attention scores from the previous subsection are used to weight the context-aware sentence embeddings. This is achieved by performing element-wise multiplication of the sentence embeddings with their attention scores. This results in a single, fixed-size document representation that captures the most relevant information from the input sentences, based on their attention scores. Subsequently, the weighted sentence embeddings are combined by summing them across the sentence dimension, as shown in Equation \ref{eq:doc_embs}:
\begin{equation}
\label{eq:doc_embs}
    \mathbf{d_{i}} = \sum^{L}_{s=1}\alpha_{s}h,
\end{equation}
\noindent where $\mathbf{d_i} \in \mathbb{R}^{attdim}$ represents the document embeddings for document i, $\alpha_s \in \mathbb{R}^{L}$ is the sentence attention scores, and $h \in \mathbb{R}^{L \times attdim}$ is the context-aware sentence embeddings. To formalize the transformation of the attended document representation into \textbf{risk logit}, we define a fully connected linear transformation applied to the document embedding \(\mathbf{d_i}\). The transformation is parameterized by a weight matrix \(\mathbf{W_d} \in \mathbb{R}^{attdim}\) and a bias vector \(\mathbf{b_d} \in \mathbb{R}^{1}\). The risk logit \(\mathbf{z_i}\) for document \(i\) are computed as follows:
\begin{equation}
\label{eq:risk_logits}
    \mathbf{z_i} = \mathbf{W_d} \mathbf{d_i} + \mathbf{b_d},
\end{equation}
\noindent where \(\mathbf{z_i} \in \mathbb{R}^{1}\) represents the risk logit associated with document \(i\). This logit will be used in the ranking loss function described in the following subsection.

\subsection{Ranking Loss}

Since we have partitioned our financial risk datasets into bins based on quartiles, it is essential to use ranking loss for training our model. A natural first choice for classification tasks is cross-entropy loss, particularly when we assume that there are three classes corresponding to the three bins. The cross-entropy loss is defined as follows:
\begin{equation}
    \mathcal{L}_{\text{CE}} = - \sum_{i=1}^{N} \sum_{j=0}^{K-1} y_{i,j} \log \hat{p}_{i,j},
\end{equation}
\noindent where \( N \) is the number of samples, \( K \) is the number of classes (bins), \( y_{i,j} \) is a one-hot encoded indicator that equals 1 if sample \( i \) belongs to class \( j \), and \( \hat{p}_{i,j} \) is the predicted probability for class \( j \). However, using a naive cross-entropy loss is suboptimal for this task because it treats each bin as an independent class. In reality, these bins possess an inherent ordinal structure (i.e., \(2 > 1 > 0\)), which cross-entropy does not account for. Ignoring this ordering could lead to inefficient learning, as the model would not be explicitly encouraged to respect the ranking relationships among the classes. Therefore, ranking loss is a more suitable choice for our model. That said, for completeness, we have included the cross-entropy loss results in our ablation study, which can be found in Section \ref{sec:Ablation}.

\subsubsection{Pairwise Ranking Loss} functions are widely used in learning-to-rank scenarios, where the goal is to optimize the ordering of items based on their relevance or importance. This approach is particularly effective in applications where the relative ordering of data points is essential, such as metric learning \citep{hoffer2015deep}, face verification \citep{chopra2005learning} and financial risk ranking \citep{10.1007/978-3-030-86514-6_16}. 

\begin{figure}[htb!]
     \FIGURE
     {\includegraphics[width=0.7\textwidth]{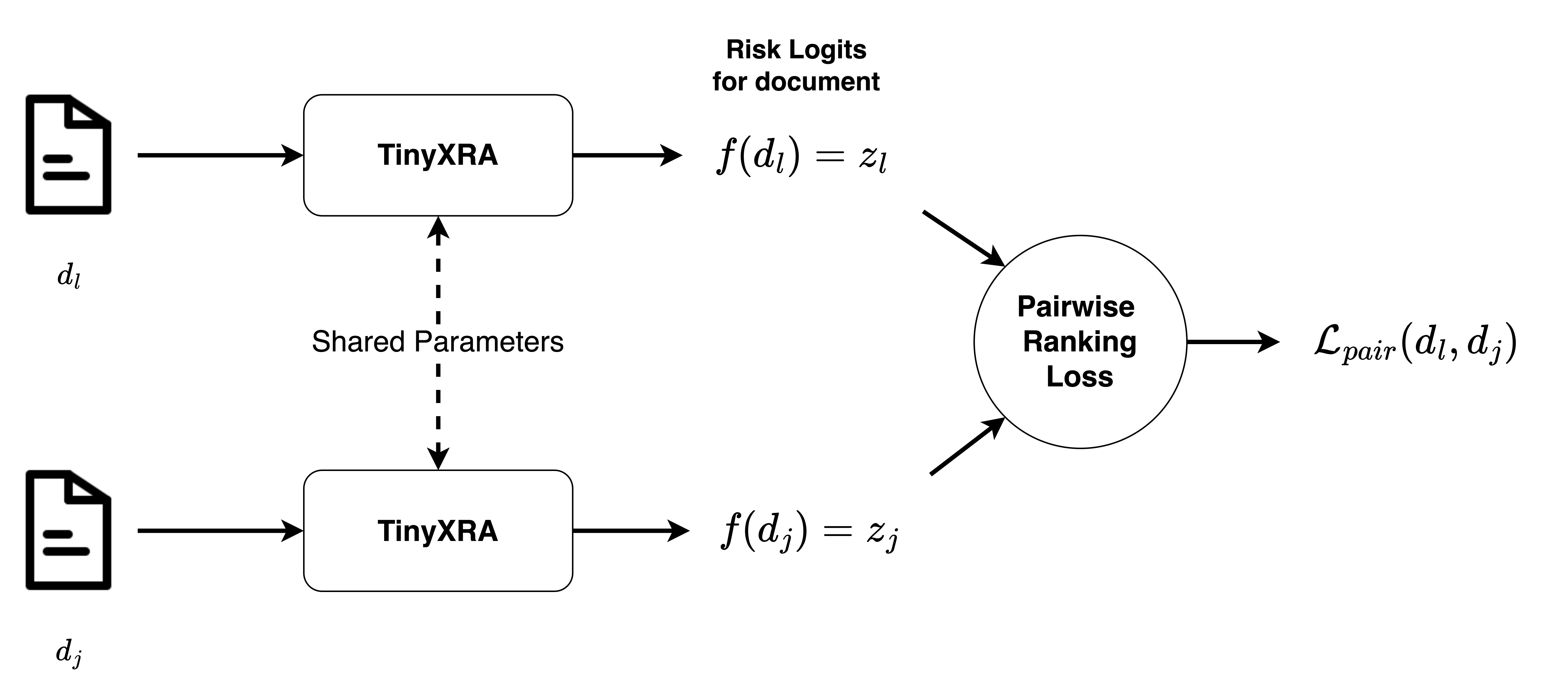}}
     {Diagram of a Siamese network with pairwise ranking loss. \label{fig:siamese}}
     {}
\end{figure}

This loss is often paired with a Siamese network architecture, which consists of two identical subnetworks that share parameters and process input pairs as seen in Figure \ref{fig:siamese}. The objective is to learn a scoring function \( f(d) \) that ranks data points based on their relative importance. Given a pair of inputs \( (d_l, d_j) \), the model assigns a higher score to \( d_l \) if it should be ranked higher than \( d_j \). The loss function promotes accurate pairwise ordering by maximizing the probability that correctly ranked pairs maintain their relative positions. The pairwise ranking loss is defined as follows:
\begin{equation}
\begin{aligned}
\mathcal{L}_{\text{pair}} &= - \sum_{(l, j) \in \mathcal{P}} 
\Big[ E(d_l, d_j) \log P_{lj} \\
&\quad + (1 - E(d_l, d_j)) \log (1 - P_{lj}) \Big],
\end{aligned}
\end{equation}
\noindent where \(\mathcal{P}\) is the set of all possible document pairs \((l, j)\),
\(E(d_l, d_j) = 1\) if \(d_l\) should be ranked higher than \(d_j\), otherwise \(0\). \(P_{lj}\) is the probability that \(d_l\) is ranked higher than \(d_j\), defined as:
\begin{equation}
    P_{lj} = \frac{\exp(f(d_l) - f(d_j))}{1 + \exp(f(d_l) - f(d_j))}.
\end{equation}
\noindent Here, \( f(d) \) represents the model's scoring function, corresponding to the risk logit produced by the TinyXRA model, where \( d \) denotes the candidate document. By leveraging the pairwise ranking loss within the Siamese network framework, the model effectively learns to distinguish between varying levels of financial risk, leading to more accurate and reliable risk assessments.
Pairwise ranking loss provides a useful way to enforce relative ordering between pairs of samples, but it does not differentiate between the severity of ranking mistakes. Specifically, it treats violations such as High Risk (2) being ranked lower than Low Risk (0), and High Risk (2) being ranked lower than Medium Risk (1) as equally incorrect. This lack of granularity can lead to suboptimal learning, as it does not account for the varying degrees of risk severity. Therefore, in the following subsubsection, we will introduce triplet ranking loss, which addresses this limitation by incorporating a finer-grained differentiation between ranking violations. The results for the pairwise ranking loss are also presented in Section \ref{sec:Ablation} as part of the ablation study.

\subsubsection{Triplet Ranking Loss} 
\begin{figure}[htb!]
     \FIGURE
     {\includegraphics[width=0.7\textwidth]{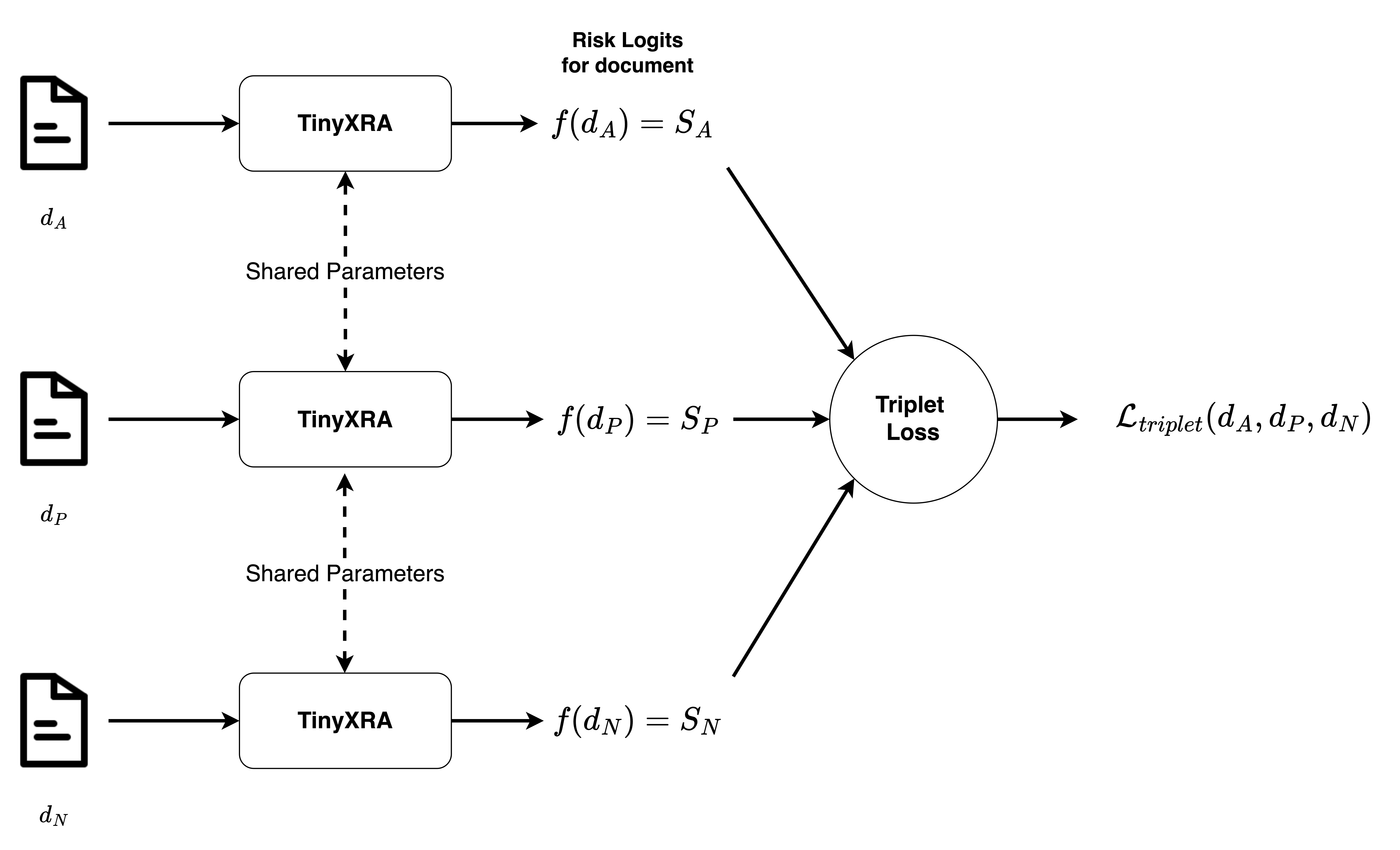}}
     {Diagram of a Triplet network with triplet ranking loss. \label{fig:triplet}}
     {}
\end{figure}

refines this by introducing an additional sample (anchor) and explicitly modeling the relationships among three different risk levels \citep{wang2014learning}. Instead of merely ensuring that a higher-risk sample is ranked above a lower-risk sample, triplet loss forces the model to learn a more structured ordering. This helps to ensure that High Risk (2) is not only ranked above Medium Risk (1) but also further away from Low Risk (0). This provides finer control over the learned representation and leads to a more robust ranking system. By leveraging triplet ranking loss, the model gains a more refined understanding of risk separation, making it particularly well-suited for financial risk ranking, where the degree of risk differentiation is critical for decision-making.The triplet ranking loss is defined as follows:
\begin{equation}
\begin{aligned}
\mathcal{L_{\text{triplet}}} &= \frac{1}{|\mathcal{T}|} \sum_{(A,P,N) \in \mathcal{T}} 
\Big[ \max(0, S_A - S_P + \delta) \\ 
&\quad + \max(0, S_N - S_A + \delta) \Big],
\end{aligned}
\end{equation}
where \(\mathcal{T}\) is the set of all valid triplets \((A, P, N)\), \(A\) \textbf{(Anchor)} is sampled from the \textbf{Medium Risk} (\(y=1\)) group, \(P\) \textbf{(Positive)} is sampled from the \textbf{High Risk} (\(y=2\)) group, \(N\) \textbf{(Negative)} is sampled from the \textbf{Low Risk} (\(y=0\)) group. To ensure the formation of valid triplets during each training epoch, we explicitly sample at least one instance from each risk group, guaranteeing a valid permutation of triplets. The remaining samples are selected randomly based on the batch size. \(S_A, S_P, S_N\) are the model's ranking scores (logits) for the anchor, positive, and negative samples respectively. \(\delta\) is the margin (default: 0.1), a larger margin enforces a stricter separation between the positive and negative samples relative to the anchor, encouraging the model to push positive examples further away from the anchor while pulling negative examples further away in the learned representation space. This helps improve ranking robustness by ensuring clearer decision boundaries between different risk levels. However, an excessively large margin can lead to overfitting, as the model may become overly confident in separating risk categories, reducing generalization to unseen data. \(\max(0, x)\) ensures that the loss is only incurred when the ordering constraint is violated. Specifically, if the model correctly ranks the samples such that \(S_P > S_A > S_N\) with a sufficient margin \(\delta\), the loss is zero. The triplet network structure is illustrated in Figure \ref{fig:triplet}.
\subsection{Attention-based Word Clouds}
To visualize the model’s focus on risk-indicative terms, we generate word clouds separately for groups of test documents. First, the model produces continuous risk predictions, which are discretized into three groups (e.g., low, medium, and high risk) using the 30th and 70th percentiles of the predicted scores. For each document, we obtain two types of attention scores: sentence-level attention, \( \alpha_s \), and word-level attention, \( \alpha_w \). Formally, let \( \alpha_{s,i} \) denote the attention weight assigned to the \( i^{th} \) sentence, and let \( \alpha_{w,i,j} \) represent the attention weight of the \( j^{th} \) token in the \( i^{th} \) sentence. The weighted importance for a token \( w_{i,j} \) (the \( j^{th} \) token in the \( i^{th} \) sentence) is computed as:
\begin{equation}
\gamma_{i,j} = \alpha_{s,i} \cdot \alpha_{w,i,j},
\end{equation}
where \( \gamma_{i,j} \) captures the combined influence of the sentence and the token within it. Rather than applying a fixed global threshold, each document \( d \) is processed individually. The set of weighted scores from all tokens in the document is flattened, and a document-specific threshold \( \tau_d \) is computed as the \( (1-K) \) quantile (with \( K = 30\% \) by default). Tokens in document \( d \) are then selected according to:
\begin{equation}
\mathcal{W}_d = \{ w_{i,j} \mid \gamma_{i,j} > \tau_d \}.
\end{equation}
Before aggregation, tokens are cleaned by removing those that are stop words or consist solely of punctuation. For each risk group, the overall “frequency” of a token is not merely its count but the sum of its attention-weighted scores across all selected occurrences in documents:
\begin{equation}
\text{freq}(w) = \sum_{d \in \mathcal{D}} \sum_{w_{i,j} \in \mathcal{W}_d} \gamma_{i,j},
\end{equation}
where \( \mathcal{D} \) denotes the set of documents in the group. To further emphasize tokens that are unique to a particular risk category, for any token appearing in multiple groups the frequency in group \( g \) is adjusted by subtracting the maximum frequency of that token found in the other groups:
\begin{equation}
\text{freq}^{new}_{g}(w) = \text{freq}_g(w) - \max_{h \neq g}( \text{freq}_h(w)).
\end{equation}
Tokens with $\text{freq}^{new}_g(w) \leq 0$ are subsequently discarded to retain only group-distinctive terms. The resulting word clouds are generated using these adjusted frequencies, with tokens receiving higher cumulative attention-weighted importance and greater group specificity displayed more prominently. This dynamic, attention-based approach provides more domain-specific and risk-aware visualizations compared to traditional frequency-based methods such as TF-IDF, which rely solely on word occurrence counts without considering contextual importance or model focus.
\section{Results \& Discussion}\label{sec:Results}

\subsection{Evaluation Metrics}

The evaluation of a model's performance is an important step in any machine learning study. Assessing the effectiveness of a model not only helps us understand its strengths and weaknesses but also enables us to identify areas for improvement and make informed decisions when selecting the most appropriate model for a given task. Furthermore, reliable evaluation metrics allow us to compare different models and approaches objectively, ensuring that we develop robust and accurate solutions for the problems we aim to solve.

\subsubsection{F1 Score}  is a widely used performance metric, with two main variants: Micro F1 and Macro F1. The Micro F1 score, which is equivalent to accuracy, measures the overall proportion of correct predictions relative to the total number of predictions made by the model. In contrast, the Macro F1 score calculates the harmonic mean of precision and recall across all classes, treating each class equally. Precision is defined as the ratio of true positive predictions to the total number of positive predictions made, while recall measures the ratio of true positive predictions to the total number of actual positive instances in the dataset. The Macro F1 score is defined as follows:
\begin{equation}
    \text{F1}_c = \frac{2 \cdot \text{Precision}_c \cdot \text{Recall}_c}{\text{Precision}_c + \text{Recall}_c},
\end{equation}
\begin{equation}
    \text{Precision}_c = \frac{\text{TP}_c}{\text{TP}_c + \text{FP}_c}, \quad
    \text{Recall}_c = \frac{\text{TP}_c}{\text{TP}_c + \text{FN}_c},
\end{equation}
\begin{equation}
\text{F1}_{\text{Macro}} = \frac{1}{C} \sum_{c=1}^{C} \text{F1}_c,
\end{equation}
\noindent where \( \text{TP}_c \) is the true Positives for class \( c \), \( \text{FP}_c \) False Positives for class \( c \), \( \text{FN}_c \) is the False Negatives for class \( c \). \( C \) is the total number of classes (\( C \) = 3 for our context). \( \text{F1}_c \) is the F1 score for class \( c \). Given that our dataset is slightly imbalanced, with the middle bin containing slightly more samples, the Macro F1 score serves as a more suitable evaluation metric for our study therefore \textbf{we will only show the Macro F1 score}. Since TinyXRA outputs risk logits rather than directly classifying documents into discrete classes, we first generate the risk logits. We then categorize them into three bins based on predefined thresholds (0-30\%, 30-70\%, 70-100\%) to classify them accordingly. However, since this is a ranking task, the F1 score alone is not sufficient to evaluate our model's performance. Therefore, we will also introduce two additional performance metrics: Spearman's Rho ($\rho$) and Kendall's Tau ($\tau$).

\subsubsection{Spearman's Rho} assesses the \textbf{\textit{monotonic relationship}} between two variables \citep{myers2013research}. It is computed by calculating the Pearson correlation coefficient between the ranked values of the variables rather than their raw values. The formula for Spearman’s Rho ($\rho$) is:
\begin{equation}
    \rho = 1 - \frac{6 \sum d_i^2}{n(n^2 - 1)},
\end{equation}
\noindent where \( d_i \) is the difference between the ranks of corresponding values in the ground truths and predicted values of each financial document. \( n \) is the number of observations. For example, if document A has a ground truth rank of 1 but a predicted rank of 3, the rank difference is calculated as \( d_A = 1 - 3 = -2 \). \( \rho = +1 \) indicates a perfect positive monotonic relationship (i.e., as one variable increases, the other also increases). \( \rho = -1 \) indicates a perfect negative monotonic relationship (i.e., as one variable increases, the other decreases). \( \rho = 0 \) suggests no monotonic correlation. Unlike Pearson’s correlation, Spearman’s Rho does not assume a linear relationship, making it more robust to non-linear associations. For the ground truth rankings will remain as 0, 1, and 2, while for the predicted values we will use the raw predicted risk logit values.

\subsubsection{Kendall's Tau} measures the \textbf{\textit{ordinal association}} between two variables based on the number of concordant and discordant pairs \citep{10.2307/2332226}. In this study, we are using the Kendall's Tau-b,  which is an extension of the standard Kendall’s Tau that adjusts for ties in rankings \citep{kendall1945treatment}. This version is particularly useful when dealing with datasets where ties (equal ranks) are present or smaller datasets where ties are more common. The Kendall's Tau-b ($\tau_b$) is defined as:
\begin{equation}
    \tau_b = \frac{P - Q}{\sqrt{(P + Q + T) (P + Q + U)}},
\end{equation}
\noindent where \( P \) (Concordant Pairs) is the number of pairs where the ranking order is the same in both the ground truth and predictions (i.e., if document A ranks higher than document B in the ground truth, it also ranks higher in the predictions). \( Q \) (Discordant Pairs) is the number of pairs where the ranking order is reversed between the ground truth and predictions (i.e., if document A ranks higher than document B in the ground truth but ranks lower in the predictions, it is a discordant pair). \( T \) (Ties in the Ground Truth) is the number of instances where two or more documents have the same rank in the ground truth.  \( U \) (Ties in the Predictions) is the number of instances where two or more documents have the same rank in the predictions.  \( \tau_b = +1 \) means there is perfect agreement in ranking (all pairs are concordant).  \( \tau_b = -1 \) means there is perfect disagreement in ranking (all pairs are discordant).  \( \tau_b = 0 \) means there is no correlation in ranking order. The inclusion of \( T \) and \( U \) in the denominator adjusts for ties, making Kendall's Tau-b more robust for datasets where ties are common. Similar to Spearman’s Rho, the ground truth rankings will remain as 0, 1, and 2, while for the predicted values we will use the raw predicted risk logit values.

\subsection{Benchmark Models}
To evaluate the performance of our TinyXRA models comprehensively, we benchmark them against a carefully selected set of comparative approaches: XRR, TF-IDF, Llama3.2-1B, and Qwen2-0.5B. Each model represents different paradigms in document analysis and classification, allowing us to situate our approach within the broader landscape of risk assessment technologies. Below, we provide detailed descriptions of each benchmark model and our methodology for comparison.

\textbf{XRR} (Explainable Risk Ranking; \citealp{10.1007/978-3-030-86514-6_16}) represents the current state-of-the-art in explainable risk ranking for financial reports. This model employs static word embeddings that have been specifically fine-tuned for risk-ranking tasks. Similar to our TinyXRA architecture, XRR implements a hierarchical structure for document processing, which allows it to capture both word-level and sentence-level features. XRR utilizes a pairwise ranking loss function that optimizes rankings by comparing document instances two at a time. 

\textbf{TF-IDF} is a traditional baseline method that offers computational efficiency and interpretability advantages. This approach relies exclusively on term frequency statistics within documents, weighted by the inverse document frequency across the corpus. While TF-IDF fails to capture semantic relationships between words or contextual meanings, it remains widely deployed in industry settings, particularly in resource-constrained environments. Its inclusion in our benchmark provides perspective on how much improvement our neural approaches offer over statistical methods that have been the industry standard for decades. We implement TF-IDF followed by a simple classifier to determine risk categories.

Given the dramatic impact of large language models on natural language processing tasks, we include two lightweight LLMs in our evaluation to ensure our research remains relevant to current technological trends. \textbf{Llama3.2-1B} is developed by Meta \citep{grattafiori2024llama}, represents a compact yet powerful iteration in the Llama model family. Despite its relatively small size (1 billion parameters), this model has demonstrated impressive performance across various NLP tasks. Llama3.2-1B employs a transformer-based architecture with optimizations for efficiency, making it suitable for deployment in environments with limited computational resources. Its inclusion allows us to compare our specialized TinyXRA models against a general-purpose LLM that has been pre-trained on diverse text corpora but not specifically optimized for financial risk assessment. \textbf{Qwen2-0.5B} is developed by Alibaba Cloud \citep{yang2024qwen2}, is an even more compact model with only 0.5 billion parameters. This model represents the frontier of ultra-lightweight LLMs that maintain reasonable performance while drastically reducing computational requirements. Qwen2-0.5B incorporates architectural innovations designed to maximize efficiency, including optimized attention mechanisms and parameter sharing techniques. We include this model to determine whether extremely compact general-purpose LLMs can compete with specialized architectures in domain-specific tasks.

To ensure fair comparison with our resource-efficient TinyXRA models, we constrain all evaluations to run within 11GB of GPU VRAM, corresponding to consumer-grade hardware (specifically, our NVIDIA GeForce RTX 2080 GPU). For the LLM-based models, we employ two distinct assessment strategies. Our first evaluation approach leverages the in-context learning capabilities of LLMs without requiring model parameter updates. For each validation document, we randomly select one example from each of the three risk ranking categories in our training dataset. These examples serve as demonstrations that provide the model with context about the task requirements. We then prompt the model to predict the risk rank of the validation document. This is also widely known as \textbf{1-shot learning}. A significant challenge in this approach is the handling of long financial documents. The average 10-K report in our dataset comprises approximately 14,000 tokens, meaning that including four documents (three examples plus the target document) would require processing 56,000 input tokens, which is well beyond the capability of our hardware-constrained setup. To address this limitation, we truncate documents to the first 7,000 tokens. While this truncation represents a compromise, it enables us to evaluate in-context learning performance within our hardware constraints.

Our second evaluation strategy employs a parameter-efficient \textbf{supervised fine-tuning} approach. Rather than updating the entire parameter set of the LLMs (which would be prohibitively expensive in terms of hardware requirements), we freeze the pre-trained models and use them solely as document encoder. Specifically, we pass each document through the frozen LLM backbone to generate token-level embeddings, which we then mean-pool across all tokens to obtain a fixed-length document representation. This document embedding is then fed into a trainable classifier head, consisting of a fully connected layer, that outputs logits corresponding to our three risk ranking categories. Despite this computationally efficient design that avoids updating the large parameter sets of the base models, the inference and training process remains resource-intensive (likewise we have to truncate the documents to the first 7,000 tokens). Our empirical measurements indicate that training with these LLM-based approaches requires approximately 6-7 times longer than training TinyXRA under identical hardware conditions, which clearly demonstrates the significant efficiency advantages of our proposed architecture.

Through these comprehensive benchmarks, we aim to provide a clear picture of how TinyXRA models perform relative to both traditional approaches and newer neural architectures when addressing the specialized task of financial risk assessment.

\subsection{Experimental Results}
\subsubsection{Experimental Setup}
The datasets used for training and evaluation are detailed in Table \ref{tab:data_count2} in Subsection \ref{subsect:Data}. To ensure a fair and comprehensive comparison, we benchmark TinyXRA against XRR, the state-of-the-art explainable ranking model \citep{10.1007/978-3-030-86514-6_16}, and TF-IDF, a classic but still widely utilized statistical approach \citep{rajaraman2011mining} that offers computational efficiency in resource-constrained environments. Additionally, we include the lightweight LLMs described in the previous section to contextualize our results against recent advancements in language model technology. For reproducibility, we provide comprehensive hyperparameter configurations for all models in Appendix \ref{appendix:hyperparam}. All experimental results represent the mean and standard deviation of test performance across five different random seeds (98, 83, 62, 42, 21).

Recognizing that financial prediction tasks inherently involve considerable data noise, on top of means and standard deviations reporting, we have added statistical analyses to more accurately quantify performance differences between models. We employed paired Cohen's d tests to measure the effect size of performance differences across different random seeds, risk measurement metrics and test years. The formula for \textbf{Cohen's d for paired samples} is: 
\begin{equation}
    d = \frac{\bar{d}}{s_d}
\end{equation}
where $\bar{d}$ is the mean of the differences between paired samples, $s_d$ is the standard deviation of those differences. In our analysis framework, a positive Cohen's d value indicates that TinyXRA outperforms the benchmark model on the corresponding metric, while a negative value signifies that TinyXRA underperforms relative to the benchmark. The magnitude of the Cohen's d value represents the effect size, with larger absolute values indicating more substantial performance differences. This statistical approach provides a more nuanced understanding of model performance beyond simple comparison of average metric values, given the inherent volatility and noise in financial data which will obscure the true performance differentials between models.

\subsubsection{Experimental Findings}

In this section, we present our experimental results, evaluating TinyXRA against the various benchmark models across multiple risk measurement metrics from 2018 to 2024. Specifically, we report results for standard deviation (Table \ref{tab:std}), skewness (Table \ref{tab:skew}), kurtosis (Table \ref{tab:kurt}), and the Sortino ratio (Table \ref{tab:sortino}). To assess model performance, we use three evaluation metrics: F1 Score, Spearman’s Rho, and Kendall’s Tau. The paired Cohen's d tests relative to TinyXRA are summarized in Table \ref{tab:cohendresults}. The datasets used for training and testing are detailed in Table \ref{tab:data_count2} in Subsection \ref{subsect:Data}.

\begin{table}[htb!]
\TABLE
{F1 Score (\%), Spearman’s Rho (\%), and Kendall’s Tau (\%) for various benchmark models across test years (2024–2018) using standard deviation risk measurement.\label{tab:std}}
{\begin{tabular}{@{}l l l l l l l l l@{}}
\hline\up
Evaluation Metrics & Models & 2024 & 2023 & 2022 & 2021 & 2020 & 2019 & 2018 \\ \hline\up
F1 Score & TinyXRA & \textbf{76.0}$_{\scalebox{0.7}{\(\pm 0.3\)}}$ & \textbf{74.5}$_{\scalebox{0.7}{\(\pm 0.2\)}}$ & \underline{75.1}$_{\scalebox{0.7}{\(\pm 0.5\)}}$ & \underline{74.1}$_{\scalebox{0.7}{\(\pm 0.3\)}}$ & \textbf{67.6}$_{\scalebox{0.7}{\(\pm 0.7\)}}$ & \textbf{76.5}$_{\scalebox{0.7}{\(\pm 0.3\)}}$ & \underline{78.1}$_{\scalebox{0.7}{\(\pm 0.2\)}}$ \\
& XRR & 73.4$_{\scalebox{0.7}{\(\pm 0.5\)}}$ & 71.7$_{\scalebox{0.7}{\(\pm 0.3\)}}$ & 72.8$_{\scalebox{0.7}{\(\pm 0.5\)}}$ & 72.2$_{\scalebox{0.7}{\(\pm 0.5\)}}$ & \underline{65.8}$_{\scalebox{0.7}{\(\pm 0.6\)}}$ & 73.1$_{\scalebox{0.7}{\(\pm 0.3\)}}$ & 75.2$_{\scalebox{0.7}{\(\pm 0.5\)}}$ \\
& TF-IDF & \underline{74.6}$_{\scalebox{0.7}{\(\pm 0.0\)}}$ & \underline{73.5}$_{\scalebox{0.7}{\(\pm 0.0\)}}$ & \textbf{75.8}$_{\scalebox{0.7}{\(\pm 0.0\)}}$ & \textbf{75.4}$_{\scalebox{0.7}{\(\pm 0.1\)}}$ & 63.9$_{\scalebox{0.7}{\(\pm 0.1\)}}$ & \underline{75.7}$_{\scalebox{0.7}{\(\pm 0.0\)}}$ & \textbf{78.3}$_{\scalebox{0.7}{\(\pm 0.0\)}}$ \\
& Llama3.2-1B (1-Shot) & 23.4$_{\scalebox{0.7}{\( \pm 0.0\)}}$ & 21.8$_{\scalebox{0.7}{\( \pm 0.0\)}}$ & 21.9$_{\scalebox{0.7}{\( \pm 0.0\)}}$ & 20.7$_{\scalebox{0.7}{\( \pm 0.0\)}}$ & 20.8$_{\scalebox{0.7}{\( \pm 0.0\)}}$ & 21.2$_{\scalebox{0.7}{\( \pm 0.0\)}}$ & 19.5$_{\scalebox{0.7}{\( \pm 0.0\)}}$ \\ 
& Llama3.2-1B (SFT) & 73.2$_{\scalebox{0.7}{\( \pm 0.0\)}}$ & 69.5$_{\scalebox{0.7}{\( \pm 0.0\)}}$ & 73.4$_{\scalebox{0.7}{\( \pm 0.0\)}}$ & 71.3$_{\scalebox{0.7}{\( \pm 0.0\)}}$ & 63.2$_{\scalebox{0.7}{\( \pm 0.0\)}}$ & 74.0$_{\scalebox{0.7}{\( \pm 0.0\)}}$ & 74.0$_{\scalebox{0.7}{\( \pm 0.0\)}}$ \\
& Qwen2-0.5B (1-Shot) & 25.3$_{\scalebox{0.7}{\(\pm 0.0\)}}$ & 24.9$_{\scalebox{0.7}{\(\pm 0.0\)}}$ & 24.0$_{\scalebox{0.7}{\(\pm 0.0\)}}$ & 24.5$_{\scalebox{0.7}{\(\pm 0.0\)}}$ & 25.0$_{\scalebox{0.7}{\(\pm 0.0\)}}$ & 24.1$_{\scalebox{0.7}{\(\pm 0.0\)}}$ & 24.5$_{\scalebox{0.7}{\(\pm 0.0\)}}$ \\
& Qwen2-0.5B (SFT) & 70.3$_{\scalebox{0.7}{\(\pm 0.0\)}}$ & 69.2$_{\scalebox{0.7}{\(\pm 0.0\)}}$ & 71.8$_{\scalebox{0.7}{\(\pm 0.0\)}}$ & 72.5$_{\scalebox{0.7}{\(\pm 0.0\)}}$ & 61.1$_{\scalebox{0.7}{\(\pm 0.0\)}}$ & 73.5$_{\scalebox{0.7}{\(\pm 0.0\)}}$ & 74.2$_{\scalebox{0.7}{\(\pm 0.0\)}}$ \\ \up

Spearman’s Rho & TinyXRA & \textbf{80.3}$_{\scalebox{0.7}{\(\pm 0.1\)}}$ & \textbf{78.8}$_{\scalebox{0.7}{\(\pm 0.3\)}}$ & \textbf{80.0}$_{\scalebox{0.7}{\(\pm 0.2\)}}$ & \underline{80.1}$_{\scalebox{0.7}{\(\pm 0.2\)}}$ & \textbf{72.7}$_{\scalebox{0.7}{\(\pm 0.9\)}}$ & \textbf{81.2}$_{\scalebox{0.7}{\(\pm 0.4\)}}$ & \textbf{82.9}$_{\scalebox{0.7}{\(\pm 0.1\)}}$ \\
& XRR & \underline{77.7}$_{\scalebox{0.7}{\(\pm 0.6\)}}$ & 76.2$_{\scalebox{0.7}{\(\pm 0.5\)}}$ & 77.4$_{\scalebox{0.7}{\(\pm 0.5\)}}$ & 77.4$_{\scalebox{0.7}{\(\pm 0.8\)}}$ & \underline{70.5}$_{\scalebox{0.7}{\(\pm 0.9\)}}$ & 78.2$_{\scalebox{0.7}{\(\pm 0.4\)}}$ & 80.1$_{\scalebox{0.7}{\(\pm 0.5\)}}$ \\
& TF-IDF & 77.4$_{\scalebox{0.7}{\(\pm 0.0\)}}$ & \underline{76.3}$_{\scalebox{0.7}{\(\pm 0.0\)}}$ & \underline{78.9}$_{\scalebox{0.7}{\(\pm 0.0\)}}$ & \textbf{80.5}$_{\scalebox{0.7}{\(\pm 0.0\)}}$ & 70.2$_{\scalebox{0.7}{\(\pm 0.0\)}}$ & \underline{79.6}$_{\scalebox{0.7}{\(\pm 0.0\)}}$ & \underline{81.1}$_{\scalebox{0.7}{\(\pm 0.0\)}}$ \\ 
& Llama3.2-1B (1-Shot) & -9.0$_{\scalebox{0.7}{\(\pm 0.0\)}}$ & -11.0$_{\scalebox{0.7}{\(\pm 0.0\)}}$ & -12.2$_{\scalebox{0.7}{\(\pm 0.0\)}}$ & -13.7$_{\scalebox{0.7}{\(\pm 0.0\)}}$ & -11.2$_{\scalebox{0.7}{\(\pm 0.0\)}}$ & -8.9$_{\scalebox{0.7}{\(\pm 0.0\)}}$ & -15.1$_{\scalebox{0.7}{\(\pm 0.0\)}}$ \\ 
& Llama3.2-1B (SFT) & 76.5$_{\scalebox{0.7}{\(\pm 0.0\)}}$ & 73.9$_{\scalebox{0.7}{\(\pm 0.0\)}}$ & 76.5$_{\scalebox{0.7}{\(\pm 0.0\)}}$ & 79.2$_{\scalebox{0.7}{\(\pm 0.0\)}}$ & 68.2$_{\scalebox{0.7}{\(\pm 0.0\)}}$ & 78.0$_{\scalebox{0.7}{\(\pm 0.0\)}}$ & 78.9$_{\scalebox{0.7}{\(\pm 0.0\)}}$ \\
& Qwen2-0.5B (1-Shot) & -2.7$_{\scalebox{0.7}{\(\pm 0.0\)}}$ & -4.6$_{\scalebox{0.7}{\(\pm 0.0\)}}$ & -5.7$_{\scalebox{0.7}{\(\pm 0.0\)}}$ & -2.7$_{\scalebox{0.7}{\(\pm 0.0\)}}$ & -3.9$_{\scalebox{0.7}{\(\pm 0.0\)}}$ & -3.6$_{\scalebox{0.7}{\(\pm 0.0\)}}$ & -3.2$_{\scalebox{0.7}{\(\pm 0.0\)}}$ \\ 
& Qwen2-0.5B (SFT) & 75.1$_{\scalebox{0.7}{\(\pm 0.0\)}}$ & 73.3$_{\scalebox{0.7}{\(\pm 0.0\)}}$ & 76.0$_{\scalebox{0.7}{\(\pm 0.0\)}}$ & 78.3$_{\scalebox{0.7}{\(\pm 0.0\)}}$ & 68.2$_{\scalebox{0.7}{\(\pm 0.0\)}}$ & 77.0$_{\scalebox{0.7}{\(\pm 0.0\)}}$ & 77.7$_{\scalebox{0.7}{\(\pm 0.0\)}}$ \\ \up

Kendall’s Tau & TinyXRA & \textbf{67.2}$_{\scalebox{0.7}{\(\pm 0.1\)}}$ & \textbf{65.7}$_{\scalebox{0.7}{\(\pm 0.3\)}}$ & \textbf{66.9}$_{\scalebox{0.7}{\(\pm 0.2\)}}$ & \underline{66.8}$_{\scalebox{0.7}{\(\pm 0.2\)}}$ & \textbf{59.4}$_{\scalebox{0.7}{\(\pm 0.9\)}}$ & \textbf{67.8}$_{\scalebox{0.7}{\(\pm 0.3\)}}$ & \textbf{69.6}$_{\scalebox{0.7}{\(\pm 0.1\)}}$ \\
& XRR & \underline{64.6}$_{\scalebox{0.7}{\(\pm 0.6\)}}$ & 63.1$_{\scalebox{0.7}{\(\pm 0.4\)}}$ & 64.2$_{\scalebox{0.7}{\(\pm 0.5\)}}$ & 64.0$_{\scalebox{0.7}{\(\pm 0.8\)}}$ & \underline{57.2}$_{\scalebox{0.7}{\(\pm 0.8\)}}$ & 64.8$_{\scalebox{0.7}{\(\pm 0.4\)}}$ & 66.7$_{\scalebox{0.7}{\(\pm 0.4\)}}$ \\
& TF-IDF & 64.5$_{\scalebox{0.7}{\(\pm 0.0\)}}$ & \underline{63.3}$_{\scalebox{0.7}{\(\pm 0.0\)}}$ & \underline{66.0}$_{\scalebox{0.7}{\(\pm 0.0\)}}$ & \textbf{67.2}$_{\scalebox{0.7}{\(\pm 0.0\)}}$ & 56.9$_{\scalebox{0.7}{\(\pm 0.0\)}}$ & \underline{66.3}$_{\scalebox{0.7}{\(\pm 0.0\)}}$ & \underline{67.8}$_{\scalebox{0.7}{\(\pm 0.0\)}}$ \\ 
& Llama3.2-1B (1-Shot) & -6.9$_{\scalebox{0.7}{\(\pm 0.0\)}}$ & -8.4$_{\scalebox{0.7}{\(\pm 0.0\)}}$ & -9.4$_{\scalebox{0.7}{\(\pm 0.0\)}}$ & -10.5$_{\scalebox{0.7}{\(\pm 0.0\)}}$ & -8.6$_{\scalebox{0.7}{\(\pm 0.0\)}}$ & -6.8$_{\scalebox{0.7}{\(\pm 0.0\)}}$ & -11.6$_{\scalebox{0.7}{\(\pm 0.0\)}}$ \\ 
& Llama3.2-1B (SFT) & 63.8$_{\scalebox{0.7}{\(\pm 0.0\)}}$ & 61.1$_{\scalebox{0.7}{\(\pm 0.0\)}}$ & 63.9$_{\scalebox{0.7}{\(\pm 0.0\)}}$ & 65.9$_{\scalebox{0.7}{\(\pm 0.0\)}}$ & 55.0$_{\scalebox{0.7}{\(\pm 0.0\)}}$ & 64.7$_{\scalebox{0.7}{\(\pm 0.0\)}}$ & 65.6$_{\scalebox{0.7}{\(\pm 0.0\)}}$ \\
& Qwen2-0.5B (1-Shot) & -2.1$_{\scalebox{0.7}{\(\pm 0.0\)}}$ & -3.5$_{\scalebox{0.7}{\(\pm 0.0\)}}$ & -4.4$_{\scalebox{0.7}{\(\pm 0.0\)}}$ & -2.1$_{\scalebox{0.7}{\(\pm 0.0\)}}$ & -3.0$_{\scalebox{0.7}{\(\pm 0.0\)}}$ & -2.8$_{\scalebox{0.7}{\(\pm 0.0\)}}$ & -2.5$_{\scalebox{0.7}{\(\pm 0.0\)}}$ \\ 
& Qwen2-0.5B (SFT) & 62.3$_{\scalebox{0.7}{\(\pm 0.0\)}}$ & 60.7$_{\scalebox{0.7}{\(\pm 0.0\)}}$ & 63.3$_{\scalebox{0.7}{\(\pm 0.0\)}}$ & 65.1$_{\scalebox{0.7}{\(\pm 0.0\)}}$ & 55.2$_{\scalebox{0.7}{\(\pm 0.0\)}}$ & 63.8$_{\scalebox{0.7}{\(\pm 0.0\)}}$ & 64.6$_{\scalebox{0.7}{\(\pm 0.0\)}}$ \down\\ \hline
\end{tabular}}{Results are reported as the mean ± standard deviation across five random seeds (98, 83, 62, 42, 21).}
\end{table}

\begin{table}[htb!]
\TABLE
{F1 Score (\%), Spearman’s Rho (\%), and Kendall’s Tau (\%) for various benchmark models across test years (2024–2018) using skewness risk measurement.\label{tab:skew}}
{\begin{tabular}{@{}l l l l l l l l l@{}}
\hline\up
Evaluation Metrics & Models & 2024 & 2023 & 2022 & 2021 & 2020 & 2019 & 2018 \\ \hline\up
F1 Score & TinyXRA & \underline{44.4}$_{\scalebox{0.7}{\(\pm 0.2\)}}$ & 44.1$_{\scalebox{0.7}{\(\pm 0.6\)}}$ & 46.8$_{\scalebox{0.7}{\(\pm 0.5\)}}$ & \textbf{47.1}$_{\scalebox{0.7}{\(\pm 0.4\)}}$ & \textbf{47.8}$_{\scalebox{0.7}{\(\pm 0.4\)}}$ & 45.1$_{\scalebox{0.7}{\(\pm 0.4\)}}$ & 46.0$_{\scalebox{0.7}{\(\pm 0.3\)}}$ \\
& XRR & 43.0$_{\scalebox{0.7}{\(\pm 0.4\)}}$ & 43.3$_{\scalebox{0.7}{\(\pm 0.5\)}}$ & 46.2$_{\scalebox{0.7}{\(\pm 0.4\)}}$ & 45.1$_{\scalebox{0.7}{\(\pm 0.2\)}}$ & \underline{46.6}$_{\scalebox{0.7}{\(\pm 0.3\)}}$ & 44.9$_{\scalebox{0.7}{\(\pm 0.5\)}}$ & 44.7$_{\scalebox{0.7}{\(\pm 0.5\)}}$ \\
& TF-IDF & \textbf{46.3}$_{\scalebox{0.7}{\(\pm 0.0\)}}$ & \textbf{46.2}$_{\scalebox{0.7}{\(\pm 0.0\)}}$ & \textbf{47.2}$_{\scalebox{0.7}{\(\pm 0.1\)}}$ & \underline{45.8}$_{\scalebox{0.7}{\(\pm 0.0\)}}$ & 45.0$_{\scalebox{0.7}{\(\pm 0.0\)}}$ & \textbf{46.0}$_{\scalebox{0.7}{\(\pm 0.1\)}}$ & \textbf{49.1}$_{\scalebox{0.7}{\(\pm 0.0\)}}$ \\ 
& Llama3.2-1B (1-Shot) & 23.3$_{\scalebox{0.7}{\(\pm 0.0\)}}$ & 23.9$_{\scalebox{0.7}{\(\pm 0.0\)}}$ & 22.5$_{\scalebox{0.7}{\(\pm 0.0\)}}$ & 21.2$_{\scalebox{0.7}{\(\pm 0.0\)}}$ & 21.1$_{\scalebox{0.7}{\(\pm 0.0\)}}$ & 21.7$_{\scalebox{0.7}{\(\pm 0.0\)}}$ & 20.9$_{\scalebox{0.7}{\(\pm 0.0\)}}$ \\ 
& Llama3.2-1B (SFT) & 43.7$_{\scalebox{0.7}{\(\pm 0.0\)}}$ & \underline{45.4}$_{\scalebox{0.7}{\(\pm 0.0\)}}$ & \underline{47.1}$_{\scalebox{0.7}{\(\pm 0.0\)}}$ & 43.7$_{\scalebox{0.7}{\(\pm 0.0\)}}$ & 46.5$_{\scalebox{0.7}{\(\pm 0.0\)}}$ & 43.5$_{\scalebox{0.7}{\(\pm 0.0\)}}$ & 46.5$_{\scalebox{0.7}{\(\pm 0.0\)}}$ \\
& Qwen2-0.5B (1-Shot) & 25.9$_{\scalebox{0.7}{\(\pm 0.0\)}}$ & 26.1$_{\scalebox{0.7}{\(\pm 0.0\)}}$ & 26.2$_{\scalebox{0.7}{\(\pm 0.0\)}}$ & 26.4$_{\scalebox{0.7}{\(\pm 0.0\)}}$ & 26.1$_{\scalebox{0.7}{\(\pm 0.0\)}}$ & 27.0$_{\scalebox{0.7}{\(\pm 0.0\)}}$ & 26.7$_{\scalebox{0.7}{\(\pm 0.0\)}}$ \\ 
& Qwen2-0.5B (SFT) & 42.5$_{\scalebox{0.7}{\(\pm 0.0\)}}$ & 44.3$_{\scalebox{0.7}{\(\pm 0.0\)}}$ & 44.4$_{\scalebox{0.7}{\(\pm 0.0\)}}$ & 45.7$_{\scalebox{0.7}{\(\pm 0.0\)}}$ & 40.6$_{\scalebox{0.7}{\(\pm 0.0\)}}$ & \underline{45.7}$_{\scalebox{0.7}{\(\pm 0.0\)}}$ & \underline{47.2}$_{\scalebox{0.7}{\(\pm 0.0\)}}$ \\ \up

Spearman’s Rho & TinyXRA & \underline{31.3}$_{\scalebox{0.7}{\(\pm 0.5\)}}$ & \underline{32.5}$_{\scalebox{0.7}{\(\pm 0.5\)}}$ & \underline{36.9}$_{\scalebox{0.7}{\(\pm 0.4\)}}$ & \textbf{36.8}$_{\scalebox{0.7}{\(\pm 0.5\)}}$ & \textbf{39.6}$_{\scalebox{0.7}{\(\pm 0.3\)}}$ & \underline{33.9}$_{\scalebox{0.7}{\(\pm 0.5\)}}$ & \underline{35.0}$_{\scalebox{0.7}{\(\pm 0.5\)}}$ \\
& XRR & 28.1$_{\scalebox{0.7}{\(\pm 0.7\)}}$ & 30.4$_{\scalebox{0.7}{\(\pm 0.7\)}}$ & 35.7$_{\scalebox{0.7}{\(\pm 0.9\)}}$ & 33.1$_{\scalebox{0.7}{\(\pm 0.5\)}}$ & 36.3$_{\scalebox{0.7}{\(\pm 0.8\)}}$ & 31.2$_{\scalebox{0.7}{\(\pm 0.9\)}}$ & 32.1$_{\scalebox{0.7}{\(\pm 0.7\)}}$ \\
& TF-IDF & \textbf{32.4}$_{\scalebox{0.7}{\(\pm 0.0\)}}$ & \textbf{32.7}$_{\scalebox{0.7}{\(\pm 0.0\)}}$ & \textbf{37.6}$_{\scalebox{0.7}{\(\pm 0.0\)}}$ & \underline{36.1}$_{\scalebox{0.7}{\(\pm 0.0\)}}$ & \underline{39.4}$_{\scalebox{0.7}{\(\pm 0.0\)}}$ & \textbf{34.6}$_{\scalebox{0.7}{\(\pm 0.0\)}}$ & \textbf{35.6}$_{\scalebox{0.7}{\(\pm 0.0\)}}$ \\
& Llama3.2-1B (1-Shot) & -4.3$_{\scalebox{0.7}{\(\pm 0.0\)}}$ & -4.5$_{\scalebox{0.7}{\(\pm 0.0\)}}$ & -6.2$_{\scalebox{0.7}{\(\pm 0.0\)}}$ & -8.3$_{\scalebox{0.7}{\(\pm 0.0\)}}$ & -8.7$_{\scalebox{0.7}{\(\pm 0.0\)}}$ & -7.0$_{\scalebox{0.7}{\(\pm 0.0\)}}$ & -10.7$_{\scalebox{0.7}{\(\pm 0.0\)}}$ \\ 
& Llama3.2-1B (SFT) & 30.9$_{\scalebox{0.7}{\(\pm 0.0\)}}$ & 31.3$_{\scalebox{0.7}{\(\pm 0.0\)}}$ & 36.0$_{\scalebox{0.7}{\(\pm 0.0\)}}$ & 35.2$_{\scalebox{0.7}{\(\pm 0.0\)}}$ & 37.8$_{\scalebox{0.7}{\(\pm 0.0\)}}$ & 33.0$_{\scalebox{0.7}{\(\pm 0.0\)}}$ & 34.2$_{\scalebox{0.7}{\(\pm 0.0\)}}$ \\
& Qwen2-0.5B (1-Shot) & -2.4$_{\scalebox{0.7}{\(\pm 0.0\)}}$ & -3.5$_{\scalebox{0.7}{\(\pm 0.0\)}}$ & -3.6$_{\scalebox{0.7}{\(\pm 0.0\)}}$ & -3.8$_{\scalebox{0.7}{\(\pm 0.0\)}}$ & -4.5$_{\scalebox{0.7}{\(\pm 0.0\)}}$ & -3.6$_{\scalebox{0.7}{\(\pm 0.0\)}}$ & -3.8$_{\scalebox{0.7}{\(\pm 0.0\)}}$ \\ 
& Qwen2-0.5B (SFT) & 29.7$_{\scalebox{0.7}{\(\pm 0.0\)}}$ & 30.6$_{\scalebox{0.7}{\(\pm 0.0\)}}$ & 35.0$_{\scalebox{0.7}{\(\pm 0.0\)}}$ & 35.1$_{\scalebox{0.7}{\(\pm 0.0\)}}$ & 37.5$_{\scalebox{0.7}{\(\pm 0.0\)}}$ & 32.0$_{\scalebox{0.7}{\(\pm 0.0\)}}$ & 32.9$_{\scalebox{0.7}{\(\pm 0.0\)}}$ \\ \up
 
Kendall’s Tau & TinyXRA & \underline{24.2}$_{\scalebox{0.7}{\(\pm 0.3\)}}$ & \underline{25.0}$_{\scalebox{0.7}{\(\pm 0.4\)}}$ & \underline{28.6}$_{\scalebox{0.7}{\(\pm 0.3\)}}$ & \textbf{28.6}$_{\scalebox{0.7}{\(\pm 0.5\)}}$ & \textbf{30.7}$_{\scalebox{0.7}{\(\pm 0.3\)}}$ & \underline{26.3}$_{\scalebox{0.7}{\(\pm 0.4\)}}$ & \underline{27.2}$_{\scalebox{0.7}{\(\pm 0.3\)}}$ \\
& XRR & 21.7$_{\scalebox{0.7}{\(\pm 0.6\)}}$ & 23.5$_{\scalebox{0.7}{\(\pm 0.5\)}}$ & 27.7$_{\scalebox{0.7}{\(\pm 0.7\)}}$ & 25.6$_{\scalebox{0.7}{\(\pm 0.5\)}}$ & 28.1$_{\scalebox{0.7}{\(\pm 0.6\)}}$ & 24.2$_{\scalebox{0.7}{\(\pm 0.7\)}}$ & 24.9$_{\scalebox{0.7}{\(\pm 0.5\)}}$ \\
& TF-IDF & \textbf{25.2}$_{\scalebox{0.7}{\(\pm 0.0\)}}$ & \textbf{25.3}$_{\scalebox{0.7}{\(\pm 0.0\)}}$ & \textbf{29.1}$_{\scalebox{0.7}{\(\pm 0.0\)}}$ & \underline{28.0}$_{\scalebox{0.7}{\(\pm 0.0\)}}$ & \underline{30.6}$_{\scalebox{0.7}{\(\pm 0.0\)}}$ & \textbf{26.8}$_{\scalebox{0.7}{\(\pm 0.0\)}}$ & \textbf{27.8}$_{\scalebox{0.7}{\(\pm 0.0\)}}$ \\ 
& Llama3.2-1B (1-Shot) & -3.3$_{\scalebox{0.7}{\(\pm 0.0\)}}$ & -3.5$_{\scalebox{0.7}{\(\pm 0.0\)}}$ & -4.7$_{\scalebox{0.7}{\(\pm 0.0\)}}$ & -6.4$_{\scalebox{0.7}{\(\pm 0.0\)}}$ & -6.7$_{\scalebox{0.7}{\(\pm 0.0\)}}$ & -5.4$_{\scalebox{0.7}{\(\pm 0.0\)}}$ & -8.2$_{\scalebox{0.7}{\(\pm 0.0\)}}$ \\ 
& Llama3.2-1B (SFT) & 24.0$_{\scalebox{0.7}{\(\pm 0.0\)}}$ & 24.1$_{\scalebox{0.7}{\(\pm 0.0\)}}$ & 27.8$_{\scalebox{0.7}{\(\pm 0.0\)}}$ & 27.3$_{\scalebox{0.7}{\(\pm 0.0\)}}$ & 29.4$_{\scalebox{0.7}{\(\pm 0.0\)}}$ & 25.5$_{\scalebox{0.7}{\(\pm 0.0\)}}$ & 26.5$_{\scalebox{0.7}{\(\pm 0.0\)}}$ \\
& Qwen2-0.5B (1-Shot) & -1.9$_{\scalebox{0.7}{\(\pm 0.0\)}}$ & -2.7$_{\scalebox{0.7}{\(\pm 0.0\)}}$ & -2.8$_{\scalebox{0.7}{\(\pm 0.0\)}}$ & -2.9$_{\scalebox{0.7}{\(\pm 0.0\)}}$ & -3.5$_{\scalebox{0.7}{\(\pm 0.0\)}}$ & -2.8$_{\scalebox{0.7}{\(\pm 0.0\)}}$ & -2.9$_{\scalebox{0.7}{\(\pm 0.0\)}}$ \\ 
& Qwen2-0.5B (SFT) & 22.9$_{\scalebox{0.7}{\(\pm 0.0\)}}$ & 23.5$_{\scalebox{0.7}{\(\pm 0.0\)}}$ & 27.2$_{\scalebox{0.7}{\(\pm 0.0\)}}$ & 27.1$_{\scalebox{0.7}{\(\pm 0.0\)}}$ & 29.1$_{\scalebox{0.7}{\(\pm 0.0\)}}$ & 24.9$_{\scalebox{0.7}{\(\pm 0.0\)}}$ & 25.7$_{\scalebox{0.7}{\(\pm 0.0\)}}$ \down\\ \hline
\end{tabular}}{Results are reported as the mean ± standard deviation across five random seeds (98, 83, 62, 42, 21).}
\end{table}

\begin{table}[htb!]
\TABLE
{F1 Score (\%), Spearman’s Rho (\%), and Kendall’s Tau (\%) for various benchmark models across test years (2024–2018) using kurtosis risk measurement.\label{tab:kurt}}
{\begin{tabular}{@{}l l l l l l l l l@{}}
\hline\up
Evaluation Metrics & Models & 2024 & 2023 & 2022 & 2021 & 2020 & 2019 & 2018 \\ \hline\up
F1 Score & TinyXRA & \textbf{44.6}$_{\scalebox{0.7}{\(\pm 0.3\)}}$ & \textbf{47.3}$_{\scalebox{0.7}{\(\pm 0.3\)}}$ & \textbf{45.0}$_{\scalebox{0.7}{\(\pm 0.3\)}}$ & \underline{47.2}$_{\scalebox{0.7}{\(\pm 0.1\)}}$ & \textbf{39.9}$_{\scalebox{0.7}{\(\pm 0.3\)}}$ & \textbf{46.3}$_{\scalebox{0.7}{\(\pm 0.7\)}}$ & \textbf{48.4}$_{\scalebox{0.7}{\(\pm 0.6\)}}$ \\
& XRR & 43.9$_{\scalebox{0.7}{\(\pm 0.2\)}}$ & 46.5$_{\scalebox{0.7}{\(\pm 0.4\)}}$ & \underline{44.4}$_{\scalebox{0.7}{\(\pm 0.3\)}}$ & 46.2$_{\scalebox{0.7}{\(\pm 0.5\)}}$ & \underline{39.3}$_{\scalebox{0.7}{\(\pm 0.6\)}}$ & 44.7$_{\scalebox{0.7}{\(\pm 0.6\)}}$ & 46.7$_{\scalebox{0.7}{\(\pm 0.6\)}}$ \\
& TF-IDF & \underline{44.0}$_{\scalebox{0.7}{\(\pm 0.2\)}}$ & \underline{47.1}$_{\scalebox{0.7}{\(\pm 0.0\)}}$ & 44.3$_{\scalebox{0.7}{\(\pm 0.0\)}}$ & \textbf{48.2}$_{\scalebox{0.7}{\(\pm 0.0\)}}$ & 38.2$_{\scalebox{0.7}{\(\pm 0.2\)}}$ & \underline{45.7}$_{\scalebox{0.7}{\(\pm 0.1\)}}$ & \underline{46.9}$_{\scalebox{0.7}{\(\pm 0.0\)}}$ \\ 
& Llama3.2-1B (1-Shot) & 24.3$_{\scalebox{0.7}{\(\pm 0.0\)}}$ & 23.9$_{\scalebox{0.7}{\(\pm 0.0\)}}$ & 22.8$_{\scalebox{0.7}{\(\pm 0.0\)}}$ & 22.1$_{\scalebox{0.7}{\(\pm 0.0\)}}$ & 22.8$_{\scalebox{0.7}{\(\pm 0.0\)}}$ & 22.0$_{\scalebox{0.7}{\(\pm 0.0\)}}$ & 23.5$_{\scalebox{0.7}{\(\pm 0.0\)}}$ \\ 
& Llama3.2-1B (SFT) & 43.4$_{\scalebox{0.7}{\(\pm 0.0\)}}$ & 44.6$_{\scalebox{0.7}{\(\pm 0.0\)}}$ & 44.0$_{\scalebox{0.7}{\(\pm 0.0\)}}$ & 45.4$_{\scalebox{0.7}{\(\pm 0.0\)}}$ & 36.1$_{\scalebox{0.7}{\(\pm 0.0\)}}$ & 42.4$_{\scalebox{0.7}{\(\pm 0.0\)}}$ & 41.7$_{\scalebox{0.7}{\(\pm 0.0\)}}$ \\
& Qwen2-0.5B (1-Shot) & 27.2$_{\scalebox{0.7}{\(\pm 0.0\)}}$ & 26.1$_{\scalebox{0.7}{\(\pm 0.0\)}}$ & 26.6$_{\scalebox{0.7}{\(\pm 0.0\)}}$ & 25.9$_{\scalebox{0.7}{\(\pm 0.0\)}}$ & 26.7$_{\scalebox{0.7}{\(\pm 0.0\)}}$ & 26.1$_{\scalebox{0.7}{\(\pm 0.0\)}}$ & 25.6$_{\scalebox{0.7}{\(\pm 0.0\)}}$ \\ 
& Qwen2-0.5B (SFT) & 42.8$_{\scalebox{0.7}{\(\pm 0.0\)}}$ & 43.3$_{\scalebox{0.7}{\(\pm 0.0\)}}$ & \underline{44.4}$_{\scalebox{0.7}{\(\pm 0.0\)}}$ & 44.8$_{\scalebox{0.7}{\(\pm 0.0\)}}$ & 38.7$_{\scalebox{0.7}{\(\pm 0.0\)}}$ & 44.2$_{\scalebox{0.7}{\(\pm 0.0\)}}$ & 43.9$_{\scalebox{0.7}{\(\pm 0.0\)}}$ \\ \up

Spearman’s Rho & TinyXRA & 30.0$_{\scalebox{0.7}{\(\pm 0.2\)}}$ & 35.2$_{\scalebox{0.7}{\(\pm 0.5\)}}$ & 31.0$_{\scalebox{0.7}{\(\pm 0.4\)}}$ & 36.0$_{\scalebox{0.7}{\(\pm 0.6\)}}$ & \textbf{17.1}$_{\scalebox{0.7}{\(\pm 1.3\)}}$ & \underline{33.1}$_{\scalebox{0.7}{\(\pm 1.1\)}}$ & \underline{39.0}$_{\scalebox{0.7}{\(\pm 0.6\)}}$ \\
& XRR & 28.2$_{\scalebox{0.7}{\(\pm 0.8\)}}$ & 32.4$_{\scalebox{0.7}{\(\pm 0.4\)}}$ & 28.5$_{\scalebox{0.7}{\(\pm 0.7\)}}$ & 33.4$_{\scalebox{0.7}{\(\pm 0.7\)}}$ & \textbf{17.1}$_{\scalebox{0.7}{\(\pm 1.2\)}}$ & 28.7$_{\scalebox{0.7}{\(\pm 1.3\)}}$ & 34.3$_{\scalebox{0.7}{\(\pm 0.8\)}}$ \\
& TF-IDF & \textbf{31.9}$_{\scalebox{0.7}{\(\pm 0.0\)}}$ & \textbf{38.0}$_{\scalebox{0.7}{\(\pm 0.0\)}}$ & \textbf{33.2}$_{\scalebox{0.7}{\(\pm 0.0\)}}$ & \textbf{39.0}$_{\scalebox{0.7}{\(\pm 0.0\)}}$ & \underline{16.5}$_{\scalebox{0.7}{\(\pm 0.0\)}}$ & \textbf{33.9}$_{\scalebox{0.7}{\(\pm 0.0\)}}$ & \textbf{39.4}$_{\scalebox{0.7}{\(\pm 0.0\)}}$ \\ 
& Llama3.2-1B (1-Shot) & -2.1$_{\scalebox{0.7}{\(\pm 0.0\)}}$ & -2.3$_{\scalebox{0.7}{\(\pm 0.0\)}}$ & -4.8$_{\scalebox{0.7}{\(\pm 0.0\)}}$ & -5.6$_{\scalebox{0.7}{\(\pm 0.0\)}}$ & -5.1$_{\scalebox{0.7}{\(\pm 0.0\)}}$ & -4.6$_{\scalebox{0.7}{\(\pm 0.0\)}}$ & -0.9$_{\scalebox{0.7}{\(\pm 0.0\)}}$ \\ 
& Llama3.2-1B (SFT) & \underline{30.6}$_{\scalebox{0.7}{\(\pm 0.0\)}}$ & \underline{35.9}$_{\scalebox{0.7}{\(\pm 0.0\)}}$ & \underline{31.9}$_{\scalebox{0.7}{\(\pm 0.0\)}}$ & \underline{36.7}$_{\scalebox{0.7}{\(\pm 0.0\)}}$ & 14.7$_{\scalebox{0.7}{\(\pm 0.0\)}}$ & 30.3$_{\scalebox{0.7}{\(\pm 0.0\)}}$ & 36.9$_{\scalebox{0.7}{\(\pm 0.0\)}}$ \\
& Qwen2-0.5B (1-Shot) & -0.2$_{\scalebox{0.7}{\(\pm 0.0\)}}$ & -4.7$_{\scalebox{0.7}{\(\pm 0.0\)}}$ & -2.8$_{\scalebox{0.7}{\(\pm 0.0\)}}$ & -4.9$_{\scalebox{0.7}{\(\pm 0.0\)}}$ & -4.8$_{\scalebox{0.7}{\(\pm 0.0\)}}$ & -2.0$_{\scalebox{0.7}{\(\pm 0.0\)}}$ & -6.0$_{\scalebox{0.7}{\(\pm 0.0\)}}$ \\ 
& Qwen2-0.5B (SFT) & 28.3$_{\scalebox{0.7}{\(\pm 0.0\)}}$ & 34.6$_{\scalebox{0.7}{\(\pm 0.0\)}}$ & 30.5$_{\scalebox{0.7}{\(\pm 0.0\)}}$ & 35.9$_{\scalebox{0.7}{\(\pm 0.0\)}}$ & 15.3$_{\scalebox{0.7}{\(\pm 0.0\)}}$ & 30.4$_{\scalebox{0.7}{\(\pm 0.0\)}}$ & 34.7$_{\scalebox{0.7}{\(\pm 0.0\)}}$ \\ \up

Kendall’s Tau & TinyXRA & 23.3$_{\scalebox{0.7}{\(\pm 0.2\)}}$ & 27.6$_{\scalebox{0.7}{\(\pm 0.4\)}}$ & 24.2$_{\scalebox{0.7}{\(\pm 0.3\)}}$ & 28.2$_{\scalebox{0.7}{\(\pm 0.5\)}}$ & \textbf{13.2}$_{\scalebox{0.7}{\(\pm 1.0\)}}$ & \underline{25.9}$_{\scalebox{0.7}{\(\pm 0.9\)}}$ & \underline{30.6}$_{\scalebox{0.7}{\(\pm 0.5\)}}$ \\
& XRR & 22.0$_{\scalebox{0.7}{\(\pm 0.6\)}}$ & 25.3$_{\scalebox{0.7}{\(\pm 0.3\)}}$ & 22.3$_{\scalebox{0.7}{\(\pm 0.5\)}}$ & 26.1$_{\scalebox{0.7}{\(\pm 0.5\)}}$ & \underline{13.1}$_{\scalebox{0.7}{\(\pm 0.9\)}}$ & 22.4$_{\scalebox{0.7}{\(\pm 1.0\)}}$ & 26.8$_{\scalebox{0.7}{\(\pm 0.6\)}}$ \\
& TF-IDF & \textbf{24.9}$_{\scalebox{0.7}{\(\pm 0.0\)}}$ & \textbf{30.0}$_{\scalebox{0.7}{\(\pm 0.0\)}}$ & \textbf{26.1}$_{\scalebox{0.7}{\(\pm 0.0\)}}$ & \textbf{30.7}$_{\scalebox{0.7}{\(\pm 0.0\)}}$ & 12.8$_{\scalebox{0.7}{\(\pm 0.0\)}}$ & \textbf{26.5}$_{\scalebox{0.7}{\(\pm 0.0\)}}$ & \textbf{30.9}$_{\scalebox{0.7}{\(\pm 0.0\)}}$ \\ 
& Llama3.2-1B (1-Shot) & -1.6$_{\scalebox{0.7}{\(\pm 0.0\)}}$ & -1.8$_{\scalebox{0.7}{\(\pm 0.0\)}}$ & -3.7$_{\scalebox{0.7}{\(\pm 0.0\)}}$ & -4.3$_{\scalebox{0.7}{\(\pm 0.0\)}}$ & -3.9$_{\scalebox{0.7}{\(\pm 0.0\)}}$ & -3.5$_{\scalebox{0.7}{\(\pm 0.0\)}}$ & -0.7$_{\scalebox{0.7}{\(\pm 0.0\)}}$ \\ 
& Llama3.2-1B (SFT) & \underline{23.9}$_{\scalebox{0.7}{\(\pm 0.0\)}}$ & \underline{28.2}$_{\scalebox{0.7}{\(\pm 0.0\)}}$ & \underline{25.0}$_{\scalebox{0.7}{\(\pm 0.0\)}}$ & \underline{28.8}$_{\scalebox{0.7}{\(\pm 0.0\)}}$ & 11.3$_{\scalebox{0.7}{\(\pm 0.0\)}}$ & 23.6$_{\scalebox{0.7}{\(\pm 0.0\)}}$ & 29.0$_{\scalebox{0.7}{\(\pm 0.0\)}}$ \\
& Qwen2-0.5B (1-Shot) & -0.1$_{\scalebox{0.7}{\(\pm 0.0\)}}$ & -3.6$_{\scalebox{0.7}{\(\pm 0.0\)}}$ & -2.1$_{\scalebox{0.7}{\(\pm 0.0\)}}$ & -3.8$_{\scalebox{0.7}{\(\pm 0.0\)}}$ & -3.7$_{\scalebox{0.7}{\(\pm 0.0\)}}$ & -1.5$_{\scalebox{0.7}{\(\pm 0.0\)}}$ & -4.6$_{\scalebox{0.7}{\(\pm 0.0\)}}$ \\ 
& Qwen2-0.5B (SFT) & 22.0$_{\scalebox{0.7}{\(\pm 0.0\)}}$ & 27.1$_{\scalebox{0.7}{\(\pm 0.0\)}}$ & 23.9$_{\scalebox{0.7}{\(\pm 0.0\)}}$ & 28.2$_{\scalebox{0.7}{\(\pm 0.0\)}}$ & 11.8$_{\scalebox{0.7}{\(\pm 0.0\)}}$ & 23.7$_{\scalebox{0.7}{\(\pm 0.0\)}}$ & 27.1$_{\scalebox{0.7}{\(\pm 0.0\)}}$ \down\\ \hline
\end{tabular}}{Results are reported as the mean ± standard deviation across five random seeds (98, 83, 62, 42, 21).}
\end{table}

\begin{table}[htb!]
\TABLE
{F1 Score (\%), Spearman’s Rho (\%), and Kendall’s Tau (\%) for various benchmark models across test years (2024–2018) using Sortino ratio risk measurement.\label{tab:sortino}}
{\begin{tabular}{@{}l l l l l l l l l@{}}
\hline\up
Evaluation Metrics & Models & 2024 & 2023 & 2022 & 2021 & 2020 & 2019 & 2018 \\ \hline\up
F1 Score & TinyXRA & \underline{48.2}$_{\scalebox{0.7}{\(\pm 0.7\)}}$ & 39.9$_{\scalebox{0.7}{\(\pm 1.1\)}}$ & \textbf{48.5}$_{\scalebox{0.7}{\(\pm 1.3\)}}$ & \textbf{35.6}$_{\scalebox{0.7}{\(\pm 1.0\)}}$ & \textbf{53.8}$_{\scalebox{0.7}{\(\pm 1.5\)}}$ & \textbf{49.0}$_{\scalebox{0.7}{\(\pm 0.8\)}}$ & \underline{59.8}$_{\scalebox{0.7}{\(\pm 0.7\)}}$ \\
& XRR & 45.8$_{\scalebox{0.7}{\(\pm 1.5\)}}$ & \textbf{43.8}$_{\scalebox{0.7}{\(\pm 4.1\)}}$ & \underline{37.2}$_{\scalebox{0.7}{\(\pm 2.3\)}}$ & \underline{33.9}$_{\scalebox{0.7}{\(\pm 0.8\)}}$ & 51.4$_{\scalebox{0.7}{\(\pm 0.9\)}}$ & \underline{47.4}$_{\scalebox{0.7}{\(\pm 0.5\)}}$ & 57.3$_{\scalebox{0.7}{\(\pm 0.9\)}}$ \\
& TF-IDF & \textbf{49.5}$_{\scalebox{0.7}{\(\pm 0.1\)}}$ & \underline{40.4}$_{\scalebox{0.7}{\(\pm 0.3\)}}$ & 32.1$_{\scalebox{0.7}{\(\pm 0.0\)}}$ & 32.6$_{\scalebox{0.7}{\(\pm 0.1\)}}$ & 45.8$_{\scalebox{0.7}{\(\pm 0.2\)}}$ & 46.6$_{\scalebox{0.7}{\(\pm 0.1\)}}$ & \textbf{60.6}$_{\scalebox{0.7}{\(\pm 0.0\)}}$ \\ 
& Llama3.2-1B (1-Shot) & 22.9$_{\scalebox{0.7}{\(\pm 0.0\)}}$ & 22.6$_{\scalebox{0.7}{\(\pm 0.0\)}}$ & 28.1$_{\scalebox{0.7}{\(\pm 0.0\)}}$ & 25.0$_{\scalebox{0.7}{\(\pm 0.0\)}}$ & 21.2$_{\scalebox{0.7}{\(\pm 0.0\)}}$ & 20.9$_{\scalebox{0.7}{\(\pm 0.0\)}}$ & 18.9$_{\scalebox{0.7}{\(\pm 0.0\)}}$ \\ 
& Llama3.2-1B (SFT) & 45.4$_{\scalebox{0.7}{\(\pm 0.0\)}}$ & 37.8$_{\scalebox{0.7}{\(\pm 0.0\)}}$ & 32.9$_{\scalebox{0.7}{\(\pm 0.0\)}}$ & 30.9$_{\scalebox{0.7}{\(\pm 0.0\)}}$ & \underline{52.6}$_{\scalebox{0.7}{\(\pm 0.0\)}}$ & 44.4$_{\scalebox{0.7}{\(\pm 0.0\)}}$ & 58.1$_{\scalebox{0.7}{\(\pm 0.0\)}}$ \\
& Qwen2-0.5B (1-Shot) & 24.7$_{\scalebox{0.7}{\(\pm 0.0\)}}$ & 24.9$_{\scalebox{0.7}{\(\pm 0.0\)}}$ & 27.9$_{\scalebox{0.7}{\(\pm 0.0\)}}$ & 27.8$_{\scalebox{0.7}{\(\pm 0.0\)}}$ & 26.1$_{\scalebox{0.7}{\(\pm 0.0\)}}$ & 25.6$_{\scalebox{0.7}{\(\pm 0.0\)}}$ & 24.9$_{\scalebox{0.7}{\(\pm 0.0\)}}$ \\ 
& Qwen2-0.5B (SFT) & 47.0$_{\scalebox{0.7}{\(\pm 0.0\)}}$ & 36.4$_{\scalebox{0.7}{\(\pm 0.0\)}}$ & 31.2$_{\scalebox{0.7}{\(\pm 0.0\)}}$ & 29.3$_{\scalebox{0.7}{\(\pm 0.0\)}}$ & 51.1$_{\scalebox{0.7}{\(\pm 0.0\)}}$ & 43.3$_{\scalebox{0.7}{\(\pm 0.0\)}}$ & 57.0$_{\scalebox{0.7}{\(\pm 0.0\)}}$ \\ \up

Spearman’s Rho & TinyXRA & \textbf{31.8}$_{\scalebox{0.7}{\(\pm 1.4\)}}$ & 15.4$_{\scalebox{0.7}{\(\pm 3.4\)}}$ & \textbf{40.6}$_{\scalebox{0.7}{\(\pm 3.1\)}}$ & \textbf{7.5}$_{\scalebox{0.7}{\(\pm 1.9\)}}$ & \underline{44.1}$_{\scalebox{0.7}{\(\pm 1.6\)}}$ & \textbf{44.6}$_{\scalebox{0.7}{\(\pm 1.5\)}}$ & \textbf{61.4}$_{\scalebox{0.7}{\(\pm 0.8\)}}$ \\
& XRR & 25.9$_{\scalebox{0.7}{\(\pm 1.7\)}}$ & \textbf{27.2}$_{\scalebox{0.7}{\(\pm 11.5\)}}$ & \underline{11.0}$_{\scalebox{0.7}{\(\pm 7.4\)}}$ & -1.9$_{\scalebox{0.7}{\(\pm 1.8\)}}$ & 37.2$_{\scalebox{0.7}{\(\pm 1.7\)}}$ & 39.8$_{\scalebox{0.7}{\(\pm 0.7\)}}$ & 56.4$_{\scalebox{0.7}{\(\pm 1.9\)}}$ \\
& TF-IDF & 22.0$_{\scalebox{0.7}{\(\pm 0.1\)}}$ & \underline{21.1}$_{\scalebox{0.7}{\(\pm 0.5\)}}$ & -15.2$_{\scalebox{0.7}{\(\pm 0.0\)}}$ & 3.0$_{\scalebox{0.7}{\(\pm 0.3\)}}$ & \textbf{44.9}$_{\scalebox{0.7}{\(\pm 0.0\)}}$ & \underline{43.6}$_{\scalebox{0.7}{\(\pm 0.0\)}}$ & \underline{60.5}$_{\scalebox{0.7}{\(\pm 0.0\)}}$ \\ 
& Llama3.2-1B (1-Shot) & -7.3$_{\scalebox{0.7}{\(\pm 0.0\)}}$ & -9.2$_{\scalebox{0.7}{\(\pm 0.0\)}}$ & 10.3$_{\scalebox{0.7}{\(\pm 0.0\)}}$ & \underline{4.4}$_{\scalebox{0.7}{\(\pm 0.0\)}}$ & -9.2$_{\scalebox{0.7}{\(\pm 0.0\)}}$ & -8.1$_{\scalebox{0.7}{\(\pm 0.0\)}}$ & -18.2$_{\scalebox{0.7}{\(\pm 0.0\)}}$ \\ 
& Llama3.2-1B (SFT) & \underline{30.2}$_{\scalebox{0.7}{\(\pm 0.0\)}}$ & 20.3$_{\scalebox{0.7}{\(\pm 0.0\)}}$ & -5.9$_{\scalebox{0.7}{\(\pm 0.0\)}}$ & -4.9$_{\scalebox{0.7}{\(\pm 0.0\)}}$ & 40.8$_{\scalebox{0.7}{\(\pm 0.0\)}}$ & 40.3$_{\scalebox{0.7}{\(\pm 0.0\)}}$ & 59.6$_{\scalebox{0.7}{\(\pm 0.0\)}}$ \\
& Qwen2-0.5B (1-Shot) & -5.6$_{\scalebox{0.7}{\(\pm 0.0\)}}$ & -5.6$_{\scalebox{0.7}{\(\pm 0.0\)}}$ & 5.6$_{\scalebox{0.7}{\(\pm 0.0\)}}$ & 3.0$_{\scalebox{0.7}{\(\pm 0.0\)}}$ & -6.0$_{\scalebox{0.7}{\(\pm 0.0\)}}$ & -7.4$_{\scalebox{0.7}{\(\pm 0.0\)}}$ & -10.6$_{\scalebox{0.7}{\(\pm 0.0\)}}$ \\ 
& Qwen2-0.5B (SFT) & 24.2$_{\scalebox{0.7}{\(\pm 0.0\)}}$ & 9.6$_{\scalebox{0.7}{\(\pm 0.0\)}}$ & -11.3$_{\scalebox{0.7}{\(\pm 0.0\)}}$ & -3.3$_{\scalebox{0.7}{\(\pm 0.0\)}}$ & 40.5$_{\scalebox{0.7}{\(\pm 0.0\)}}$ & 39.1$_{\scalebox{0.7}{\(\pm 0.0\)}}$ & 59.3$_{\scalebox{0.7}{\(\pm 0.0\)}}$ \\ \up

Kendall’s Tau & TinyXRA & \textbf{25.2}$_{\scalebox{0.7}{\(\pm 1.0\)}}$ & 11.9$_{\scalebox{0.7}{\(\pm 2.7\)}}$ & \textbf{31.8}$_{\scalebox{0.7}{\(\pm 2.6\)}}$ & \textbf{5.6}$_{\scalebox{0.7}{\(\pm 1.5\)}}$ & \underline{35.0}$_{\scalebox{0.7}{\(\pm 1.5\)}}$ & \textbf{34.7}$_{\scalebox{0.7}{\(\pm 1.3\)}}$ & \textbf{49.2}$_{\scalebox{0.7}{\(\pm 0.7\)}}$ \\
& XRR & 20.4$_{\scalebox{0.7}{\(\pm 1.4\)}}$ & \textbf{21.3}$_{\scalebox{0.7}{\(\pm 9.1\)}}$ & \underline{8.5}$_{\scalebox{0.7}{\(\pm 5.7\)}}$ & -1.6$_{\scalebox{0.7}{\(\pm 1.4\)}}$ & 29.5$_{\scalebox{0.7}{\(\pm 1.5\)}}$ & 30.9$_{\scalebox{0.7}{\(\pm 0.5\)}}$ & 45.1$_{\scalebox{0.7}{\(\pm 1.5\)}}$ \\
& TF-IDF & 17.1$_{\scalebox{0.7}{\(\pm 0.1\)}}$ & \underline{16.3}$_{\scalebox{0.7}{\(\pm 0.4\)}}$ & -11.6$_{\scalebox{0.7}{\(\pm 0.0\)}}$ & 2.3$_{\scalebox{0.7}{\(\pm 0.2\)}}$ & \textbf{36.2}$_{\scalebox{0.7}{\(\pm 0.0\)}}$ & \underline{33.9}$_{\scalebox{0.7}{\(\pm 0.0\)}}$ & \underline{48.5}$_{\scalebox{0.7}{\(\pm 0.0\)}}$ \\ 
& Llama3.2-1B (1-Shot) & -5.6$_{\scalebox{0.7}{\(\pm 0.0\)}}$ & -7.1$_{\scalebox{0.7}{\(\pm 0.0\)}}$ & 7.9$_{\scalebox{0.7}{\(\pm 0.0\)}}$ & \underline{3.4}$_{\scalebox{0.7}{\(\pm 0.0\)}}$ & -7.0$_{\scalebox{0.7}{\(\pm 0.0\)}}$ & -6.2$_{\scalebox{0.7}{\(\pm 0.0\)}}$ & -14.0$_{\scalebox{0.7}{\(\pm 0.0\)}}$ \\ 
& Llama3.2-1B (SFT) & \underline{24.1}$_{\scalebox{0.7}{\(\pm 0.0\)}}$ & 15.8$_{\scalebox{0.7}{\(\pm 0.0\)}}$ & -4.5$_{\scalebox{0.7}{\(\pm 0.0\)}}$ & -3.9$_{\scalebox{0.7}{\(\pm 0.0\)}}$ & 32.6$_{\scalebox{0.7}{\(\pm 0.0\)}}$ & 31.2$_{\scalebox{0.7}{\(\pm 0.0\)}}$ & 47.8$_{\scalebox{0.7}{\(\pm 0.0\)}}$ \\
& Qwen2-0.5B (1-Shot) & -4.3$_{\scalebox{0.7}{\(\pm 0.0\)}}$ & -4.3$_{\scalebox{0.7}{\(\pm 0.0\)}}$ & 4.3$_{\scalebox{0.7}{\(\pm 0.0\)}}$ & 2.3$_{\scalebox{0.7}{\(\pm 0.0\)}}$ & -4.7$_{\scalebox{0.7}{\(\pm 0.0\)}}$ & -5.7$_{\scalebox{0.7}{\(\pm 0.0\)}}$ & -8.2$_{\scalebox{0.7}{\(\pm 0.0\)}}$ \\ 
& Qwen2-0.5B (SFT) & 19.1$_{\scalebox{0.7}{\(\pm 0.0\)}}$ & 7.4$_{\scalebox{0.7}{\(\pm 0.0\)}}$ & -8.6$_{\scalebox{0.7}{\(\pm 0.0\)}}$ & -2.6$_{\scalebox{0.7}{\(\pm 0.0\)}}$ & 32.4$_{\scalebox{0.7}{\(\pm 0.0\)}}$ & 30.2$_{\scalebox{0.7}{\(\pm 0.0\)}}$ & 47.5$_{\scalebox{0.7}{\(\pm 0.0\)}}$ \down\\ \hline
\end{tabular}}{Results are reported as the mean ± standard deviation across five random seeds (98, 83, 62, 42, 21).}
\end{table}

\noindent \textbf{1. Overall Robustness and Superior Performance of TinyXRA}. TinyXRA consistently demonstrates superior predictive performance across a comprehensive set of evaluation metrics, including F1 Score, Spearman's Rho, and Kendall's Tau, as well as across various financial risk measures such as standard deviation, skewness, kurtosis, and the Sortino ratio, ranking within the top two across most experimental settings. This superior performance is substantiated through rigorous statistical validation. As shown in Table~\ref{tab:cohendresults}, effect size analysis using paired Cohen’s d tests yield positive values across all comparisons, confirming that TinyXRA outperforms all benchmark models, with particularly pronounced improvements over recent lightweight LLM-based methods. The observed medium to large effect sizes provide robust statistical evidence of TinyXRA’s superior efficacy in financial risk prediction tasks.

\noindent \textbf{2. Comparative Difficulty of Risk Measurement Tasks}. Among the risk measurements evaluated, standard deviation stands out as the least challenging task, primarily because it does not explicitly account for directionality or asymmetry in returns. In contrast, skewness, kurtosis, and Sortino ratio involve more complex risk dynamics that present significant modeling challenges. The Sortino ratio proves especially demanding as it predicts expected returns with an emphasis on downside volatility, explaining why models generally struggle more with this particular risk measurement compared to others.

\noindent \textbf{3. Observations on Model Performance Variability by Year}. Across test years (2024–2018), there is noticeable variability in performance metrics with certain periods presenting distinct challenges. The year 2020 consistently shows lower performance scores across all models in the standard deviation and kurtosis risk measures, likely attributable to unprecedented market conditions triggered by the COVID-19 pandemic. This period was characterized by heightened volatility and abrupt changes in financial reporting patterns, complicating model generalization. Despite these adverse conditions, TinyXRA demonstrates remarkable resilience compared to benchmarks, highlighting its robustness. The years 2022 and 2021 also exhibit performance variability for Sortino ratio risk measure, probably reflecting the post COVID-19 recovery during this period. Performance fluctuations may arise from potential discrepancies between training datasets (consisting of financial reports from the five years preceding the test year) and validation datasets (the test year's financial reports), especially when structural market shifts, regulatory changes, or evolving financial reporting practices significantly alter the nature and distribution of textual features.

\noindent \textbf{4. TF-IDF's Surprising Competitiveness}. The traditional TF-IDF baseline performs remarkably well, particularly under the skewness and kurtosis measures, consistently ranking close to or better than more advanced neural models. However, it is important to note that TF-IDF still struggles with the Sortino ratio risk measurement, which pulls down its overall performance. TF-IDF's strong performance may be attributed to its direct utilization of term frequency and document specificity, effectively capturing important and discriminative financial terms and concepts relevant to financial risk prediction. Its simplicity allows it to robustly handle noisy or sparse data often encountered in financial reports. This effectiveness demonstrates the sustained relevance of classical text-representation methods and emphasizes the importance of preserving rigorous baseline comparisons in financial text analysis, rather than presuming that newer and more complex models inherently surpass traditional approaches.

\noindent \textbf{5. Supervised Finetuning vs. One-Shot Performance}. The empirical results reveal pronounced performance disparities between one-shot learning approaches and supervised finetuning (SFT). These findings emphasize the inadequacy of one-shot paradigms in high-stakes financial forecasting contexts. One-shot models exhibit performance levels that are near-random or, in some cases, negatively correlated with ranking-based evaluation metrics, indicating substantial limitations in their capacity to capture complex financial signals from limited examples. In contrast, supervised finetuning yields significant improvements in predictive accuracy by leveraging large, task-specific datasets, thereby facilitating more effective representation learning and enabling the model to capture subtle financial patterns. Among the SFT models evaluated, Llama3.2-1B outperformed Qwen2-0.5B, plausibly attributable to its larger parameter count and enhanced capacity for extracting salient features from financial documents. While Llama3.2-1B also outperforms the XRR model in terms of prediction quality, achieving a lower Cohen’s d value (relative to TinyXRA). It is important to consider the trade-off in computational efficiency where XRR trains approximately 20 times faster, potentially offering a more practical balance of performance and efficiency in real-world deployments.

\noindent \textbf{6. Stability and Reliability Across Time and Risk Measures}. TinyXRA consistently demonstrates low performance variance across multiple years and risk metrics, indicating strong generalizability and stable predictive capacity, which are critical for practical application in financial contexts. In contrast to alternative models that exhibit marked performance fluctuations under varying market conditions, TinyXRA maintains robust and reliable accuracy over time. Notably, evaluation using the Sortino ratio reveals considerable challenges for all models, particularly during 2021 and 2022. These difficulties highlight the inherent complexity of modeling downside risk and the current methodological limitations in effectively capturing asymmetric risk profiles. In some instances, competing models exhibit negative correlations with actual outcomes, highlighting their instability during periods of market stress.  Our TinyXRA, by comparison, maintains relatively stable performance and consistently outperforms benchmark models, accentuates its robustness and reliability in modeling downside financial risk.
\section{Ablation Studies}
\label{sec:Ablation}
\subsection{Triplet Ranking Loss vs Pairwise Ranking Loss vs Cross Entropy Loss}
In this section, we present ablation studies to evaluate the effectiveness of triplet ranking loss in comparison to pairwise ranking loss and cross-entropy loss. The goal is to assess whether triplet ranking loss offers a significant advantage over the alternative loss functions. A decline in performance when substituting triplet ranking loss with pairwise ranking loss or cross-entropy loss would suggest that the triplet approach is more effective in capturing fine-grained ranking distinctions. On the other hand, if performance remains stable across all loss functions, it would imply that the choice of loss function plays a limited role in this task.
\begin{table}[htb!]
  \TABLE
  {F1 Score (\%), Spearman’s Rho (\%), and Kendall’s Tau (\%) for various loss functions across test years (2024–2018) under different risk measurements.\label{tab:abl1_all}}
  {\begin{tabular}{@{}l l l l l l l l l@{}}
    \toprule
    Evaluation Metrics & Loss Functions & 2024 & 2023 & 2022 & 2021 & 2020 & 2019 & 2018 \\
    \midrule
    \multicolumn{9}{c}{\textbf{Standard Deviation}} \\
    \midrule
    \multirow{3}{*}{F1 Score}
      & Triplet Ranking & \textbf{76.0}$_{\scalebox{0.7}{\(\pm 0.3\)}}$ & 74.5$_{\scalebox{0.7}{\(\pm 0.2\)}}$ & 75.1$_{\scalebox{0.7}{\(\pm 0.5\)}}$ & \textbf{74.1}$_{\scalebox{0.7}{\(\pm 0.3\)}}$ & \textbf{67.6}$_{\scalebox{0.7}{\(\pm 0.7\)}}$ & \textbf{76.5}$_{\scalebox{0.7}{\(\pm 0.3\)}}$ & 78.1$_{\scalebox{0.7}{\(\pm 0.2\)}}$ \\
      & Pairwise Ranking & 75.4$_{\scalebox{0.7}{\(\pm 0.2\)}}$ & \textbf{74.8}$_{\scalebox{0.7}{\(\pm 0.4\)}}$ & \textbf{75.5}$_{\scalebox{0.7}{\(\pm 0.4\)}}$ & 73.6$_{\scalebox{0.7}{\(\pm 0.4\)}}$ & 67.2$_{\scalebox{0.7}{\(\pm 0.3\)}}$ & 76.3$_{\scalebox{0.7}{\(\pm 0.7\)}}$ & \textbf{78.3}$_{\scalebox{0.7}{\(\pm 0.2\)}}$ \\
      & Cross-Entropy & 75.3$_{\scalebox{0.7}{\(\pm 0.4\)}}$ & 73.8$_{\scalebox{0.7}{\(\pm 0.2\)}}$ & 75.0$_{\scalebox{0.7}{\(\pm 0.2\)}}$ & 69.5$_{\scalebox{0.7}{\(\pm 2.0\)}}$ & 64.5$_{\scalebox{0.7}{\(\pm 0.5\)}}$ & 75.7$_{\scalebox{0.7}{\(\pm 0.6\)}}$ & 77.7$_{\scalebox{0.7}{\(\pm 0.4\)}}$ \\ \up

    \multirow{3}{*}{Spearman’s Rho}
      & Triplet Ranking & \textbf{80.3}$_{\scalebox{0.7}{\(\pm 0.1\)}}$ & 78.8$_{\scalebox{0.7}{\(\pm 0.3\)}}$ & \textbf{80.0}$_{\scalebox{0.7}{\(\pm 0.2\)}}$ & \textbf{80.1}$_{\scalebox{0.7}{\(\pm 0.2\)}}$ & \textbf{72.7}$_{\scalebox{0.7}{\(\pm 0.9\)}}$ & 81.2$_{\scalebox{0.7}{\(\pm 0.4\)}}$ & 82.9$_{\scalebox{0.7}{\(\pm 0.1\)}}$ \\
      & Pairwise Ranking & 80.0$_{\scalebox{0.7}{\(\pm 0.1\)}}$ & \textbf{78.9}$_{\scalebox{0.7}{\(\pm 0.2\)}}$ & 79.9$_{\scalebox{0.7}{\(\pm 0.2\)}}$ & 79.8$_{\scalebox{0.7}{\(\pm 0.4\)}}$ & 72.0$_{\scalebox{0.7}{\(\pm 0.1\)}}$ & \textbf{81.3}$_{\scalebox{0.7}{\(\pm 0.5\)}}$ & \textbf{83.0}$_{\scalebox{0.7}{\(\pm 0.1\)}}$ \\
      & Cross-Entropy & 79.1$_{\scalebox{0.7}{\(\pm 0.3\)}}$ & 77.7$_{\scalebox{0.7}{\(\pm 0.3\)}}$ & 79.1$_{\scalebox{0.7}{\(\pm 0.3\)}}$ & 78.5$_{\scalebox{0.7}{\(\pm 0.4\)}}$ & 70.3$_{\scalebox{0.7}{\(\pm 0.5\)}}$ & 80.4$_{\scalebox{0.7}{\(\pm 0.3\)}}$ & 81.9$_{\scalebox{0.7}{\(\pm 0.5\)}}$ \\ \up

    \multirow{3}{*}{Kendall’s Tau}
      & Triplet Ranking & \textbf{67.2}$_{\scalebox{0.7}{\(\pm 0.1\)}}$ & \textbf{65.7}$_{\scalebox{0.7}{\(\pm 0.3\)}}$ & \textbf{66.9}$_{\scalebox{0.7}{\(\pm 0.2\)}}$ & \textbf{66.8}$_{\scalebox{0.7}{\(\pm 0.2\)}}$ & \textbf{59.4}$_{\scalebox{0.7}{\(\pm 0.9\)}}$ & 67.8$_{\scalebox{0.7}{\(\pm 0.3\)}}$ & 69.6$_{\scalebox{0.7}{\(\pm 0.1\)}}$ \\
      & Pairwise Ranking & 66.9$_{\scalebox{0.7}{\(\pm 0.1\)}}$ & \textbf{65.7}$_{\scalebox{0.7}{\(\pm 0.1\)}}$ & \textbf{66.9}$_{\scalebox{0.7}{\(\pm 0.2\)}}$ & 66.5$_{\scalebox{0.7}{\(\pm 0.4\)}}$ & 58.8$_{\scalebox{0.7}{\(\pm 0.1\)}}$ & \textbf{67.9}$_{\scalebox{0.7}{\(\pm 0.4\)}}$ & \textbf{69.7}$_{\scalebox{0.7}{\(\pm 0.1\)}}$ \\
      & Cross-Entropy & 66.0$_{\scalebox{0.7}{\(\pm 0.3\)}}$ & 64.6$_{\scalebox{0.7}{\(\pm 0.3\)}}$ & 66.0$_{\scalebox{0.7}{\(\pm 0.3\)}}$ & 65.1$_{\scalebox{0.7}{\(\pm 0.4\)}}$ & 57.0$_{\scalebox{0.7}{\(\pm 0.4\)}}$ & 67.1$_{\scalebox{0.7}{\(\pm 0.4\)}}$ & 68.6$_{\scalebox{0.7}{\(\pm 0.5\)}}$ \\ 
    \midrule

    \multicolumn{9}{c}{\textbf{Skewness}} \\
    \midrule
    \multirow{3}{*}{F1 Score}
      & Triplet Ranking & 44.4$_{\scalebox{0.7}{\(\pm 0.2\)}}$ & 44.1$_{\scalebox{0.7}{\(\pm 0.6\)}}$ & \textbf{46.8}$_{\scalebox{0.7}{\(\pm 0.5\)}}$ & \textbf{47.1}$_{\scalebox{0.7}{\(\pm 0.4\)}}$ & 47.8$_{\scalebox{0.7}{\(\pm 0.4\)}}$ & 45.1$_{\scalebox{0.7}{\(\pm 0.4\)}}$ & 46.0$_{\scalebox{0.7}{\(\pm 0.3\)}}$ \\
      & Pairwise Ranking & 43.9$_{\scalebox{0.7}{\(\pm 0.3\)}}$ & 44.1$_{\scalebox{0.7}{\(\pm 0.3\)}}$ & 46.3$_{\scalebox{0.7}{\(\pm 0.2\)}}$ & 46.4$_{\scalebox{0.7}{\(\pm 0.2\)}}$ & \textbf{48.0}$_{\scalebox{0.7}{\(\pm 0.4\)}}$ & 45.2$_{\scalebox{0.7}{\(\pm 0.6\)}}$ & 45.4$_{\scalebox{0.7}{\(\pm 0.4\)}}$ \\
      & Cross-Entropy & \textbf{45.4}$_{\scalebox{0.7}{\(\pm 0.3\)}}$ & \textbf{46.0}$_{\scalebox{0.7}{\(\pm 0.6\)}}$ & 46.6$_{\scalebox{0.7}{\(\pm 0.3\)}}$ & \textbf{47.1}$_{\scalebox{0.7}{\(\pm 0.3\)}}$ & 46.4$_{\scalebox{0.7}{\(\pm 0.5\)}}$ & \textbf{46.2}$_{\scalebox{0.7}{\(\pm 0.3\)}}$ & \textbf{47.7}$_{\scalebox{0.7}{\(\pm 0.4\)}}$ \\ \up

    \multirow{3}{*}{Spearman’s Rho}
      & Triplet Ranking & \textbf{31.3}$_{\scalebox{0.7}{\(\pm 0.5\)}}$ & \textbf{32.5}$_{\scalebox{0.7}{\(\pm 0.5\)}}$ & \textbf{36.9}$_{\scalebox{0.7}{\(\pm 0.4\)}}$ & \textbf{36.8}$_{\scalebox{0.7}{\(\pm 0.5\)}}$ & \textbf{39.6}$_{\scalebox{0.7}{\(\pm 0.3\)}}$ & 33.9$_{\scalebox{0.7}{\(\pm 0.5\)}}$ & \textbf{35.0}$_{\scalebox{0.7}{\(\pm 0.5\)}}$ \\
      & Pairwise Ranking & 30.7$_{\scalebox{0.7}{\(\pm 0.7\)}}$ & 32.2$_{\scalebox{0.7}{\(\pm 0.2\)}}$ & 36.5$_{\scalebox{0.7}{\(\pm 0.9\)}}$ & 35.5$_{\scalebox{0.7}{\(\pm 0.6\)}}$ & 39.5$_{\scalebox{0.7}{\(\pm 1.0\)}}$ & \textbf{34.0}$_{\scalebox{0.7}{\(\pm 0.8\)}}$ & 34.8$_{\scalebox{0.7}{\(\pm 0.5\)}}$ \\
      & Cross-Entropy & 29.2$_{\scalebox{0.7}{\(\pm 0.7\)}}$ & 30.8$_{\scalebox{0.7}{\(\pm 0.6\)}}$ & 34.6$_{\scalebox{0.7}{\(\pm 0.7\)}}$ & 34.5$_{\scalebox{0.7}{\(\pm 0.6\)}}$ & 36.9$_{\scalebox{0.7}{\(\pm 1.0\)}}$ & 33.1$_{\scalebox{0.7}{\(\pm 0.6\)}}$ & 32.5$_{\scalebox{0.7}{\(\pm 1.0\)}}$ \\ \up

    \multirow{3}{*}{Kendall’s Tau}
      & Triplet Ranking & \textbf{24.2}$_{\scalebox{0.7}{\(\pm 0.3\)}}$ & \textbf{25.0}$_{\scalebox{0.7}{\(\pm 0.4\)}}$ & \textbf{28.6}$_{\scalebox{0.7}{\(\pm 0.3\)}}$ & \textbf{28.6}$_{\scalebox{0.7}{\(\pm 0.5\)}}$ & \textbf{30.7}$_{\scalebox{0.7}{\(\pm 0.3\)}}$ & 26.3$_{\scalebox{0.7}{\(\pm 0.4\)}}$ & \textbf{27.2}$_{\scalebox{0.7}{\(\pm 0.3\)}}$ \\
      & Pairwise Ranking & 23.8$_{\scalebox{0.7}{\(\pm 0.5\)}}$ & 24.8$_{\scalebox{0.7}{\(\pm 0.0\)}}$ & 28.2$_{\scalebox{0.7}{\(\pm 0.7\)}}$ & 27.5$_{\scalebox{0.7}{\(\pm 0.5\)}}$ & \textbf{30.7}$_{\scalebox{0.7}{\(\pm 0.8\)}}$ & \textbf{26.4}$_{\scalebox{0.7}{\(\pm 0.7\)}}$ & 27.0$_{\scalebox{0.7}{\(\pm 0.3\)}}$ \\
      & Cross-Entropy & 22.7$_{\scalebox{0.7}{\(\pm 0.5\)}}$ & 23.7$_{\scalebox{0.7}{\(\pm 0.5\)}}$ & 26.8$_{\scalebox{0.7}{\(\pm 0.6\)}}$ & 26.8$_{\scalebox{0.7}{\(\pm 0.4\)}}$ & 28.8$_{\scalebox{0.7}{\(\pm 0.8\)}}$ & 25.7$_{\scalebox{0.7}{\(\pm 0.5\)}}$ & 25.3$_{\scalebox{0.7}{\(\pm 0.8\)}}$ \\ 
    \midrule

    \multicolumn{9}{c}{\textbf{Kurtosis}} \\
    \midrule
    \multirow{3}{*}{F1 Score}
      & Triplet Ranking & \textbf{44.6}$_{\scalebox{0.7}{\(\pm 0.3\)}}$ & \textbf{47.3}$_{\scalebox{0.7}{\(\pm 0.3\)}}$ & \textbf{45.0}$_{\scalebox{0.7}{\(\pm 0.3\)}}$ & \textbf{47.2}$_{\scalebox{0.7}{\(\pm 0.1\)}}$ & \textbf{39.9}$_{\scalebox{0.7}{\(\pm 0.3\)}}$ & \textbf{46.3}$_{\scalebox{0.7}{\(\pm 0.7\)}}$ & 48.4$_{\scalebox{0.7}{\(\pm 0.6\)}}$ \\
      & Pairwise Ranking & 44.4$_{\scalebox{0.7}{\(\pm 0.4\)}}$ & 46.8$_{\scalebox{0.7}{\(\pm 0.3\)}}$ & 44.9$_{\scalebox{0.7}{\(\pm 0.5\)}}$ & 46.6$_{\scalebox{0.7}{\(\pm 0.1\)}}$ & 39.6$_{\scalebox{0.7}{\(\pm 0.8\)}}$ & 45.7$_{\scalebox{0.7}{\(\pm 0.6\)}}$ & \textbf{48.5}$_{\scalebox{0.7}{\(\pm 0.9\)}}$ \\
      & Cross-Entropy & 44.0$_{\scalebox{0.7}{\(\pm 0.5\)}}$ & 46.2$_{\scalebox{0.7}{\(\pm 0.4\)}}$ & 44.5$_{\scalebox{0.7}{\(\pm 0.3\)}}$ & 45.8$_{\scalebox{0.7}{\(\pm 0.2\)}}$ & 39.5$_{\scalebox{0.7}{\(\pm 0.5\)}}$ & 45.2$_{\scalebox{0.7}{\(\pm 0.5\)}}$ & 47.0$_{\scalebox{0.7}{\(\pm 0.4\)}}$ \\ \up

    \multirow{3}{*}{Spearman’s Rho}
      & Triplet Ranking & 30.0$_{\scalebox{0.7}{\(\pm 0.2\)}}$ & \textbf{35.2}$_{\scalebox{0.7}{\(\pm 0.5\)}}$ & \textbf{31.0}$_{\scalebox{0.7}{\(\pm 0.4\)}}$ & \textbf{36.0}$_{\scalebox{0.7}{\(\pm 0.6\)}}$ & \textbf{17.1}$_{\scalebox{0.7}{\(\pm 1.3\)}}$ & \textbf{33.1}$_{\scalebox{0.7}{\(\pm 1.1\)}}$ & \textbf{39.0}$_{\scalebox{0.7}{\(\pm 0.6\)}}$ \\
      & Pairwise Ranking & \textbf{30.1}$_{\scalebox{0.7}{\(\pm 0.8\)}}$ & 34.6$_{\scalebox{0.7}{\(\pm 0.8\)}}$ & 30.6$_{\scalebox{0.7}{\(\pm 0.8\)}}$ & 35.1$_{\scalebox{0.7}{\(\pm 0.5\)}}$ & 16.5$_{\scalebox{0.7}{\(\pm 1.2\)}}$ & 32.7$_{\scalebox{0.7}{\(\pm 1.1\)}}$ & 38.7$_{\scalebox{0.7}{\(\pm 1.3\)}}$ \\
      & Cross-Entropy & 28.1$_{\scalebox{0.7}{\(\pm 1.1\)}}$ & 33.3$_{\scalebox{0.7}{\(\pm 0.9\)}}$ & 29.3$_{\scalebox{0.7}{\(\pm 0.7\)}}$ & 33.8$_{\scalebox{0.7}{\(\pm 0.7\)}}$ & 16.1$_{\scalebox{0.7}{\(\pm 1.3\)}}$ & 30.9$_{\scalebox{0.7}{\(\pm 0.9\)}}$ & 35.0$_{\scalebox{0.7}{\(\pm 0.8\)}}$ \\ \up

    \multirow{3}{*}{Kendall’s Tau}
      & Triplet Ranking & 23.3$_{\scalebox{0.7}{\(\pm 0.2\)}}$ & \textbf{27.6}$_{\scalebox{0.7}{\(\pm 0.4\)}}$ & \textbf{24.2}$_{\scalebox{0.7}{\(\pm 0.3\)}}$ & \textbf{28.2}$_{\scalebox{0.7}{\(\pm 0.5\)}}$ & \textbf{13.2}$_{\scalebox{0.7}{\(\pm 1.0\)}}$ & \textbf{25.9}$_{\scalebox{0.7}{\(\pm 0.9\)}}$ & \textbf{30.6}$_{\scalebox{0.7}{\(\pm 0.5\)}}$ \\
      & Pairwise Ranking & \textbf{23.4}$_{\scalebox{0.7}{\(\pm 0.6\)}}$ & 27.1$_{\scalebox{0.7}{\(\pm 0.6\)}}$ & 23.9$_{\scalebox{0.7}{\(\pm 0.6\)}}$ & 27.5$_{\scalebox{0.7}{\(\pm 0.3\)}}$ & 12.7$_{\scalebox{0.7}{\(\pm 1.0\)}}$ & 25.5$_{\scalebox{0.7}{\(\pm 0.9\)}}$ & 30.3$_{\scalebox{0.7}{\(\pm 1.0\)}}$ \\
      & Cross-Entropy & 21.9$_{\scalebox{0.7}{\(\pm 0.9\)}}$ & 26.1$_{\scalebox{0.7}{\(\pm 0.7\)}}$ & 22.9$_{\scalebox{0.7}{\(\pm 0.6\)}}$ & 26.4$_{\scalebox{0.7}{\(\pm 0.6\)}}$ & 12.4$_{\scalebox{0.7}{\(\pm 1.1\)}}$ & 24.1$_{\scalebox{0.7}{\(\pm 0.7\)}}$ & 27.5$_{\scalebox{0.7}{\(\pm 0.6\)}}$ \\ 
    \midrule

    \multicolumn{9}{c}{\textbf{Sortino Ratio}} \\
    \midrule
    \multirow{3}{*}{F1 Score}
      & Triplet Ranking & 48.2$_{\scalebox{0.7}{\(\pm 0.7\)}}$ & \textbf{39.9}$_{\scalebox{0.7}{\(\pm 1.1\)}}$ & 48.5$_{\scalebox{0.7}{\(\pm 1.3\)}}$ & \textbf{35.6}$_{\scalebox{0.7}{\(\pm 1.0\)}}$ & \textbf{53.8}$_{\scalebox{0.7}{\(\pm 1.5\)}}$ & 49.0$_{\scalebox{0.7}{\(\pm 0.8\)}}$ & \textbf{59.8}$_{\scalebox{0.7}{\(\pm 0.7\)}}$ \\
      & Pairwise Ranking & 48.1$_{\scalebox{0.7}{\(\pm 0.3\)}}$ & 39.5$_{\scalebox{0.7}{\(\pm 1.9\)}}$ & 48.3$_{\scalebox{0.7}{\(\pm 2.0\)}}$ & 35.2$_{\scalebox{0.7}{\(\pm 0.5\)}}$ & 52.2$_{\scalebox{0.7}{\(\pm 1.3\)}}$ & \textbf{50.2}$_{\scalebox{0.7}{\(\pm 0.4\)}}$ & 59.5$_{\scalebox{0.7}{\(\pm 1.2\)}}$ \\
      & Cross-Entropy &  \textbf{51.3}$_{\scalebox{0.7}{\(\pm 0.6\)}}$ & 30.5$_{\scalebox{0.7}{\(\pm 6.4\)}}$ & \textbf{53.6}$_{\scalebox{0.7}{\(\pm 2.1\)}}$ & 31.9$_{\scalebox{0.7}{\(\pm 0.9\)}}$ & 47.3$_{\scalebox{0.7}{\(\pm 1.9\)}}$ & 47.9$_{\scalebox{0.7}{\(\pm 1.0\)}}$ & 53.9$_{\scalebox{0.7}{\(\pm 1.7\)}}$ \\ \up

    \multirow{3}{*}{Spearman’s Rho}
      & Triplet Ranking & 31.8$_{\scalebox{0.7}{\(\pm 1.4\)}}$ & 15.4$_{\scalebox{0.7}{\(\pm 3.4\)}}$ & 40.6$_{\scalebox{0.7}{\(\pm 3.1\)}}$ & \textbf{7.5}$_{\scalebox{0.7}{\(\pm 1.9\)}}$ & \textbf{44.1}$_{\scalebox{0.7}{\(\pm 1.6\)}}$ & 44.6$_{\scalebox{0.7}{\(\pm 1.5\)}}$ & \textbf{61.4}$_{\scalebox{0.7}{\(\pm 0.8\)}}$ \\
      & Pairwise Ranking & 32.7$_{\scalebox{0.7}{\(\pm 0.7\)}}$ & 18.4$_{\scalebox{0.7}{\(\pm 7.0\)}}$ & 41.1$_{\scalebox{0.7}{\(\pm 5.0\)}}$ & 6.8$_{\scalebox{0.7}{\(\pm 1.7\)}}$ & 43.2$_{\scalebox{0.7}{\(\pm 2.5\)}}$ & 46.0$_{\scalebox{0.7}{\(\pm 1.3\)}}$ & 61.1$_{\scalebox{0.7}{\(\pm 2.1\)}}$ \\
      & Cross-Entropy & \textbf{33.1}$_{\scalebox{0.7}{\(\pm 0.9\)}}$ & \textbf{28.3}$_{\scalebox{0.7}{\(\pm 8.5\)}}$ & \textbf{45.8}$_{\scalebox{0.7}{\(\pm 4.5\)}}$ & -0.3$_{\scalebox{0.7}{\(\pm 1.3\)}}$ & 38.1$_{\scalebox{0.7}{\(\pm 1.7\)}}$ & \textbf{44.9}$_{\scalebox{0.7}{\(\pm 0.8\)}}$ & 56.0$_{\scalebox{0.7}{\(\pm 0.8\)}}$ \\ \up

    \multirow{3}{*}{Kendall’s Tau}
      & Triplet Ranking & 25.2$_{\scalebox{0.7}{\(\pm 1.0\)}}$ & 11.9$_{\scalebox{0.7}{\(\pm 2.7\)}}$ & 31.8$_{\scalebox{0.7}{\(\pm 2.6\)}}$ & \textbf{5.6}$_{\scalebox{0.7}{\(\pm 1.5\)}}$ & \textbf{35.0}$_{\scalebox{0.7}{\(\pm 1.5\)}}$ & 34.7$_{\scalebox{0.7}{\(\pm 1.3\)}}$ & \textbf{49.2}$_{\scalebox{0.7}{\(\pm 0.7\)}}$ \\
      & Pairwise Ranking & 25.8$_{\scalebox{0.7}{\(\pm 0.5\)}}$ & 14.3$_{\scalebox{0.7}{\(\pm 5.6\)}}$ & 32.1$_{\scalebox{0.7}{\(\pm 3.9\)}}$ & 5.1$_{\scalebox{0.7}{\(\pm 1.3\)}}$ & 34.0$_{\scalebox{0.7}{\(\pm 2.2\)}}$ & \textbf{35.8}$_{\scalebox{0.7}{\(\pm 1.1\)}}$ & 49.0$_{\scalebox{0.7}{\(\pm 1.8\)}}$ \\
      & Cross-Entropy & \textbf{26.3}$_{\scalebox{0.7}{\(\pm 0.8\)}}$ & \textbf{22.2}$_{\scalebox{0.7}{\(\pm 6.9\)}}$ & \textbf{36.3}$_{\scalebox{0.7}{\(\pm 3.8\)}}$ & -0.4$_{\scalebox{0.7}{\(\pm 1.0\)}}$ & 30.1$_{\scalebox{0.7}{\(\pm 1.5\)}}$ & 35.0$_{\scalebox{0.7}{\(\pm 0.8\)}}$ & 44.5$_{\scalebox{0.7}{\(\pm 0.7\)}}$ \down\\ \hline
  \end{tabular}}
  {Results are reported as the mean ± standard deviation across five random seeds (98, 83, 62, 42, 21).}
\end{table}

When using cross-entropy loss for a ranking task with ordinal labels (i.e., \(2 > 1 > 0\)), the model inherently treats the problem as a classification task, predicting one of the three discrete categories. While this approach is well-suited for classification metrics such as F1-score, it poses challenges for ranking-based evaluations. Specifically, metrics like Spearman’s rho and Kendall’s tau require continuous or ordinal predictions rather than categorical class outputs. A straightforward approach to obtain ranked predictions from a classification model is to apply a hard assignment via the argmax function, which selects the most probable class. However, this method discards the probability distribution across the three classes, potentially losing valuable ranking information. To better preserve this information, we adapt the cross-entropy loss by converting probabilities into ranking scores. Since the model outputs a probability distribution over the three ordinal classes, we derive a ranking score by computing the expected value of the predicted distribution as follows:  
\begin{equation}
    \text{Risk Score} = \sum_{i=0}^{2} i \cdot P(y=i),
\end{equation}
This formulation interprets the prediction as a weighted sum of class probabilities, producing a continuous score rather than a discrete category. By using this score approximation, we preserve ranking nuances while ensuring compatibility with ranking-based evaluation metrics. This approach is intuitive, as the ordinal classes inherently carry a notion of magnitude i.e., higher-class labels should contribute more weight reflecting their relative importance in the ranking.

Overall, the results presented in Table \ref{tab:abl1_all}, along with the paired Cohen’s d values reported in Table \ref{tab:cohendresults}, indicate that incorporating triplet ranking loss leads to improved predictive performance. This enhancement suggests that triplet ranking loss is more effective at modeling the relative ordering of risk levels compared to both pairwise ranking loss and cross-entropy loss. While the gains in ranking metrics are modest when comparing to the pairwise ranking loss, the Cohen’s d values for the F1 scores exhibit statistically significant differences, with small but non-negligible effect sizes (greater than 0.2). By explicitly enforcing structured comparisons among anchor, positive, and negative samples, triplet loss enables the model to learn more refined distinctions in risk levels, thereby improving its ability to produce accurate ranking predictions.

\subsection{Replacing hierarchical structure with Mean or Max Pooling}

In this subsection, we investigate the impact of replacing the hierarchical structure with mean or max pooling. Our objective is to assess not only the interpretability benefits of hierarchical attention but also its contribution to overall model performance. To conduct this study, we entirely substitute the hierarchical model with mean or max pooling. The document embeddings are computed as follows:
\begin{equation}
\mathbf{d}_{\text{mean}} = \frac{1}{L} \sum_{i=1}^{L} \mathbf{h}_i,
\end{equation}
\begin{equation}
\mathbf{d}_{\text{max}} = \max_{1 \leq i \leq L} \mathbf{h}_i.
\end{equation}
Here, $\mathbf{h}_i \in \mathbb{R}^{m}$ denotes the token-level embedding at position $i$, and $L$ is the total number of tokens. Since TinyBERT accepts a maximum of 512 tokens, we first encode the document using a hierarchical structure. After obtaining token-level embeddings, we flatten the sentence-level structure originally in $\mathbb{R}^{L \times l \times m}$ into a single sequence of tokens in $\mathbb{R}^{Ll \times m}$. We then apply mean or max pooling over the token dimension, resulting in a fixed-size document representation (i.e., $\mathbf{d}_{\text{mean}}$ or $\mathbf{d}_{\text{max}}$) in $\mathbb{R}^{m}$.

As shown in Table~\ref{tab:abl2_all}, the hierarchical structure generally outperforms both mean and max pooling methods. Max pooling demonstrates relatively strong performance in some cases, especially in the standard deviation and kurtosis risk measurements. Mean pooling, on the other hand, performs slightly better in the skewness and Sortino ratio settings. However, a closer look at the paired Cohen's d test relative to the hierarchical model (TinyXRA) reveals that the hierarchical approach still outperforms both pooling methods, especially in the F1 score domain, where it achieves positive effect sizes across the board. Among the two pooling strategies, mean pooling outperforms max pooling, with lower Cohen’s d values relative to the hierarchical baseline. This suggests that mean pooling provides more stable performance than max pooling.

In terms of predictive performance, the hierarchical model demonstrates robustness across most settings, particularly under the challenging Sortino ratio metric. Notably, max pooling performs poorly under this metric, failing completely in 2021 with an Kendall's Tau score of 0\%, while the hierarchical model maintains strong performance. Similarly, in the skewness setting, the hierarchical structure again outperforms both pooling alternatives. These performance gains are likely attributable to the hierarchical attention mechanism, which enables the model to assign greater weight to more informative sentences and key words. This selective focus is especially beneficial in scenarios where relevant signals are subtle or dispersed throughout the document, necessitating a more sophisticated model to extract fine-grained information.

\begin{table}[htb!]
  \TABLE
  {F1 Score (\%), Spearman’s Rho (\%), and Kendall’s Tau (\%) for various document embeddings aggregation methods across test years (2024–2018) under different risk measurements.\label{tab:abl2_all}}
  {\begin{tabular}{@{}l l l l l l l l l@{}}
    \toprule
    Evaluation Metrics & Aggregation Methods & 2024 & 2023 & 2022 & 2021 & 2020 & 2019 & 2018 \\
    \midrule
    \multicolumn{9}{c}{\textbf{Standard Deviation}} \\
    \midrule
    \multirow{3}{*}{F1 Score}
      & Hierarchical & \underline{76.0}$_{\scalebox{0.7}{\(\pm 0.3\)}}$ & \textbf{74.5}$_{\scalebox{0.7}{\(\pm 0.2\)}}$ & \underline{75.1}$_{\scalebox{0.7}{\(\pm 0.5\)}}$ & \underline{74.1}$_{\scalebox{0.7}{\(\pm 0.3\)}}$ & \underline{67.6}$_{\scalebox{0.7}{\(\pm 0.7\)}}$ & \textbf{76.5}$_{\scalebox{0.7}{\(\pm 0.3\)}}$ & \underline{78.1}$_{\scalebox{0.7}{\(\pm 0.2\)}}$ \\
      & Mean Pooling & 74.5$_{\scalebox{0.7}{\(\pm 0.8\)}}$ & 73.1$_{\scalebox{0.7}{\(\pm 0.2\)}}$ & 73.1$_{\scalebox{0.7}{\(\pm 0.5\)}}$ & 73.0$_{\scalebox{0.7}{\(\pm 0.8\)}}$ & 65.5$_{\scalebox{0.7}{\(\pm 0.8\)}}$ & 75.0$_{\scalebox{0.7}{\(\pm 0.3\)}}$ & 76.7$_{\scalebox{0.7}{\(\pm 0.3\)}}$  \\
      & Max Pooling & \textbf{76.1}$_{\scalebox{0.7}{\(\pm 0.3\)}}$ & \underline{74.4}$_{\scalebox{0.7}{\(\pm 0.4\)}}$ & \textbf{75.5}$_{\scalebox{0.7}{\(\pm 0.5\)}}$ & \textbf{74.7}$_{\scalebox{0.7}{\(\pm 0.3\)}}$ & \textbf{68.1}$_{\scalebox{0.7}{\(\pm 0.7\)}}$ & \underline{76.0}$_{\scalebox{0.7}{\(\pm 0.4\)}}$ & \textbf{78.6}$_{\scalebox{0.7}{\(\pm 0.3\)}}$ \\ \up

    \multirow{3}{*}{Spearman’s Rho}
      & Hierarchical & \textbf{80.3}$_{\scalebox{0.7}{\(\pm 0.1\)}}$ & \textbf{78.8}$_{\scalebox{0.7}{\(\pm 0.3\)}}$ & \textbf{80.0}$_{\scalebox{0.7}{\(\pm 0.2\)}}$ & \underline{80.1}$_{\scalebox{0.7}{\(\pm 0.2\)}}$ & \underline{72.7}$_{\scalebox{0.7}{\(\pm 0.9\)}}$ & \underline{81.2}$_{\scalebox{0.7}{\(\pm 0.4\)}}$ & \underline{82.9}$_{\scalebox{0.7}{\(\pm 0.1\)}}$ \\
      & Mean Pooling & \underline{78.7}$_{\scalebox{0.7}{\(\pm 0.6\)}}$ & 77.5$_{\scalebox{0.7}{\(\pm 0.1\)}}$ & 77.8$_{\scalebox{0.7}{\(\pm 0.4\)}}$ & 78.9$_{\scalebox{0.7}{\(\pm 0.6\)}}$ & 70.7$_{\scalebox{0.7}{\(\pm 0.6\)}}$ & 80.0$_{\scalebox{0.7}{\(\pm 0.1\)}}$ & 81.7$_{\scalebox{0.7}{\(\pm 0.1\)}}$ \\
      & Max Pooling & \textbf{80.3}$_{\scalebox{0.7}{\(\pm 0.2\)}}$ & \underline{78.7}$_{\scalebox{0.7}{\(\pm 0.3\)}}$ & \underline{79.9}$_{\scalebox{0.7}{\(\pm 0.3\)}}$ & \textbf{80.9}$_{\scalebox{0.7}{\(\pm 0.2\)}}$ & \textbf{73.3}$_{\scalebox{0.7}{\(\pm 0.5\)}}$ & \textbf{81.4}$_{\scalebox{0.7}{\(\pm 0.2\)}}$ & \textbf{83.4}$_{\scalebox{0.7}{\(\pm 0.2\)}}$ \\ \up

    \multirow{3}{*}{Kendall’s Tau}
      & Hierarchical & \textbf{67.2}$_{\scalebox{0.7}{\(\pm 0.1\)}}$ & \textbf{65.7}$_{\scalebox{0.7}{\(\pm 0.3\)}}$ & \textbf{66.9}$_{\scalebox{0.7}{\(\pm 0.2\)}}$ & \underline{66.8}$_{\scalebox{0.7}{\(\pm 0.2\)}}$ & \underline{59.4}$_{\scalebox{0.7}{\(\pm 0.9\)}}$ & \underline{67.8}$_{\scalebox{0.7}{\(\pm 0.3\)}}$ & \underline{69.6}$_{\scalebox{0.7}{\(\pm 0.1\)}}$ \\
      & Mean Pooling & 65.6$_{\scalebox{0.7}{\(\pm 0.6\)}}$ & 64.4$_{\scalebox{0.7}{\(\pm 0.1\)}}$ & 64.7$_{\scalebox{0.7}{\(\pm 0.4\)}}$ & 65.6$_{\scalebox{0.7}{\(\pm 0.6\)}}$ & 57.4$_{\scalebox{0.7}{\(\pm 0.6\)}}$ & 66.6$_{\scalebox{0.7}{\(\pm 0.1\)}}$ & 68.3$_{\scalebox{0.7}{\(\pm 0.1\)}}$ \\
      & Max Pooling & \underline{67.1}$_{\scalebox{0.7}{\(\pm 0.2\)}}$ & \underline{65.4}$_{\scalebox{0.7}{\(\pm 0.2\)}}$ & \underline{66.7}$_{\scalebox{0.7}{\(\pm 0.4\)}}$ & \textbf{67.4}$_{\scalebox{0.7}{\(\pm 0.2\)}}$ & \textbf{59.9}$_{\scalebox{0.7}{\(\pm 0.4\)}}$ & \textbf{68.0}$_{\scalebox{0.7}{\(\pm 0.2\)}}$ & \textbf{70.1}$_{\scalebox{0.7}{\(\pm 0.2\)}}$ \\ 
    \midrule

    \multicolumn{9}{c}{\textbf{Skewness}} \\
    \midrule
    \multirow{3}{*}{F1 Score}
      & Hierarchical & \underline{44.4}$_{\scalebox{0.7}{\(\pm 0.2\)}}$ & \textbf{44.1}$_{\scalebox{0.7}{\(\pm 0.6\)}}$ & \textbf{46.8}$_{\scalebox{0.7}{\(\pm 0.5\)}}$ & \textbf{47.1}$_{\scalebox{0.7}{\(\pm 0.4\)}}$ & \textbf{47.8}$_{\scalebox{0.7}{\(\pm 0.4\)}}$ & \underline{45.1}$_{\scalebox{0.7}{\(\pm 0.4\)}}$ & \textbf{46.0}$_{\scalebox{0.7}{\(\pm 0.3\)}}$ \\
      & Mean Pooling & \textbf{44.5}$_{\scalebox{0.7}{\(\pm 0.2\)}}$ & \underline{44.0}$_{\scalebox{0.7}{\(\pm 0.2\)}}$ & \underline{46.1}$_{\scalebox{0.7}{\(\pm 0.2\)}}$ & 46.1$_{\scalebox{0.7}{\(\pm 0.4\)}}$ & \textbf{47.8}$_{\scalebox{0.7}{\(\pm 0.1\)}}$ & \textbf{45.4}$_{\scalebox{0.7}{\(\pm 0.3\)}}$ & \underline{45.4}$_{\scalebox{0.7}{\(\pm 0.5\)}}$ \\
      & Max Pooling & 44.1$_{\scalebox{0.7}{\(\pm 0.5\)}}$ & 43.7$_{\scalebox{0.7}{\(\pm 0.4\)}}$ & 45.8$_{\scalebox{0.7}{\(\pm 0.5\)}}$ & \underline{46.2}$_{\scalebox{0.7}{\(\pm 0.4\)}}$ & \underline{47.4}$_{\scalebox{0.7}{\(\pm 0.5\)}}$ & \textbf{45.4}$_{\scalebox{0.7}{\(\pm 0.5\)}}$ & \underline{45.4}$_{\scalebox{0.7}{\(\pm 0.7\)}}$ \\ \up

    \multirow{3}{*}{Spearman’s Rho}
      & Hierarchical & \textbf{31.3}$_{\scalebox{0.7}{\(\pm 0.5\)}}$ & \textbf{32.5}$_{\scalebox{0.7}{\(\pm 0.5\)}}$ & \textbf{36.9}$_{\scalebox{0.7}{\(\pm 0.4\)}}$ & \textbf{36.8}$_{\scalebox{0.7}{\(\pm 0.5\)}}$ & \textbf{39.6}$_{\scalebox{0.7}{\(\pm 0.3\)}}$ & \textbf{33.9}$_{\scalebox{0.7}{\(\pm 0.5\)}}$ & \textbf{35.0}$_{\scalebox{0.7}{\(\pm 0.5\)}}$ \\
      & Mean Pooling & \underline{30.8}$_{\scalebox{0.7}{\(\pm 0.3\)}}$ & \underline{30.9}$_{\scalebox{0.7}{\(\pm 0.1\)}}$ & \underline{35.8}$_{\scalebox{0.7}{\(\pm 0.2\)}}$ & \underline{35.5}$_{\scalebox{0.7}{\(\pm 0.8\)}}$ & \underline{37.9}$_{\scalebox{0.7}{\(\pm 0.4\)}}$ & \underline{33.2}$_{\scalebox{0.7}{\(\pm 0.2\)}}$ & 32.5$_{\scalebox{0.7}{\(\pm 0.3\)}}$ \\
      & Max Pooling & 29.3$_{\scalebox{0.7}{\(\pm 0.5\)}}$ & 30.7$_{\scalebox{0.7}{\(\pm 1.0\)}}$ & 34.9$_{\scalebox{0.7}{\(\pm 0.7\)}}$ & 35.3$_{\scalebox{0.7}{\(\pm 0.7\)}}$ & 37.1$_{\scalebox{0.7}{\(\pm 0.7\)}}$ & 32.9$_{\scalebox{0.7}{\(\pm 1.4\)}}$ & \underline{32.9}$_{\scalebox{0.7}{\(\pm 1.1\)}}$ \\ \up

    \multirow{3}{*}{Kendall’s Tau}
      & Hierarchical & \textbf{24.2}$_{\scalebox{0.7}{\(\pm 0.3\)}}$ & \textbf{25.0}$_{\scalebox{0.7}{\(\pm 0.4\)}}$ & \textbf{28.6}$_{\scalebox{0.7}{\(\pm 0.3\)}}$ & \textbf{28.6}$_{\scalebox{0.7}{\(\pm 0.5\)}}$ & \textbf{30.7}$_{\scalebox{0.7}{\(\pm 0.3\)}}$ & \textbf{26.3}$_{\scalebox{0.7}{\(\pm 0.4\)}}$ & \textbf{27.2}$_{\scalebox{0.7}{\(\pm 0.3\)}}$ \\
      & Mean Pooling & \underline{23.9}$_{\scalebox{0.7}{\(\pm 0.3\)}}$ & \underline{23.9}$_{\scalebox{0.7}{\(\pm 0.1\)}}$ & \underline{27.7}$_{\scalebox{0.7}{\(\pm 0.2\)}}$ & \underline{27.5}$_{\scalebox{0.7}{\(\pm 0.6\)}}$ & \underline{29.5}$_{\scalebox{0.7}{\(\pm 0.4\)}}$ & \underline{25.7}$_{\scalebox{0.7}{\(\pm 0.2\)}}$ & 25.3$_{\scalebox{0.7}{\(\pm 0.2\)}}$ \\
      & Max Pooling & 22.8$_{\scalebox{0.7}{\(\pm 0.4\)}}$ & 23.7$_{\scalebox{0.7}{\(\pm 0.8\)}}$ & 27.0$_{\scalebox{0.7}{\(\pm 0.6\)}}$ & 27.4$_{\scalebox{0.7}{\(\pm 0.6\)}}$ & 28.8$_{\scalebox{0.7}{\(\pm 0.6\)}}$ & 25.6$_{\scalebox{0.7}{\(\pm 1.1\)}}$ & \underline{25.7}$_{\scalebox{0.7}{\(\pm 0.9\)}}$ \\ 
    \midrule

    \multicolumn{9}{c}{\textbf{Kurtosis}} \\
    \midrule
    \multirow{3}{*}{F1 Score}
      & Hierarchical & \underline{44.6}$_{\scalebox{0.7}{\(\pm 0.3\)}}$ & \textbf{47.3}$_{\scalebox{0.7}{\(\pm 0.3\)}}$ & 45.0$_{\scalebox{0.7}{\(\pm 0.3\)}}$ & \textbf{47.2}$_{\scalebox{0.7}{\(\pm 0.1\)}}$ & \textbf{39.9}$_{\scalebox{0.7}{\(\pm 0.3\)}}$ & \underline{46.3}$_{\scalebox{0.7}{\(\pm 0.7\)}}$ & \underline{48.4}$_{\scalebox{0.7}{\(\pm 0.6\)}}$ \\
      & Mean Pooling & 44.5$_{\scalebox{0.7}{\(\pm 0.4\)}}$ & 46.5$_{\scalebox{0.7}{\(\pm 0.4\)}}$ & \underline{45.1}$_{\scalebox{0.7}{\(\pm 0.2\)}}$ & \underline{47.0}$_{\scalebox{0.7}{\(\pm 0.2\)}}$ & \underline{39.6}$_{\scalebox{0.7}{\(\pm 0.2\)}}$ & 45.0$_{\scalebox{0.7}{\(\pm 0.3\)}}$ & 47.9$_{\scalebox{0.7}{\(\pm 0.7\)}}$ \\
      & Max Pooling & \textbf{45.1}$_{\scalebox{0.7}{\(\pm 0.4\)}}$ & \underline{46.9}$_{\scalebox{0.7}{\(\pm 0.1\)}}$ & 45.2$_{\scalebox{0.7}{\(\pm 0.5\)}}$ & 46.5$_{\scalebox{0.7}{\(\pm 0.8\)}}$ & 38.9$_{\scalebox{0.7}{\(\pm 0.8\)}}$ & \textbf{46.6}$_{\scalebox{0.7}{\(\pm 0.6\)}}$ & \textbf{48.8}$_{\scalebox{0.7}{\(\pm 0.3\)}}$ \\ \up

    \multirow{3}{*}{Spearman’s Rho}
      & Hierarchical & \underline{30.0}$_{\scalebox{0.7}{\(\pm 0.2\)}}$ & \underline{35.2}$_{\scalebox{0.7}{\(\pm 0.5\)}}$ & \textbf{31.0}$_{\scalebox{0.7}{\(\pm 0.4\)}}$ & \textbf{36.0}$_{\scalebox{0.7}{\(\pm 0.6\)}}$ & \textbf{17.1}$_{\scalebox{0.7}{\(\pm 1.3\)}}$ & \underline{33.1}$_{\scalebox{0.7}{\(\pm 1.1\)}}$ & \underline{39.0}$_{\scalebox{0.7}{\(\pm 0.6\)}}$ \\
      & Mean Pooling & 29.9$_{\scalebox{0.7}{\(\pm 0.5\)}}$ & 34.8$_{\scalebox{0.7}{\(\pm 0.9\)}}$ & \underline{30.9}$_{\scalebox{0.7}{\(\pm 0.5\)}}$ & \underline{35.5}$_{\scalebox{0.7}{\(\pm 0.6\)}}$ & \underline{16.7}$_{\scalebox{0.7}{\(\pm 1.1\)}}$ & 32.0$_{\scalebox{0.7}{\(\pm 0.5\)}}$ & 37.2$_{\scalebox{0.7}{\(\pm 0.5\)}}$ \\
      & Max Pooling & \textbf{31.1}$_{\scalebox{0.7}{\(\pm 0.6\)}}$ & \textbf{35.4}$_{\scalebox{0.7}{\(\pm 0.2\)}}$ & 30.7$_{\scalebox{0.7}{\(\pm 0.3\)}}$ & 35.4$_{\scalebox{0.7}{\(\pm 1.3\)}}$ & 15.5$_{\scalebox{0.7}{\(\pm 0.5\)}}$ & \textbf{34.6}$_{\scalebox{0.7}{\(\pm 0.9\)}}$ & \textbf{40.4}$_{\scalebox{0.7}{\(\pm 0.8\)}}$ \\ \up

    \multirow{3}{*}{Kendall’s Tau}
      & Hierarchical & \underline{23.3}$_{\scalebox{0.7}{\(\pm 0.2\)}}$ & \underline{27.6}$_{\scalebox{0.7}{\(\pm 0.4\)}}$ & \textbf{24.2}$_{\scalebox{0.7}{\(\pm 0.3\)}}$ & \textbf{28.2}$_{\scalebox{0.7}{\(\pm 0.5\)}}$ & \textbf{13.2}$_{\scalebox{0.7}{\(\pm 1.0\)}}$ & \underline{25.9}$_{\scalebox{0.7}{\(\pm 0.9\)}}$ & \textbf{30.6}$_{\scalebox{0.7}{\(\pm 0.5\)}}$ \\
      & Mean Pooling & \underline{23.3}$_{\scalebox{0.7}{\(\pm 0.4\)}}$ & 27.2$_{\scalebox{0.7}{\(\pm 0.7\)}}$ & \underline{24.1}$_{\scalebox{0.7}{\(\pm 0.4\)}}$ & \underline{27.8}$_{\scalebox{0.7}{\(\pm 0.5\)}}$ & \underline{12.9}$_{\scalebox{0.7}{\(\pm 0.9\)}}$ & 24.8$_{\scalebox{0.7}{\(\pm 0.4\)}}$ & 29.1$_{\scalebox{0.7}{\(\pm 0.4\)}}$ \\
      & Max Pooling & 24.2$_{\scalebox{0.7}{\(\pm 0.4\)}}$ & 27.7$_{\scalebox{0.7}{\(\pm 0.2\)}}$ & 24.0$_{\scalebox{0.7}{\(\pm 0.3\)}}$ & 27.7$_{\scalebox{0.7}{\(\pm 1.1\)}}$ & 12.0$_{\scalebox{0.7}{\(\pm 0.4\)}}$ & \textbf{27.1}$_{\scalebox{0.7}{\(\pm 0.7\)}}$ & \textbf{31.7}$_{\scalebox{0.7}{\(\pm 0.6\)}}$ \\ 
    \midrule

    \multicolumn{9}{c}{\textbf{Sortino Ratio}} \\
    \midrule
    \multirow{3}{*}{F1 Score}
      & Hierarchical & \underline{48.2}$_{\scalebox{0.7}{\(\pm 0.7\)}}$ & \underline{39.9}$_{\scalebox{0.7}{\(\pm 1.1\)}}$ & \underline{48.5}$_{\scalebox{0.7}{\(\pm 1.3\)}}$ & \textbf{35.6}$_{\scalebox{0.7}{\(\pm 1.0\)}}$ & \textbf{53.8}$_{\scalebox{0.7}{\(\pm 1.5\)}}$ & \textbf{49.0}$_{\scalebox{0.7}{\(\pm 0.8\)}}$ & \textbf{59.8}$_{\scalebox{0.7}{\(\pm 0.7\)}}$ \\
      & Mean Pooling & \textbf{48.5}$_{\scalebox{0.7}{\(\pm 0.5\)}}$ & \textbf{44.0}$_{\scalebox{0.7}{\(\pm 0.7\)}}$ & \textbf{52.0}$_{\scalebox{0.7}{\(\pm 1.6\)}}$ & \underline{34.8}$_{\scalebox{0.7}{\(\pm 0.7\)}}$ & \underline{52.1}$_{\scalebox{0.7}{\(\pm 0.8\)}}$ & \underline{47.6}$_{\scalebox{0.7}{\(\pm 0.5\)}}$ & \underline{59.5}$_{\scalebox{0.7}{\(\pm 1.2\)}}$ \\
      & Max Pooling & 47.8$_{\scalebox{0.7}{\(\pm 0.8\)}}$ & 37.7$_{\scalebox{0.7}{\(\pm 2.1\)}}$ & 45.1$_{\scalebox{0.7}{\(\pm 0.9\)}}$ & 33.9$_{\scalebox{0.7}{\(\pm 0.6\)}}$ & 49.9$_{\scalebox{0.7}{\(\pm 0.9\)}}$ & 47.2$_{\scalebox{0.7}{\(\pm 0.6\)}}$ & 58.4$_{\scalebox{0.7}{\(\pm 0.6\)}}$ \\ \up

    \multirow{3}{*}{Spearman’s Rho}
      & Hierarchical & 31.8$_{\scalebox{0.7}{\(\pm 1.4\)}}$ & \underline{15.4}$_{\scalebox{0.7}{\(\pm 3.4\)}}$ & \underline{40.6}$_{\scalebox{0.7}{\(\pm 3.1\)}}$ & \textbf{7.5}$_{\scalebox{0.7}{\(\pm 1.9\)}}$ & \textbf{44.1}$_{\scalebox{0.7}{\(\pm 1.6\)}}$ & \textbf{44.6}$_{\scalebox{0.7}{\(\pm 1.5\)}}$ & \textbf{61.4}$_{\scalebox{0.7}{\(\pm 0.8\)}}$ \\
      & Mean Pooling & \textbf{33.4}$_{\scalebox{0.7}{\(\pm 0.4\)}}$ & \textbf{28.3}$_{\scalebox{0.7}{\(\pm 2.7\)}}$ & \textbf{46.6}$_{\scalebox{0.7}{\(\pm 3.1\)}}$ & \underline{7.0}$_{\scalebox{0.7}{\(\pm 1.9\)}}$ & \underline{43.4}$_{\scalebox{0.7}{\(\pm 0.7\)}}$ & \underline{42.4}$_{\scalebox{0.7}{\(\pm 0.5\)}}$ & \underline{61.2}$_{\scalebox{0.7}{\(\pm 1.6\)}}$ \\
      & Max Pooling & \underline{32.5}$_{\scalebox{0.7}{\(\pm 1.6\)}}$ & 10.3$_{\scalebox{0.7}{\(\pm 6.7\)}}$ & 32.0$_{\scalebox{0.7}{\(\pm 1.6\)}}$ & 0.0$_{\scalebox{0.7}{\(\pm 2.0\)}}$ & 38.7$_{\scalebox{0.7}{\(\pm 0.5\)}}$ & 39.6$_{\scalebox{0.7}{\(\pm 0.9\)}}$ & 60.2$_{\scalebox{0.7}{\(\pm 0.1\)}}$ \\ \up

    \multirow{3}{*}{Kendall’s Tau}
      & Hierarchical & 25.2$_{\scalebox{0.7}{\(\pm 1.0\)}}$ & \underline{11.9}$_{\scalebox{0.7}{\(\pm 2.7\)}}$ & \underline{31.8}$_{\scalebox{0.7}{\(\pm 2.6\)}}$ & \textbf{5.6}$_{\scalebox{0.7}{\(\pm 1.5\)}}$ & \textbf{35.0}$_{\scalebox{0.7}{\(\pm 1.5\)}}$ & \textbf{34.7}$_{\scalebox{0.7}{\(\pm 1.3\)}}$ & \textbf{49.2}$_{\scalebox{0.7}{\(\pm 0.7\)}}$ \\
      & Mean Pooling & \textbf{26.3}$_{\scalebox{0.7}{\(\pm 0.4\)}}$ & \textbf{22.1}$_{\scalebox{0.7}{\(\pm 2.1\)}}$ & 36.9$_{\scalebox{0.7}{\(\pm 2.5\)}}$ & \underline{5.1}$_{\scalebox{0.7}{\(\pm 1.5\)}}$ & \underline{34.3}$_{\scalebox{0.7}{\(\pm 0.6\)}}$ & \underline{32.7}$_{\scalebox{0.7}{\(\pm 0.5\)}}$ & \underline{49.0}$_{\scalebox{0.7}{\(\pm 1.4\)}}$ \\
      & Max Pooling & \underline{25.8}$_{\scalebox{0.7}{\(\pm 1.3\)}}$ & 8.0$_{\scalebox{0.7}{\(\pm 5.3\)}}$ & 25.0$_{\scalebox{0.7}{\(\pm 1.3\)}}$ & -0.1$_{\scalebox{0.7}{\(\pm 1.6\)}}$ & 30.5$_{\scalebox{0.7}{\(\pm 0.5\)}}$ & 30.7$_{\scalebox{0.7}{\(\pm 0.7\)}}$ & 48.1$_{\scalebox{0.7}{\(\pm 0.2\)}}$ \down\\ \hline
  \end{tabular}}
  {Results are reported as the mean ± standard deviation across five random seeds (98, 83, 62, 42, 21).}
\end{table}

\begin{table}[htb!]
\TABLE
{Consolidated paired Cohen's d tests relative to TinyXRA. \label{tab:cohendresults}}
{\begin{tabular}{@{}l c c c c@{}}
    \hline\up
    Model & $\text{Cohen's d}_{\text{F1}}$ & $\text{Cohen's d}_{\text{Rho}}$ & $\text{Cohen's d}_{\text{Tau}}$ & $\text{Cohen's d}_{\text{all}}$ \\
    \hline 
    \multicolumn{5}{c}{\textbf{Benchmarks}} \\
    \hline
    XRR & 0.74 & 0.59 & 0.63 & 0.59 \\
    TF-IDF & 0.31 & 0.21 & 0.22 & 0.22 \\
    Llama3.2-1B (1-Shot) & 2.25 & 2.03 & 1.92 & 1.84 \\
    Llama3.2-1B (SFT) & 0.95 & 0.41 & 0.43 & 0.47 \\
    Qwen2-0.5B (1-Shot) & 2.00 & 2.15 & 2.02 & 1.85 \\
    Qwen2-0.5B (SFT) & 0.99 & 0.54 & 0.57 & 0.59 \\
    \hline
    \multicolumn{5}{c}{\textbf{Ablation Studies}} \\
    \hline 
    Pairwise Ranking & 0.26 & 0.06 & 0.08 & 0.11 \\
    Cross-Entropy & 0.35 & 0.35 & 0.35 & 0.35 \\
    Mean Pooling & 0.29 & 0.08 & 0.11 & 0.13 \\
    Max Pooling & 0.45 & 0.48 & 0.48 & 0.45 \down\\ \hline
\end{tabular}}
{General interpretation of Cohen's d effect size: 0 = no effect, 0.2 = small, 0.5 = medium, 0.8 = large. Paired Cohen's d tests across all seeds, risk measurements and test years relative to TinyXRA. Positive values indicate that our TinyXRA model outperforms the benchmarks; negative values indicate the benchmarks perform better than our TinyXRA model.}
\end{table}

\subsection{Faithfulness of TinyXRA's Explanation}
Explanations are only useful if they actually reflect what the model is doing. In our context, faithfulness means that the words and sentences the model pays most attention to should genuinely be the ones driving its predictions, if we remove these highly attended elements, performance should suffer. To test whether TinyXRA's explanations are faithful, we ran two experiments: removing the top k\% of words with the highest attention scores, and removing the top k\% of sentences with the highest attention scores. If the explanations are meaningful, we should expect performance to drop when we remove these supposedly important elements.

The results strongly support TinyXRA's faithfulness (see Appendix \ref{appendix:expabl}). When we removed highly attended words, performance consistently dropped across all metrics and time periods we tested. Figures \ref{fig:exp_word_std}-\ref{fig:exp_word_sortino} show this pattern holds regardless of the risk measurement settings. F1 Score, Spearman's Rho, and Kendall's Tau all declined as we removed more top-attended words, and this happened every single year from 2018 to 2024. The performance drop typically shows a steep decline initially when removing the most highly attended words, then levels off to a more gradual decline as we remove additional content. The sentence-level removal experiments tell a similar story (Figures \ref{fig:exp_sent_std}-\ref{fig:exp_sent_sortino}). We see the same basic pattern: steep performance drops when removing the most highly attended sentences, followed by more gradual declines. However, for skewness and kurtosis measures, F1 scores show a second sharp drop when we remove around 80\% of highly attended sentences. This suggests we have reached a breaking point where the documents become so incomplete that classification becomes nearly impossible and the model's performance essentially becomes random at this extreme level of removal.

While some risk measures and years exhibited more pronounced performance drops than others, the overall pattern remained consistent: removing the words or sentences that TinyXRA attends to leads to a decline in performance. This is not the behavior we would expect if attention scores were arbitrary or uninformative. Instead, it provides compelling evidence that TinyXRA's attention mechanism reliably highlights content that is genuinely influential in shaping the model’s predictions. For practitioners, this reinforces the trustworthiness of TinyXRA’s explanations, indicating that the model's attention scores can be relied upon to reveal which parts of the input text are most critical to its decision-making process.
\section{Explainability Insights}

Before diving into the explanations, we like to remind readers that for standard deviation and kurtosis, a higher bin number indicates greater risk associated with the company. In contrast, for skewness and the Sortino ratio, a higher bin number is generally more favorable for investors. In this section, we qualitatively evaluate the explanations on selected examples. For additional explanation examples, we encourage readers to refer to the Appendices.

\subsection{Top 5 Sentences Word Attention Heatmap}

\begin{figure}[hbt!] 
\FIGURE 
{\includegraphics[width=\textwidth]{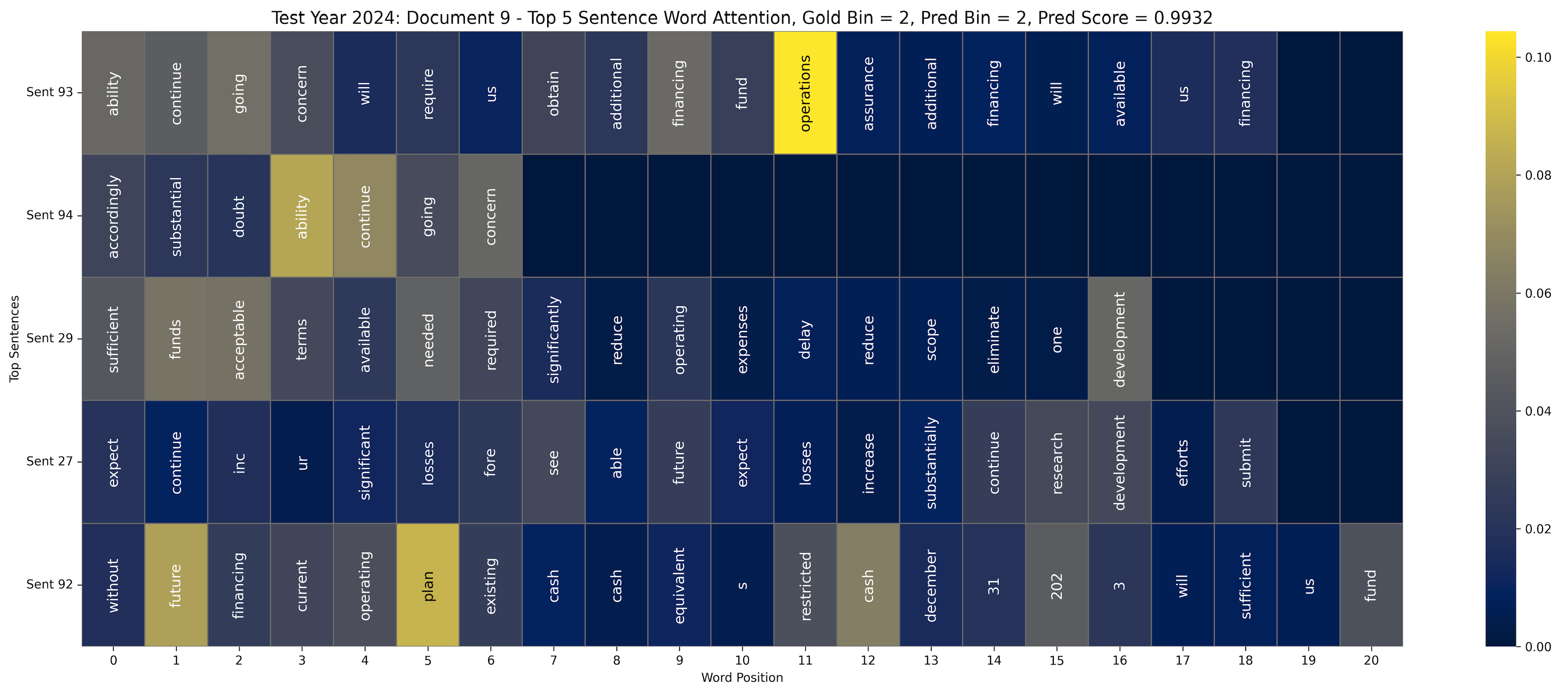}} {Top 5 sentence word attention explanation for financial document 9 for standard deviation risk measurement in test year 2024.
\label{fig:explain_main1}} 
{} 
\end{figure}

To demonstrate our model's interpretability, we present a comprehensive explanation generated from a 2024 financial report using standard deviation as the risk measurement metric, as shown in Figure \ref{fig:explain_main1}.

This visualization enhances model transparency by identifying the specific words and sentences that most significantly influenced the risk score prediction. The horizontal axis displays individual words within sentences, while the vertical axis presents the five most influential sentences from the document, ranked by importance. Color intensity represents the attention weight assigned by our TinyXRA model to each word: brighter yellow indicates higher attention (greater influence on the prediction), while darker shades represent lower attention values. In this example, the document's ground truth risk category is 2 (highest risk), which our model correctly predicted. Consequently, we expect the model to emphasize words reflecting negative sentiment or financial uncertainty.

The most influential sentence (Sent 93) assigns the highest attention score to \texttt{operations}, suggesting the model detected significant uncertainty regarding the company's core operational activities. This likely relates to concerns about securing additional financing or funding, as evidenced by the surrounding context. The second-ranked sentence (Sent 94) prominently features critical phrases such as \texttt{ability continue going concern}, highlighting fundamental doubts about the company's operational sustainability. The presence of \texttt{substantial doubt} at the sentence's beginning further reinforces concerns about financial viability. Sentence 29 emphasizes terms related to \texttt{funds}, \texttt{acceptable} and \texttt{needed}, strongly indicating potential liquidity challenges or difficulties in securing favorable financial conditions. This pattern highlights risks associated with capital accessibility and funding constraints. While Sentence 27 does not highlight individual tokens prominently, the phrase \texttt{significant losses foreseeable future} conveys clear negative sentiment, explaining its inclusion among the top influential sentences and its contribution to the high-risk prediction. Finally, Sentence 92 focuses on \texttt{future}, \texttt{financing}, and \texttt{plan}, suggesting inadequacies in the company's strategic financial planning or funding strategies. These highlighted terms collectively indicate insufficient preparation for maintaining financial stability.

The attention patterns revealed by our TinyXRA model consistently point to substantial uncertainty surrounding the company's operational and financial health. The model's focus on terms related to going concern, funding difficulties, operational disruptions, and strategic planning inadequacies provides compelling evidence supporting its high-risk classification. This alignment between the model's attention mechanism and domain-relevant risk indicators demonstrates the effectiveness of our approach in capturing meaningful financial risk signals. For additional examples of Top 5 Sentences Word-Level attention heatmaps across different risk measurement settings and test years (2018-2024), please refer to Appendix \ref{appendix:wordexp}.

\subsection{Attention-based Word Clouds}

\begin{figure}[hbt!] 
\FIGURE 
{\includegraphics[width=\textwidth]{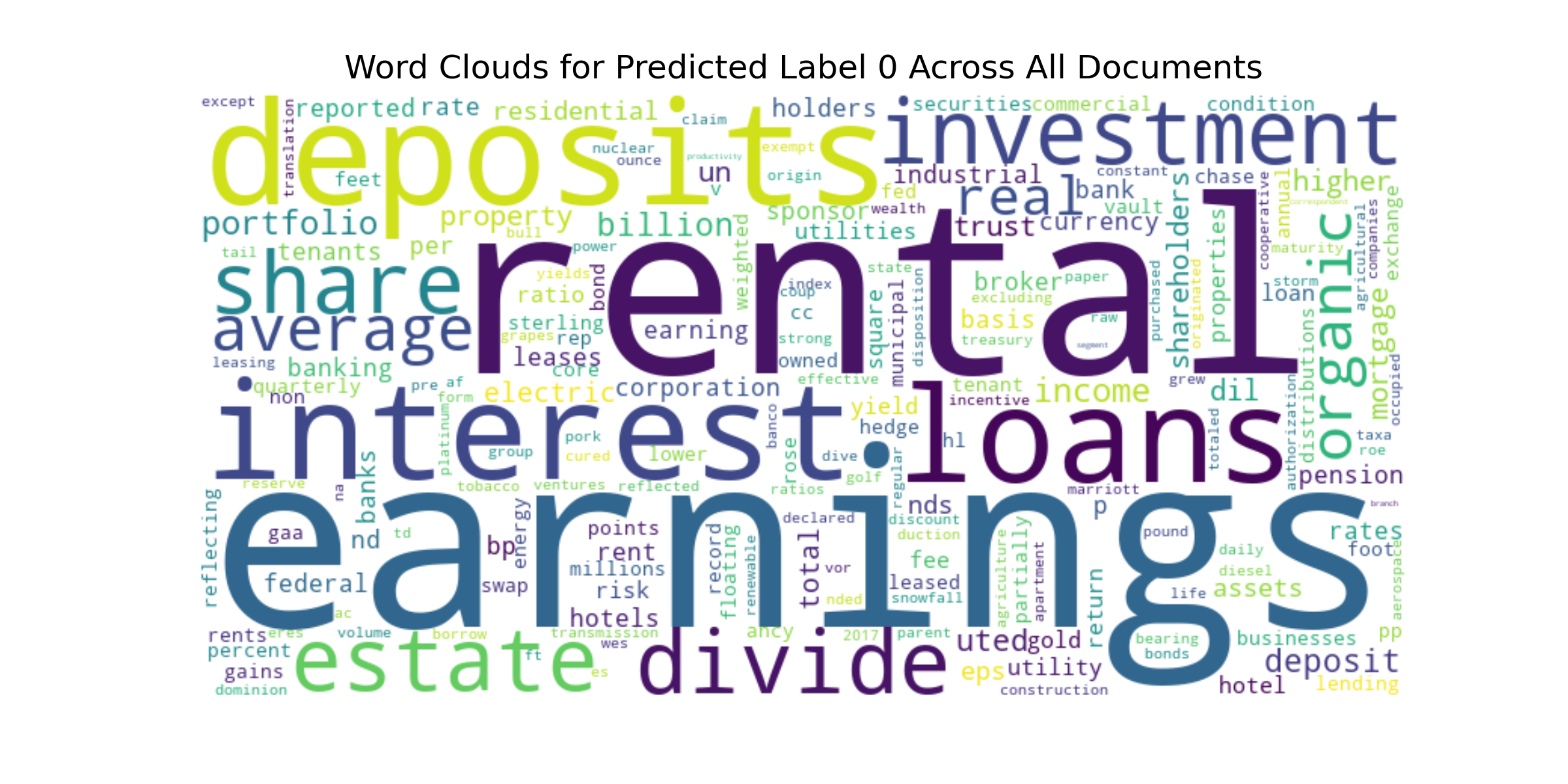}} {Low risk (bin=0) word cloud for test year 2023, using the standard deviation risk measurement.
\label{fig:explain_cloudmain1}}
{} 
\end{figure}

Word clouds provide an intuitive visualization of the most frequently identified terms by our TinyXRA model across documents within each risk category. The size of each word corresponds to its frequency and relative importance in the model's attention mechanism. Figure \ref{fig:explain_cloudmain1} presents a representative word cloud for low-risk companies using the standard deviation risk measurement.

\begin{figure}[hbt!] 
\FIGURE 
{\includegraphics[width=\textwidth]{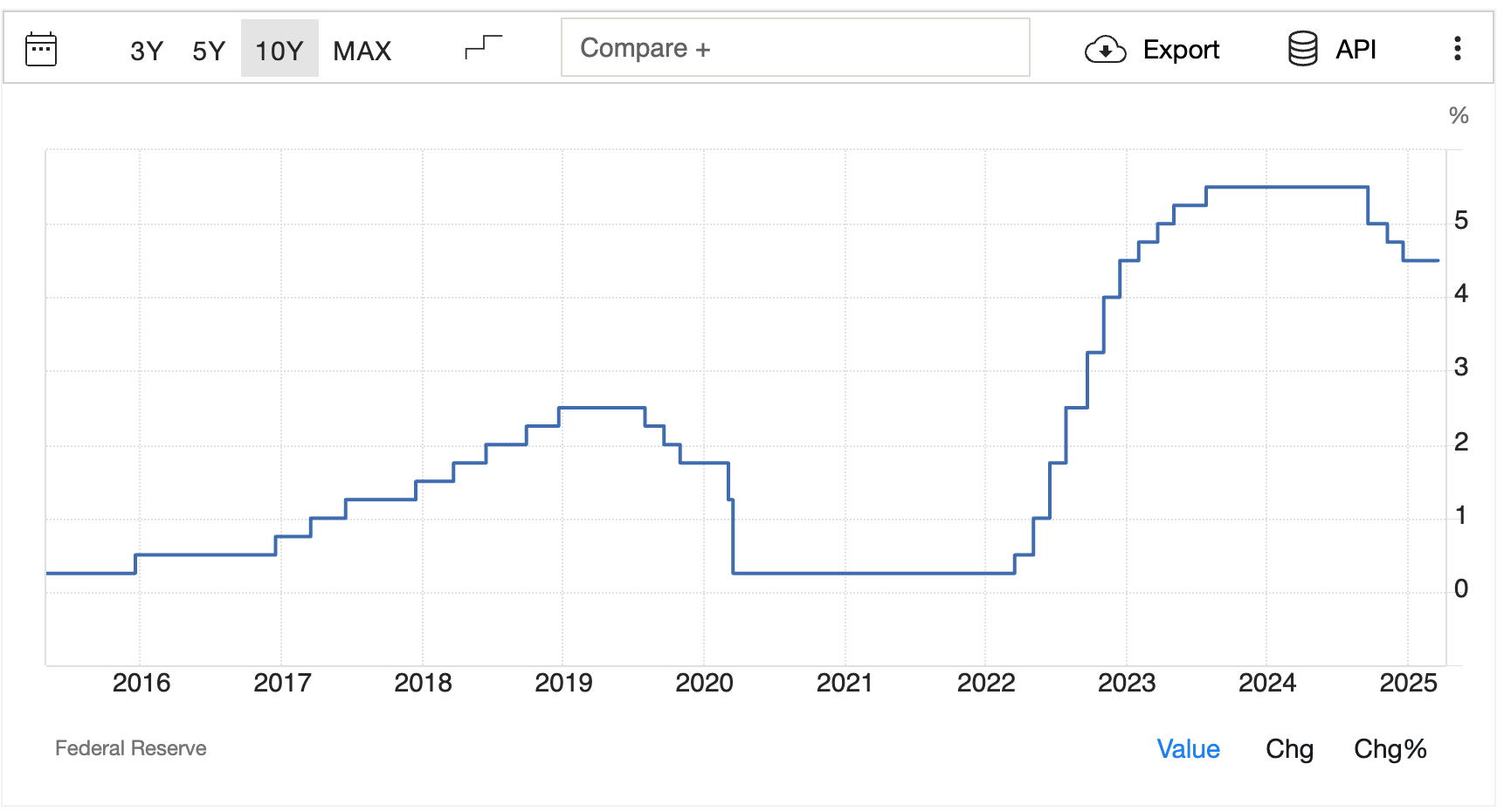}} {United States fed funds interest rate from 2016 to 2025.
\label{fig:interestrate}} 
{} 
\end{figure}

The 2023 word cloud prominently features terms such as \texttt{earnings}, \texttt{rental}, \texttt{deposits}, \texttt{loans}, \texttt{interest}, \texttt{investment}, \texttt{estate}, \texttt{share}, and \texttt{dividend}. These terms collectively characterize a financial profile centered on stable, recurring revenue streams including rental income, dividend payments, and interest earnings, alongside conservative investment strategies. This pattern aligns remarkably well with the prevailing economic conditions of 2023. As illustrated in Figure \ref{fig:interestrate}, the Federal Reserve implemented aggressive interest rate increases throughout this period to combat inflation. In this high interest rate environment, investors and market participants naturally gravitated toward financially stable, low-risk companies that demonstrated consistent earnings capacity, robust dividend policies, and secure cash flow generation. The prominence of these terms reflects several key advantages that low-risk companies possess during periods of rising interest rates. 
\begin{enumerate}
    \item Companies with substantial \textbf{deposit} bases or well-managed lending portfolios directly benefit from higher \textbf{interest} rates through expanded net interest margins, as the widening spread between borrowing and lending rates enhances profitability for traditional banking and financial institutions, making them particularly resilient during interest rate hikes.
    \item Real \textbf{estate} companies emphasizing \textbf{rental} income maintain relatively stable cash flows that provide consistent returns despite increased borrowing costs, with steady \textbf{rental} yields helping to offset rising financing expenses and effectively insulating these firms from market volatility and interest rate sensitivity.
    \item Firms with established \textbf{dividend} policies and strong \textbf{earnings} generation typically exhibit lower sensitivity to interest rate fluctuations due to their financial strength to self-finance operations and investments without heavy reliance on external debt, thereby reducing exposure to higher borrowing costs.
    \item The emphasis on prudent \textbf{investment} approaches becomes particularly valuable during rising rate periods, as conservative strategies focusing on income-generating assets, \textbf{fixed-income} securities and \textbf{gold} become increasingly attractive when returns on interest-sensitive instruments such as bonds and treasury securities improve.
\end{enumerate}

The alignment between our model's attention-based word clouds and established economic principles provides strong validation for the TinyXRA approach. The model successfully captures the fundamental characteristics that distinguish low-risk companies during specific economic cycles, demonstrating its ability to incorporate both firm-specific attributes and broader macroeconomic context in risk assessment. Additional word cloud visualizations spanning different years, risk measurement settings, and predicted risk levels are available in Appendix \ref{appendix:cloud} for further examination.
\section{Conclusion}\label{sec:Conclusion}
\subsection{Summary of Findings}
We introduce TinyXRA, a lightweight and explainable transformer-based model for comprehensive financial risk assessment using the MD\&A sections of 10-K reports. TinyXRA advances beyond traditional approaches that rely solely on volatility by incorporating additional risk measures such as skewness, kurtosis, and the Sortino ratio, to provide a more subtle understanding of company risk profiles. Extensive experiments conducted over seven years of financial data demonstrate that TinyXRA consistently achieves state-of-the-art predictive accuracy across multiple ranking metrics, while also offering interpretable, attention-driven explanations that have been validated quantitatively and qualitatively.

\subsection{Practical Implications}
The TinyXRA framework offers substantial practical value for businesses, investors, and policymakers. By automating the extraction and interpretation of risk signals from unstructured financial texts, organizations can significantly improve the speed and consistency of their risk evaluations. The model's transparent explanations support regulatory compliance and help build trust with stakeholders, while the use of lightweight neural architectures ensures that advanced AI-driven risk assessment can be deployed efficiently even on resource-constrained hardware. Policymakers and regulators may leverage TinyXRA's interpretable outputs to better monitor systemic risk and promote transparency in capital markets, ultimately enabling users to make more informed decisions based on both the model's predictions and its explainability features.

\subsection{Limitations \& Future Work}
Despite these advances, our study faces several important limitations. Earlier portions of the dataset are affected by missing historical prices for delisted or merged companies, as such information is not available via the Yahoo Finance API. Similarly, the SEC CIK to ticker mapping only maintains the latest version, which means companies lacking a current mapping cannot be linked to stock price data which is an issue that disproportionately affects earlier years in the dataset. These constraints may lead to sample omissions and potentially bias results for the earlier test periods. Future research should focus on supplementing missing data sources or integrating alternative financial databases to address historical data gaps

Additionally, our explainability approach faces technical challenges that impact practical usability. For the word clouds, when using BERT tokenization, it fragments meaningful words into constituent parts (e.g., ``uncertainty" may become ``un", ``\#\#certain", ``\#\#ty"). This fragmentation results in word clouds that display incomplete subword tokens rather than coherent, interpretable terms, potentially confusing users who expect to see complete business-relevant vocabulary. Future implementations should consider post-processing techniques to reconstruct full words from subword tokens or employ alternative visualization methods that better accommodate transformer-based tokenization schemes.

Furthermore, the attention-driven word-level explanations may not provide intuitive insights for end users, particularly those without technical backgrounds in natural language processing. Raw attention weights often highlight function words, punctuation, or contextually important but semantically unclear tokens, rather than the business-critical terms that users expect to see emphasized. This disconnect between technical attention mechanisms and user expectations for meaningful explanations suggests the need for more sophisticated interpretation layers that can translate low-level attention patterns into business-relevant insights.


\clearpage

%
%
%
\begin{APPENDICES}
\section{Top 5 Sentences Word-Level Attention Heatmaps for Test Year 2018-2024}
\label{appendix:wordexp}
\begin{figure}[H]
     \FIGURE
     {\includegraphics[width=0.8\textwidth]{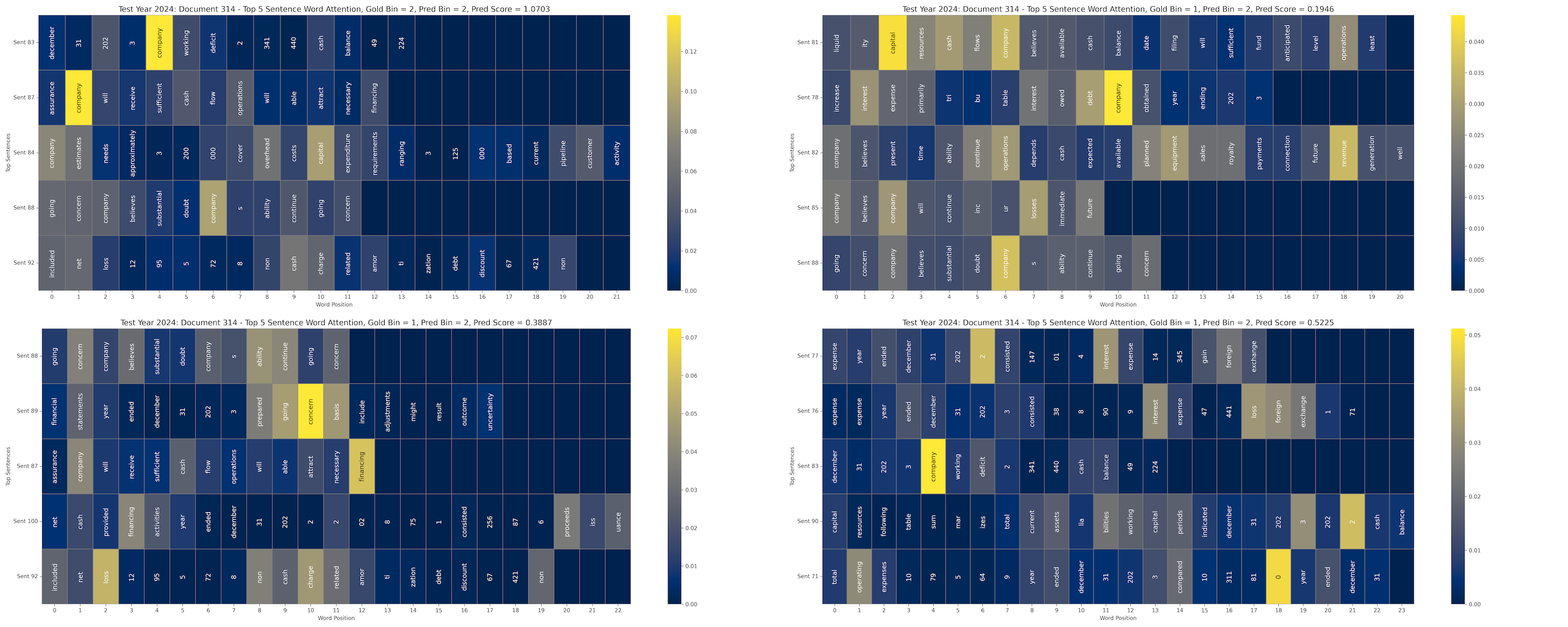}}
     {Top 5 sentence word attention explanation for financial document 314 in test year 2024.
     \label{fig:explain_word1}}
     {\textbf{Top Left:} Standard Deviation (Volatility), \textbf{Top Right:} Skewness, \textbf{Bottom Left:} Kurtosis, \textbf{Bottom Right:} Sortino Ratio.}
\end{figure}

\begin{figure}[H]
     \FIGURE
     {\includegraphics[width=0.8\textwidth]{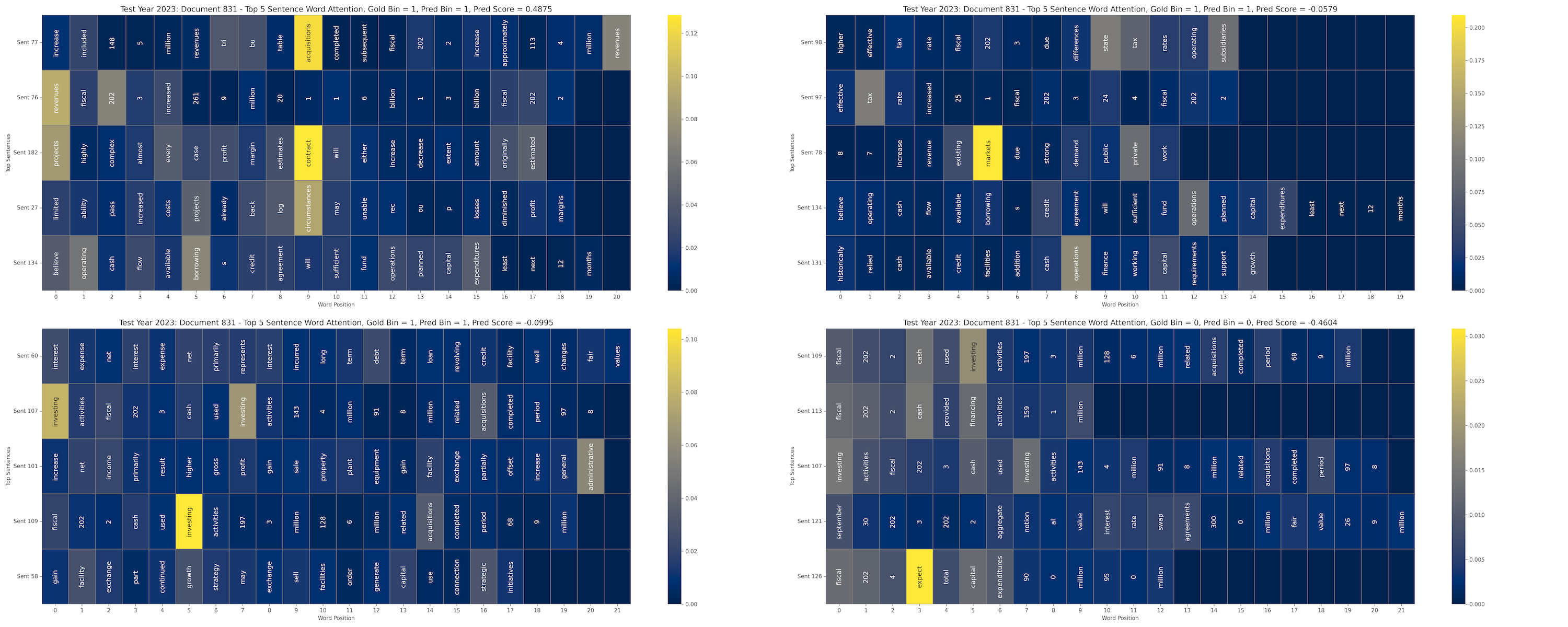}}
     {Top 5 sentence word attention explanation for financial document 831 in test year 2023.
     \label{fig:explain_word2}}
     {\textbf{Top Left:} Standard Deviation (Volatility), \textbf{Top Right:} Skewness, \textbf{Bottom Left:} Kurtosis, \textbf{Bottom Right:} Sortino Ratio.}
\end{figure}

\begin{figure}[H]
     \FIGURE
     {\includegraphics[width=0.8\textwidth]{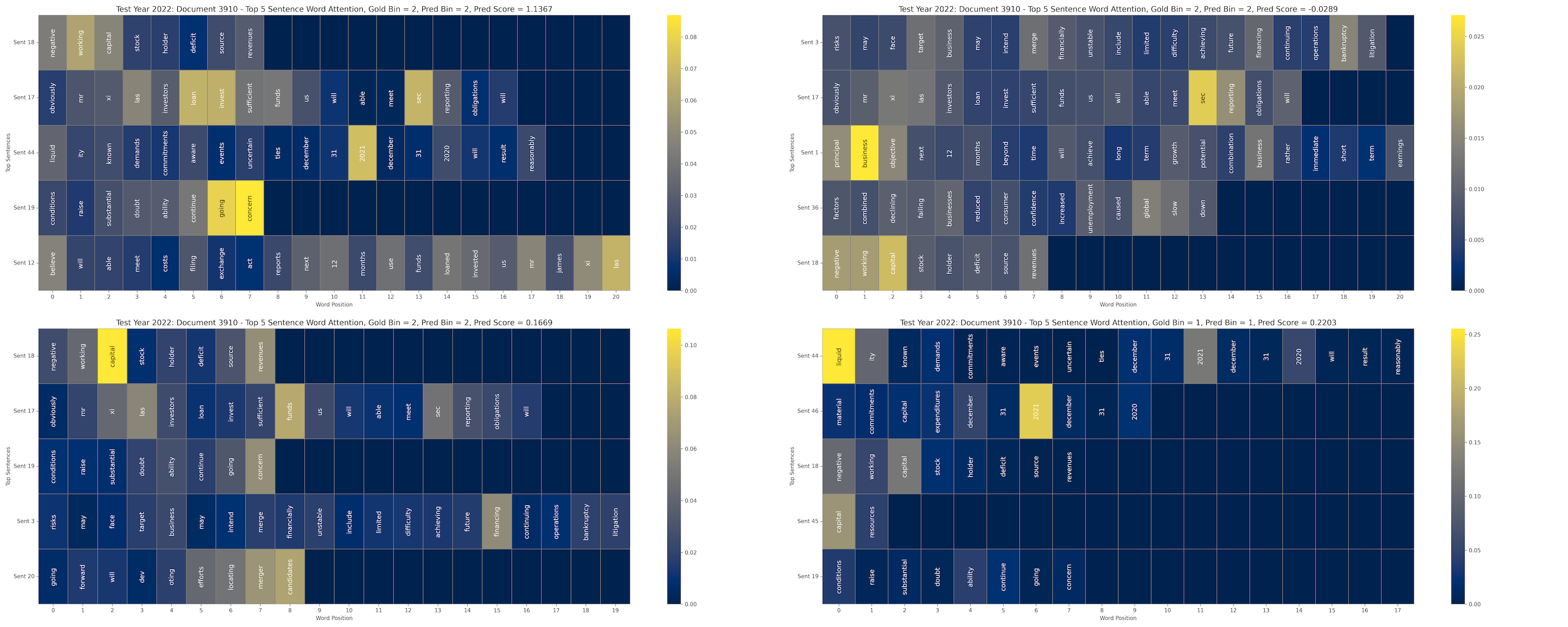}}
     {Top 5 sentence word attention explanation for financial document 3910 in test year 2022.
     \label{fig:explain_word3}}
     {\textbf{Top Left:} Standard Deviation (Volatility), \textbf{Top Right:} Skewness, \textbf{Bottom Left:} Kurtosis, \textbf{Bottom Right:} Sortino Ratio.}
\end{figure}

\begin{figure}[H]
     \FIGURE
     {\includegraphics[width=0.8\textwidth]{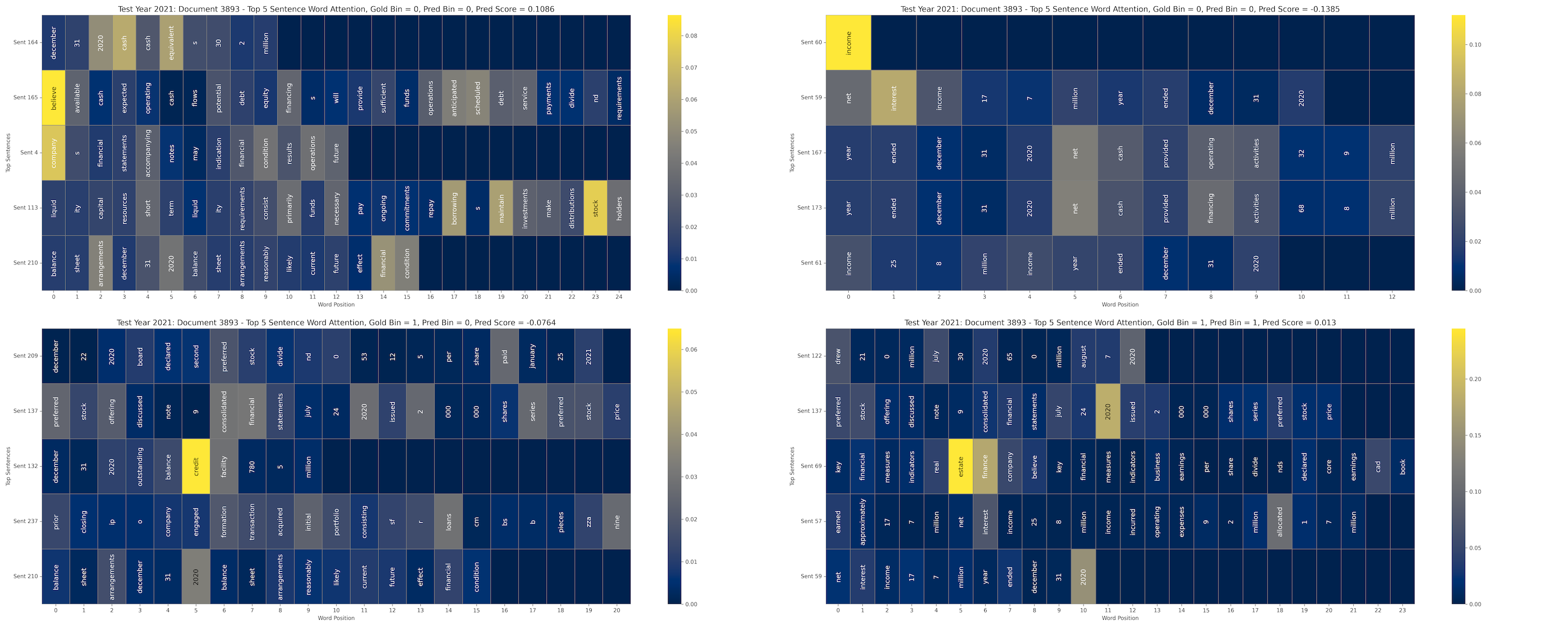}}
     {Top 5 sentence word attention explanation for financial document 3893 in test year 2021.
     \label{fig:explain_word4}}
     {\textbf{Top Left:} Standard Deviation (Volatility), \textbf{Top Right:} Skewness, \textbf{Bottom Left:} Kurtosis, \textbf{Bottom Right:} Sortino Ratio.}
\end{figure}

\begin{figure}[H]
     \FIGURE
     {\includegraphics[width=0.8\textwidth]{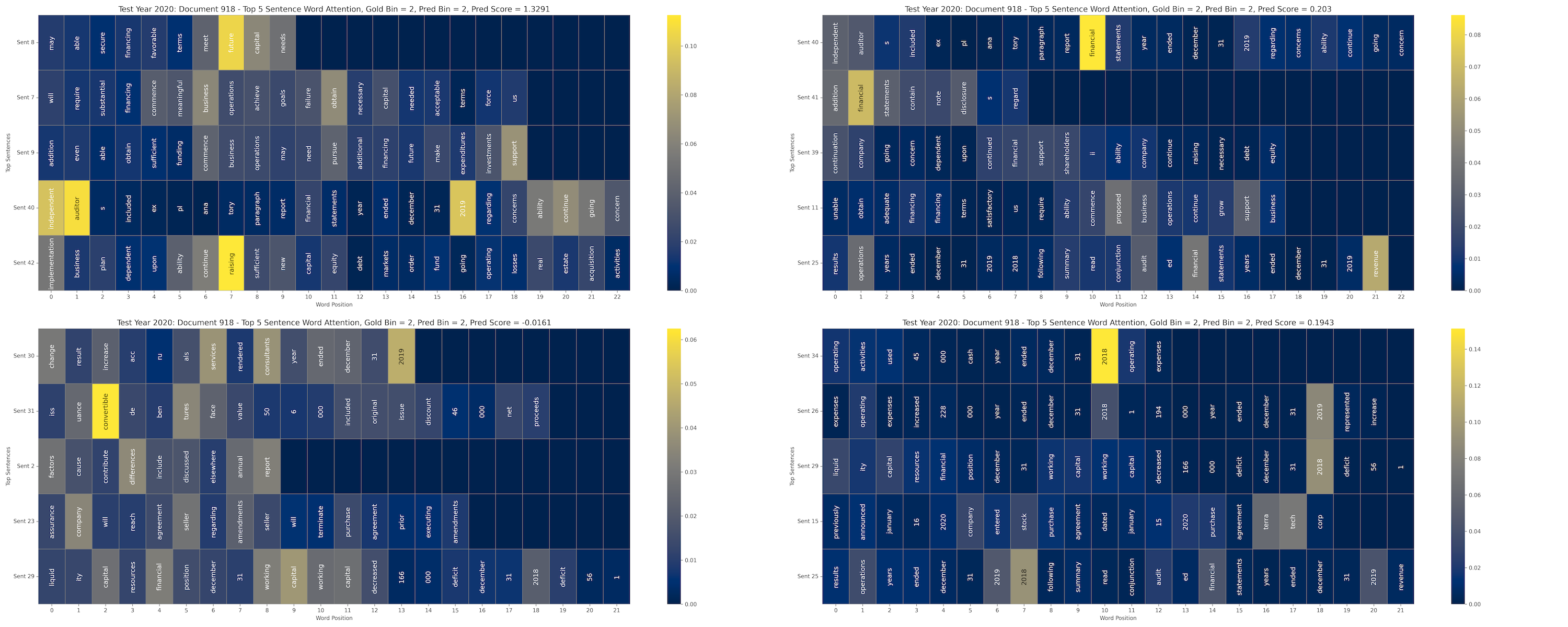}}
     {Top 5 sentence word attention explanation for financial document 918 in test year 2020.
     \label{fig:explain_word5}}
     {\textbf{Top Left:} Standard Deviation (Volatility), \textbf{Top Right:} Skewness, \textbf{Bottom Left:} Kurtosis, \textbf{Bottom Right:} Sortino Ratio.}
\end{figure}

\begin{figure}[H]
     \FIGURE
     {\includegraphics[width=0.8\textwidth]{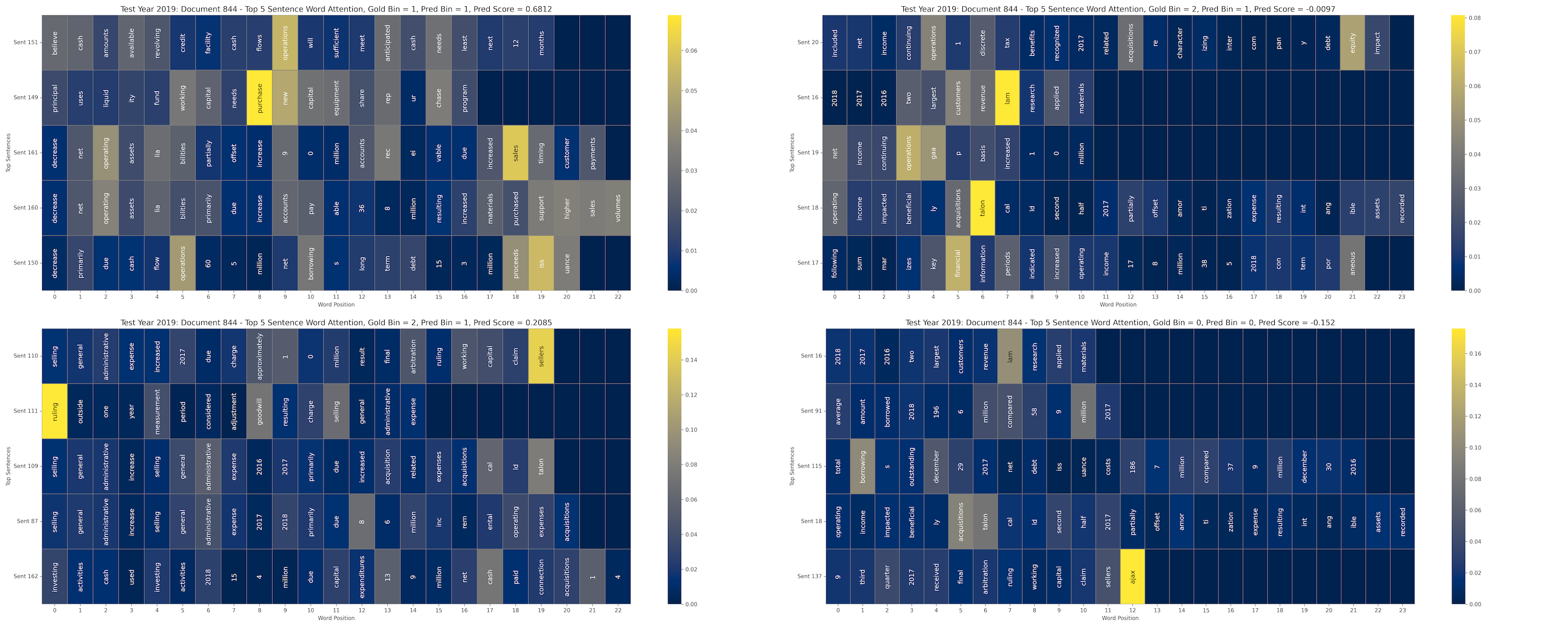}}
     {Top 5 sentence word attention explanation for financial document 844 in test year 2019.
     \label{fig:explain_word6}}
     {\textbf{Top Left:} Standard Deviation (Volatility), \textbf{Top Right:} Skewness, \textbf{Bottom Left:} Kurtosis, \textbf{Bottom Right:} Sortino Ratio.}
\end{figure}

\begin{figure}[H]
     \FIGURE
     {\includegraphics[width=0.8\textwidth]{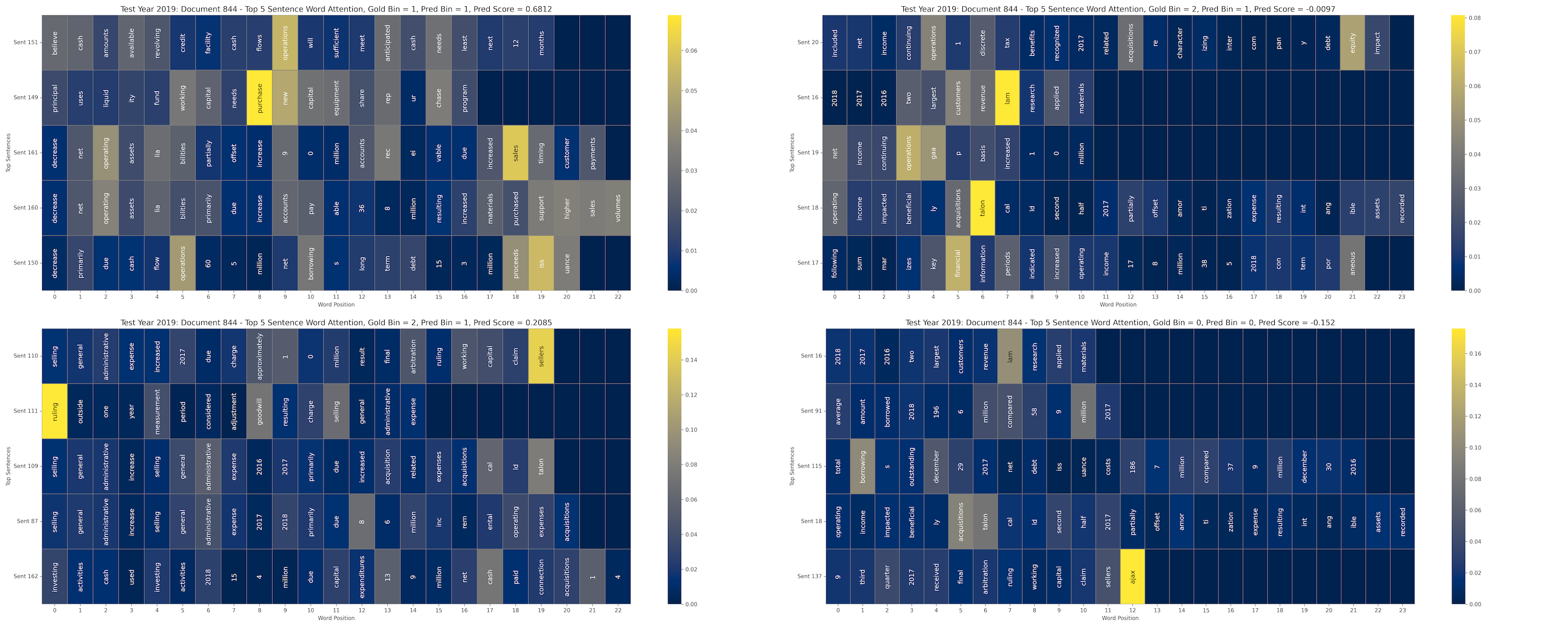}}
     {Top 5 sentence word attention explanation for financial document 844 in test year 2018.
     \label{fig:explain_word7}}
     {\textbf{Top Left:} Standard Deviation (Volatility), \textbf{Top Right:} Skewness, \textbf{Bottom Left:} Kurtosis, \textbf{Bottom Right:} Sortino Ratio.}
\end{figure}

\clearpage
\section{Attention-based Word Clouds for Test Year 2018-2024}
\label{appendix:cloud}
\begin{figure}[H]
     \FIGURE
     {\includegraphics[width=0.72\textwidth]{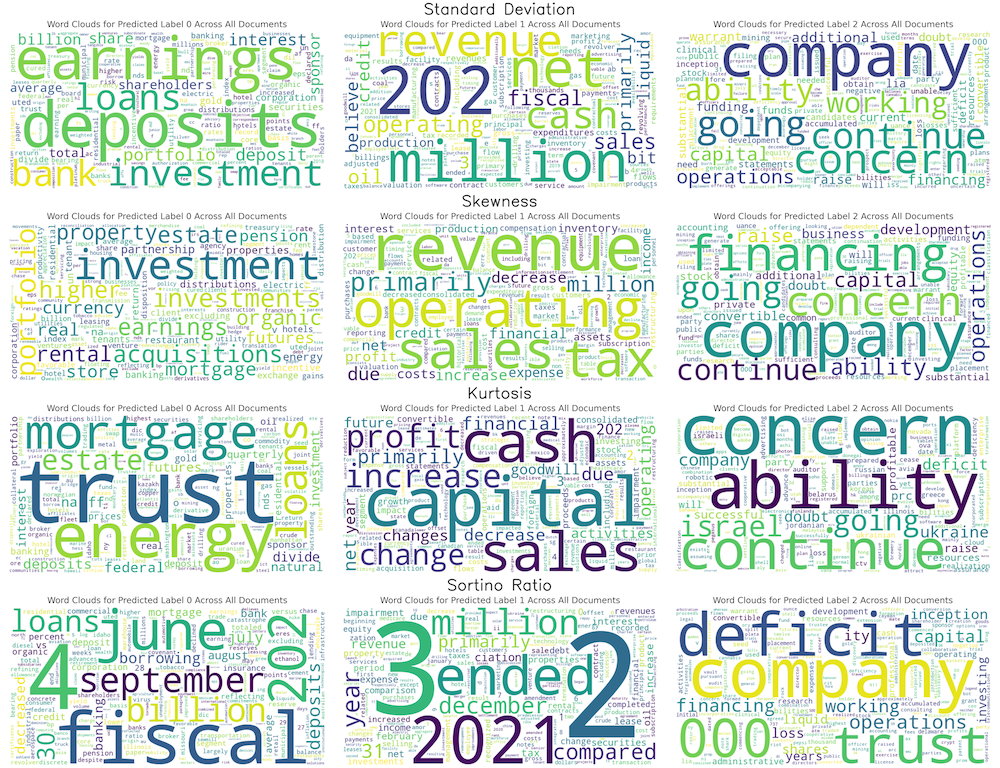}}
     {Word clouds representing the most prominent terms associated with predicted risk levels (Labels 0, 1, and 2) for financial reports in test year 2024, across four risk measures. \label{fig:cloud2024}}
     {}
\end{figure}

\begin{figure}[H]
     \FIGURE
     {\includegraphics[width=0.72\textwidth]{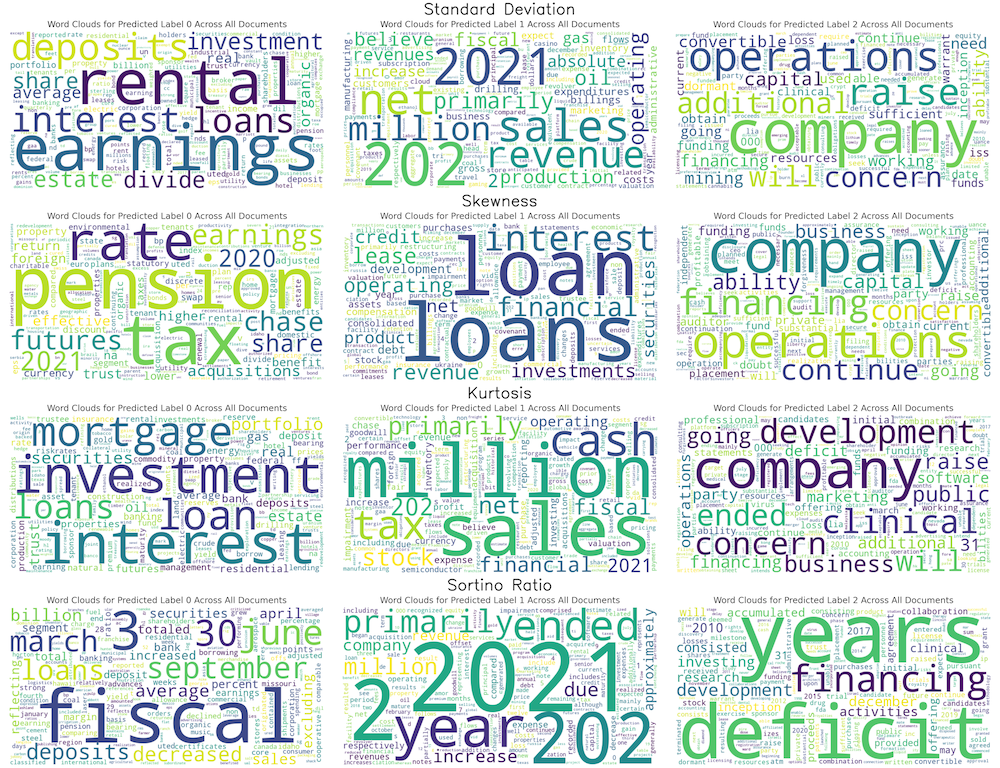}}
     {Word clouds representing the most prominent terms associated with predicted risk levels (Labels 0, 1, and 2) for financial reports in test year 2023, across four risk measures. \label{fig:cloud2023}}
     {}
\end{figure}

\begin{figure}[H]
     \FIGURE
     {\includegraphics[width=0.72\textwidth]{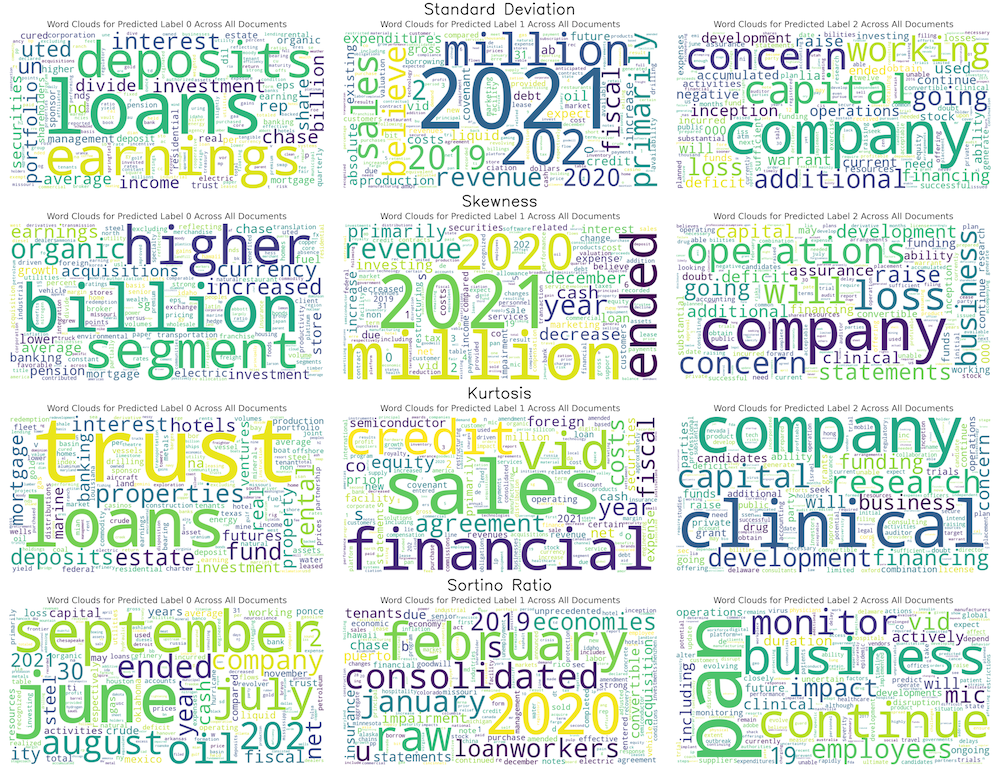}}
     {Word clouds representing the most prominent terms associated with predicted risk levels (Labels 0, 1, and 2) for financial reports in test year 2022, across four risk measures. \label{fig:cloud2022}}
     {}
\end{figure}

\begin{figure}[H]
     \FIGURE
     {\includegraphics[width=0.72\textwidth]{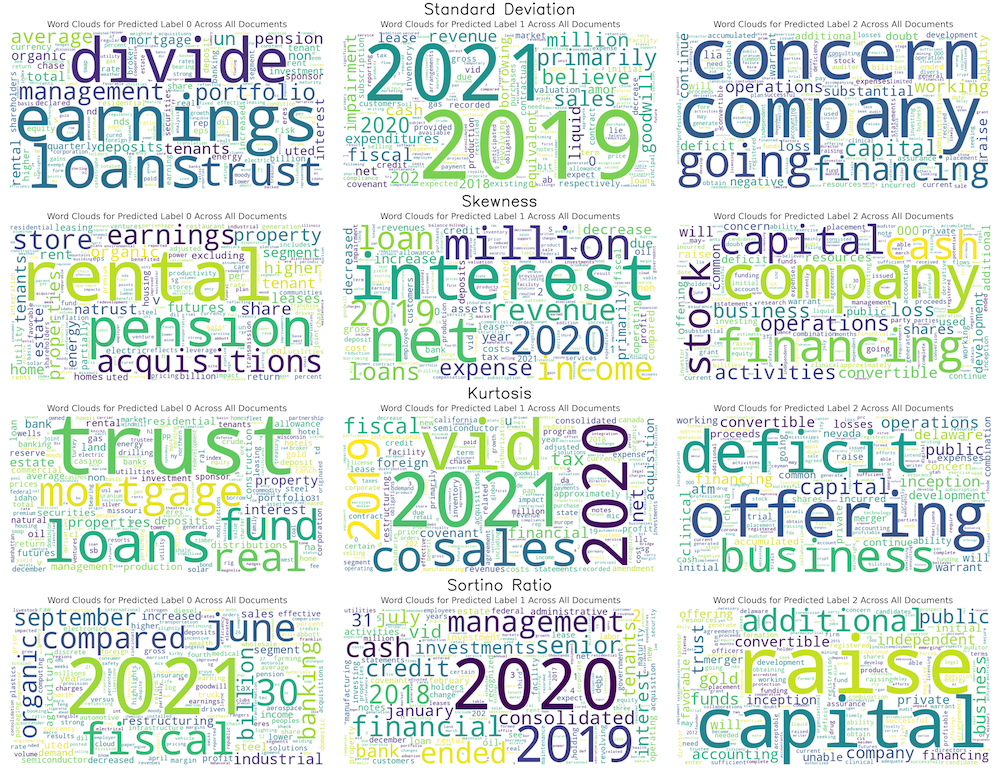}}
     {Word clouds representing the most prominent terms associated with predicted risk levels (Labels 0, 1, and 2) for financial reports in test year 2021, across four risk measures. \label{fig:cloud2021}}
     {}
\end{figure}

\begin{figure}[H]
     \FIGURE
     {\includegraphics[width=0.72\textwidth]{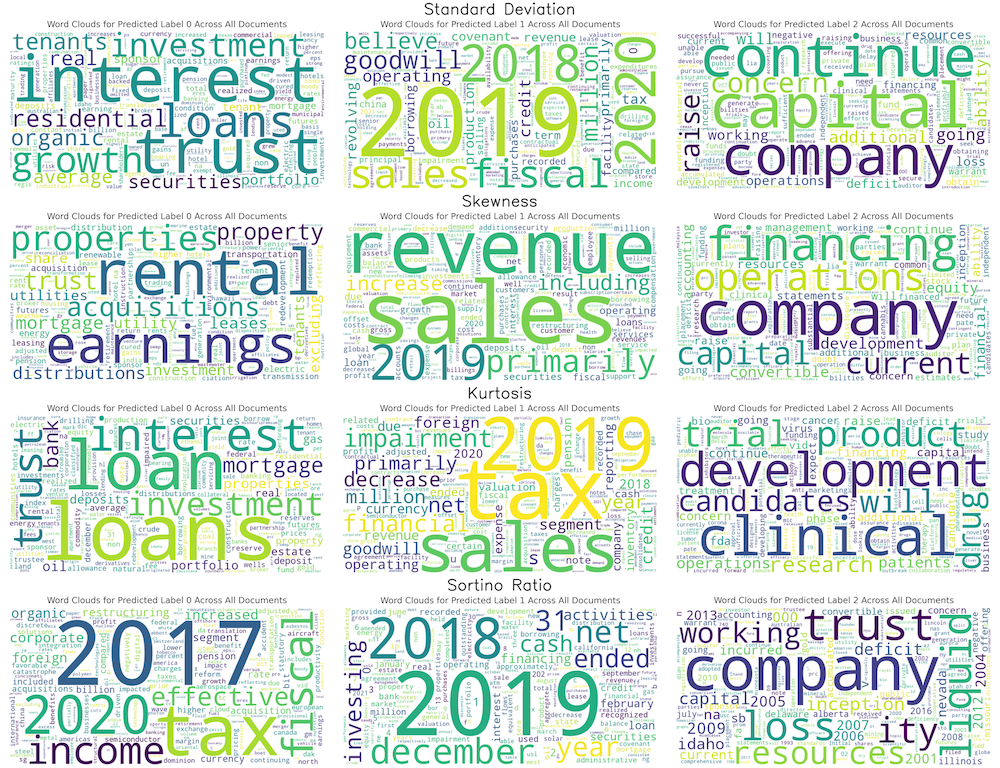}}
     {Word clouds representing the most prominent terms associated with predicted risk levels (Labels 0, 1, and 2) for financial reports in test year 2020, across four risk measures. \label{fig:cloud2020}}
     {}
\end{figure}

\begin{figure}[H]
     \FIGURE
     {\includegraphics[width=0.72\textwidth]{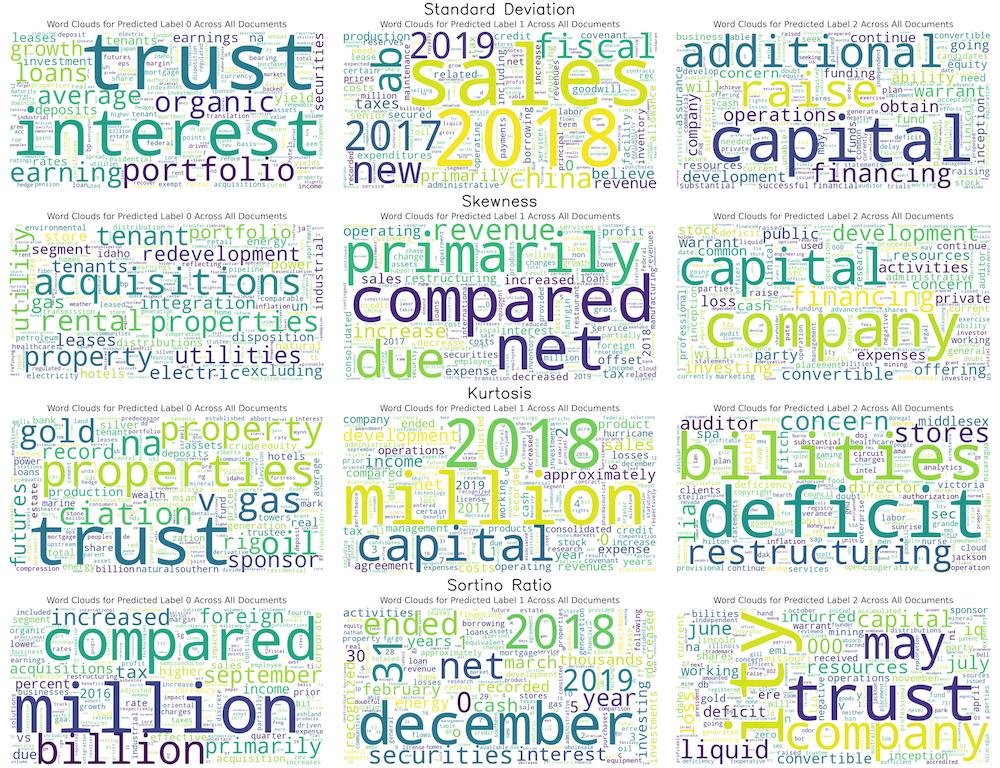}}
     {Word clouds representing the most prominent terms associated with predicted risk levels (Labels 0, 1, and 2) for financial reports in test year 2019, across four risk measures. \label{fig:cloud2019}}
     {}
\end{figure}

\begin{figure}[H]
     \FIGURE
     {\includegraphics[width=0.72\textwidth]{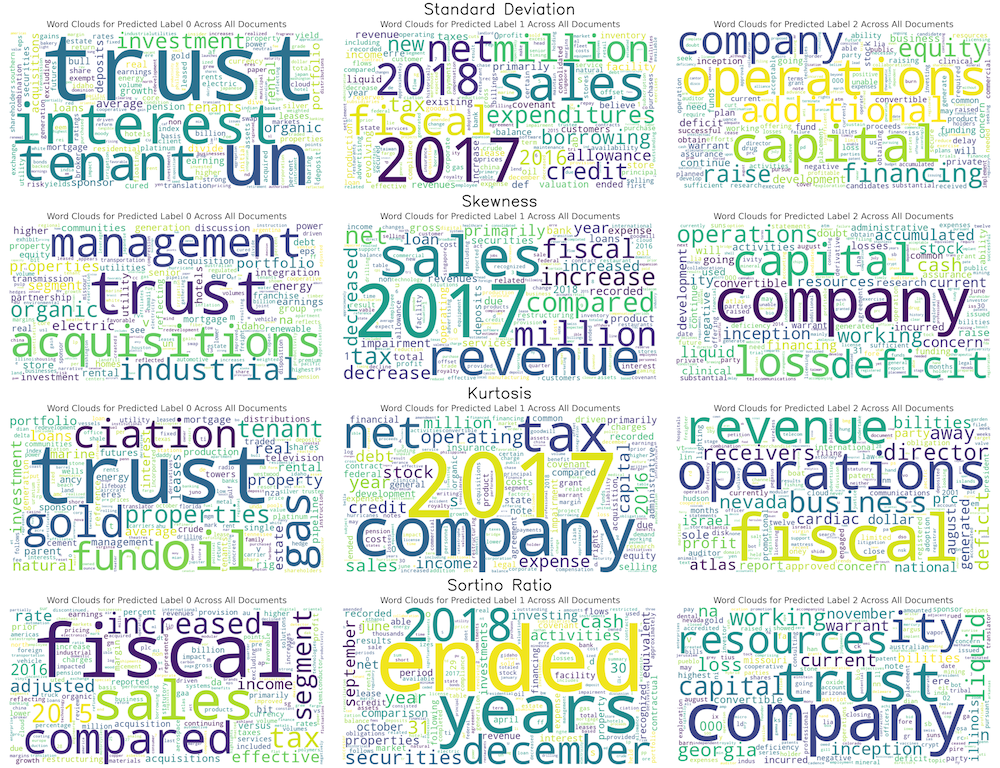}}
     {Word clouds representing the most prominent terms associated with predicted risk levels (Labels 0, 1, and 2) for financial reports in test year 2018, across four risk measures. \label{fig:cloud2018}}
     {}
\end{figure}
\clearpage
\section{TinyXRA's Explanation Faithfulness Experiments}
\label{appendix:expabl}
\subsection{Removal of Highly Attended Words}
\begin{figure}[htb!]
    \centering
    \begin{subfigure}[b]{0.32\textwidth}
        \includegraphics[width=\textwidth]{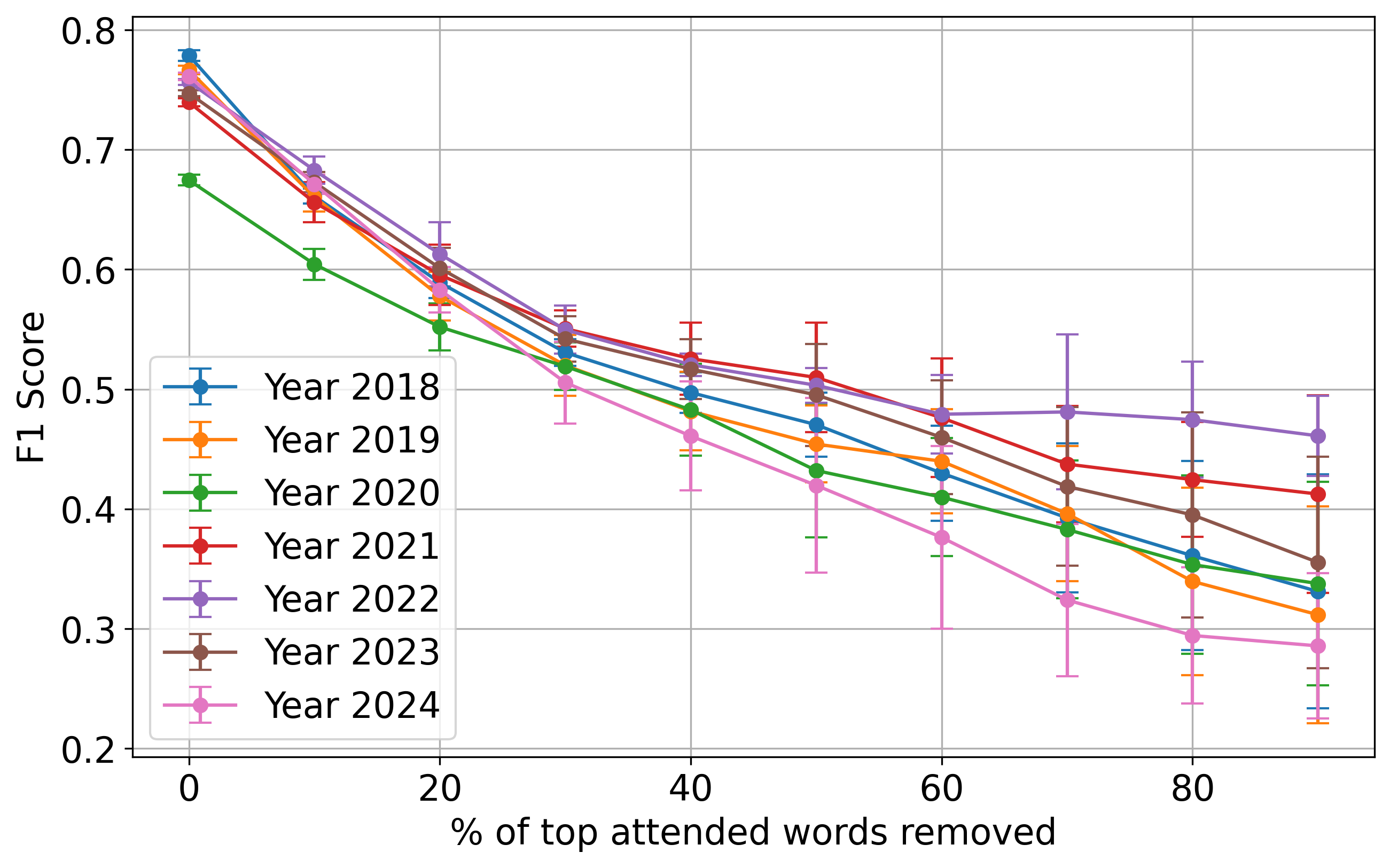}
        \caption[size=small]{}
    \end{subfigure}
    \hfill
    \begin{subfigure}[b]{0.32\textwidth}
        \includegraphics[width=\textwidth]{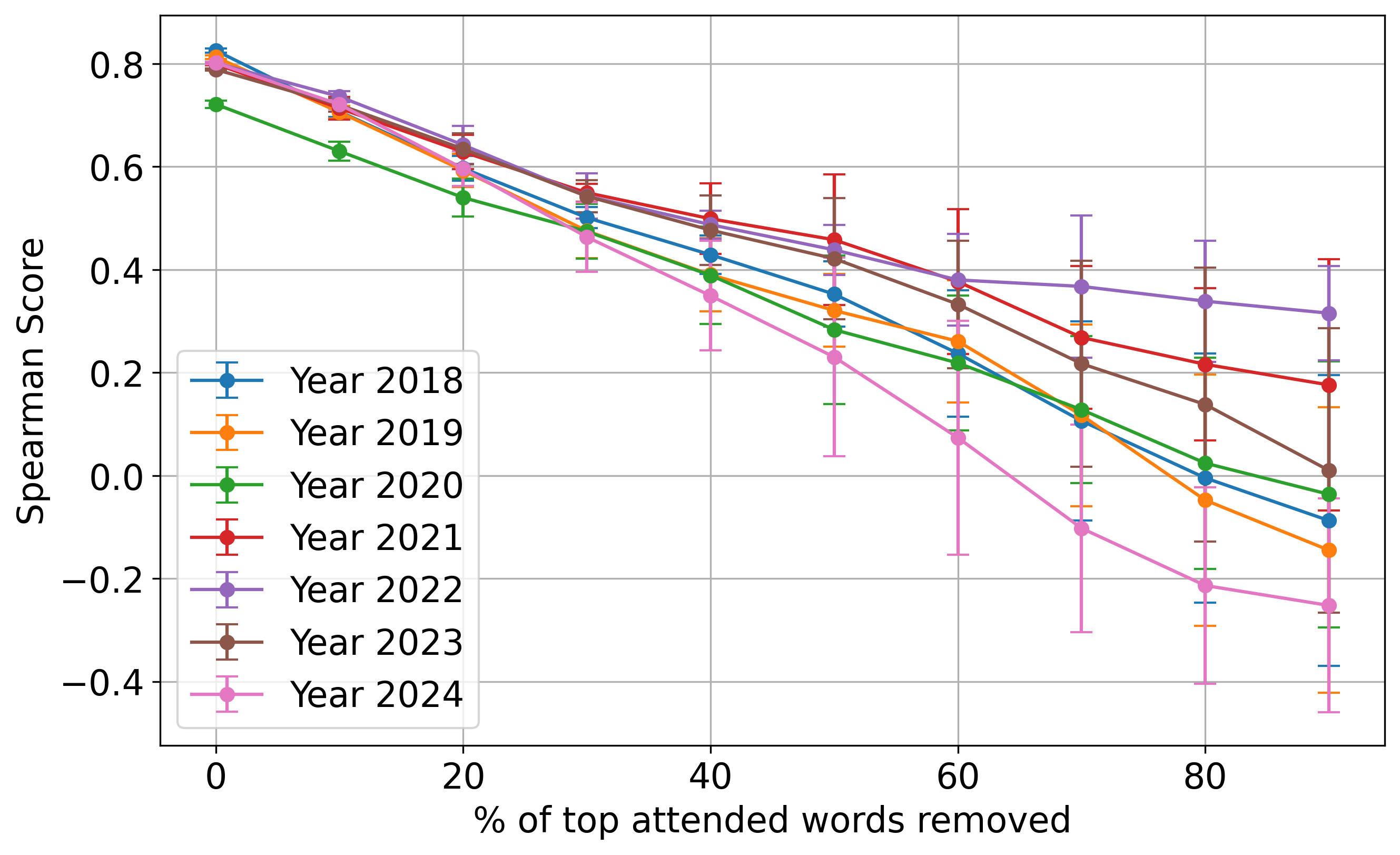}
        \caption[size=small]{}
    \end{subfigure}
    \hfill
    \begin{subfigure}[b]{0.32\textwidth}
        \includegraphics[width=\textwidth]{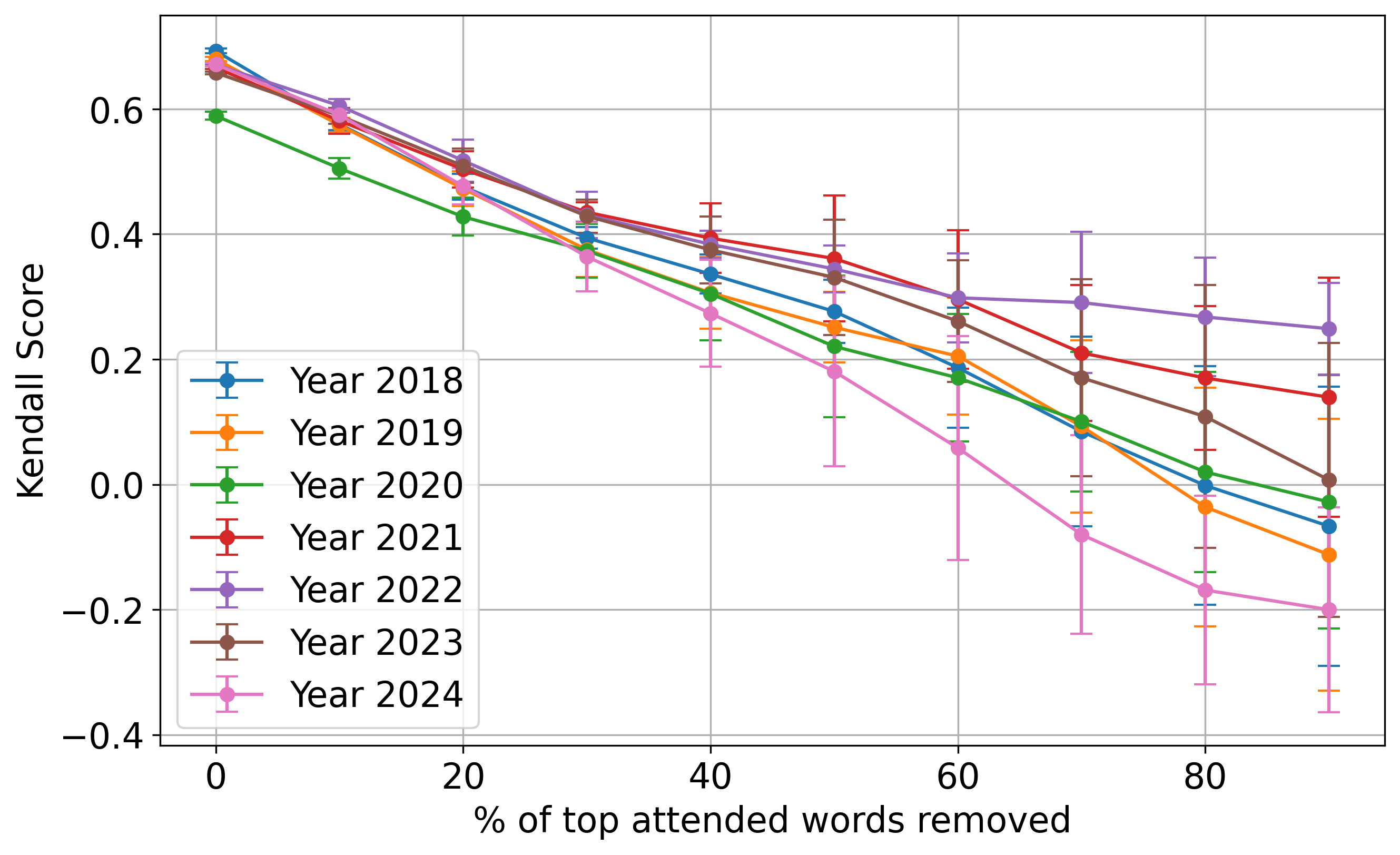}
        \caption[size=small]{}
    \end{subfigure}
    \caption{Performance metrics versus the percentage of top-\(k\%\) attended words removed, evaluated using standard deviation as the risk measure from 2018 to 2024. The plots show the mean and standard deviation (as error bars) across five random seeds for F1 Score (a), Spearman’s Rho (b), and Kendall’s Tau (c).}
    \label{fig:exp_word_std}
\end{figure}

\begin{figure}[htb!]
    \centering
    \begin{subfigure}[b]{0.32\textwidth}
        \includegraphics[width=\textwidth]{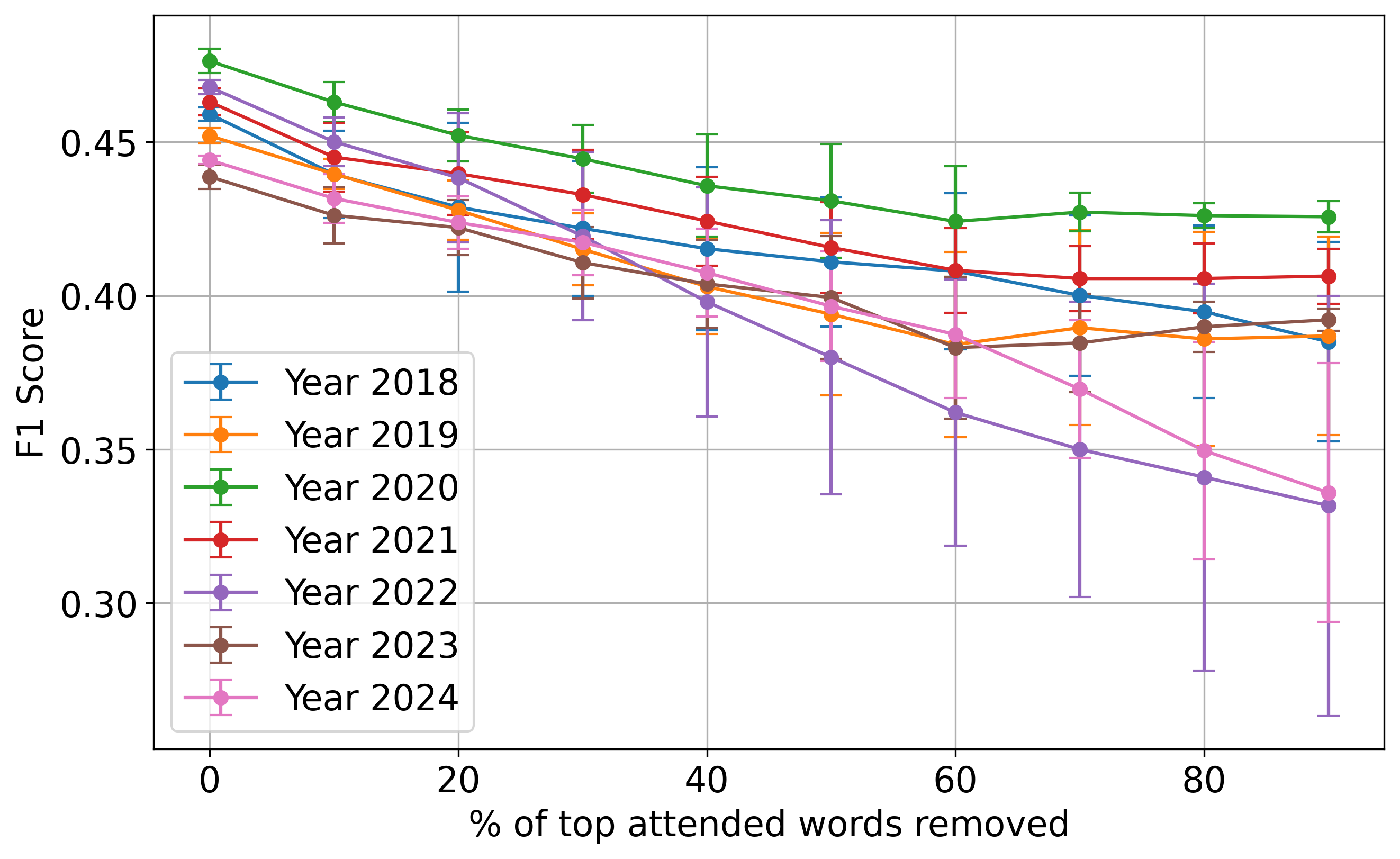}
        \caption[size=small]{}
    \end{subfigure}
    \hfill
    \begin{subfigure}[b]{0.32\textwidth}
        \includegraphics[width=\textwidth]{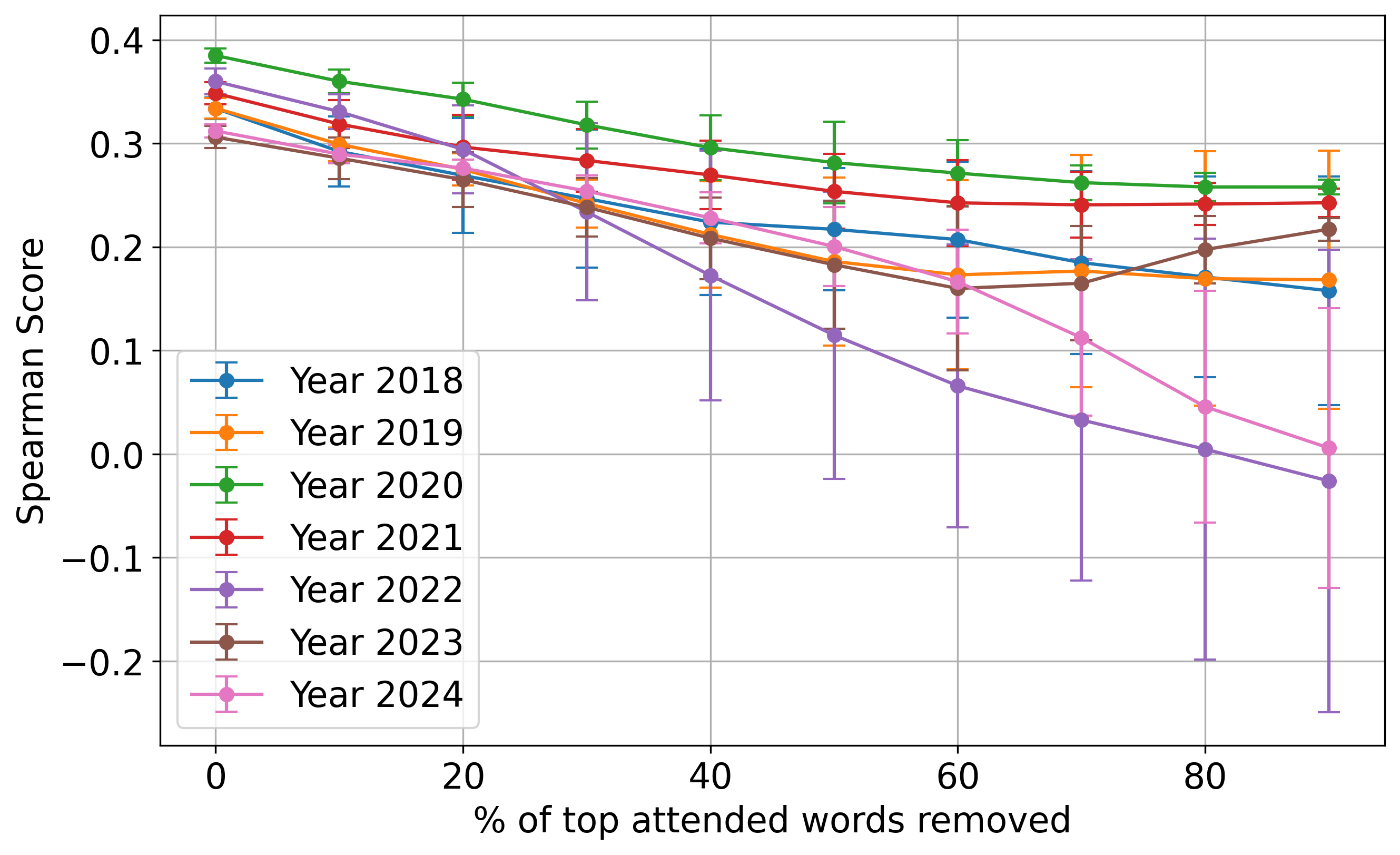}
        \caption[size=small]{}
    \end{subfigure}
    \hfill
    \begin{subfigure}[b]{0.32\textwidth}
        \includegraphics[width=\textwidth]{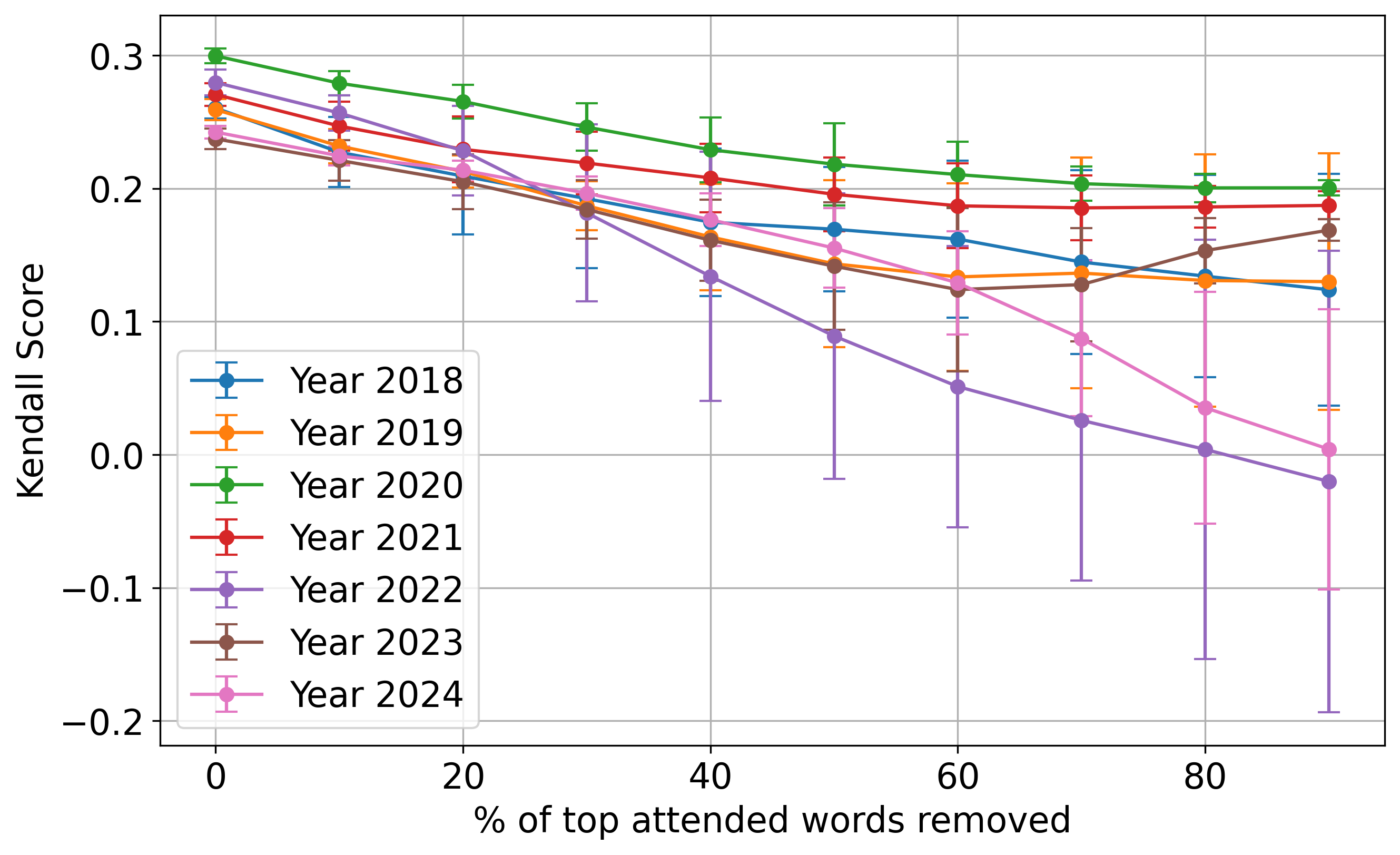}
        \caption[size=small]{}
    \end{subfigure}
    \caption{Performance metrics versus the percentage of top-\(k\%\) attended words removed, evaluated using skewness as the risk measure from 2018 to 2024. The plots show the mean and standard deviation (as error bars) across five random seeds for F1 Score (a), Spearman’s Rho (b), and Kendall’s Tau (c).}
    \label{fig:exp_word_skew}
\end{figure}

\begin{figure}[htb!]
    \centering
    \begin{subfigure}[b]{0.32\textwidth}
        \includegraphics[width=\textwidth]{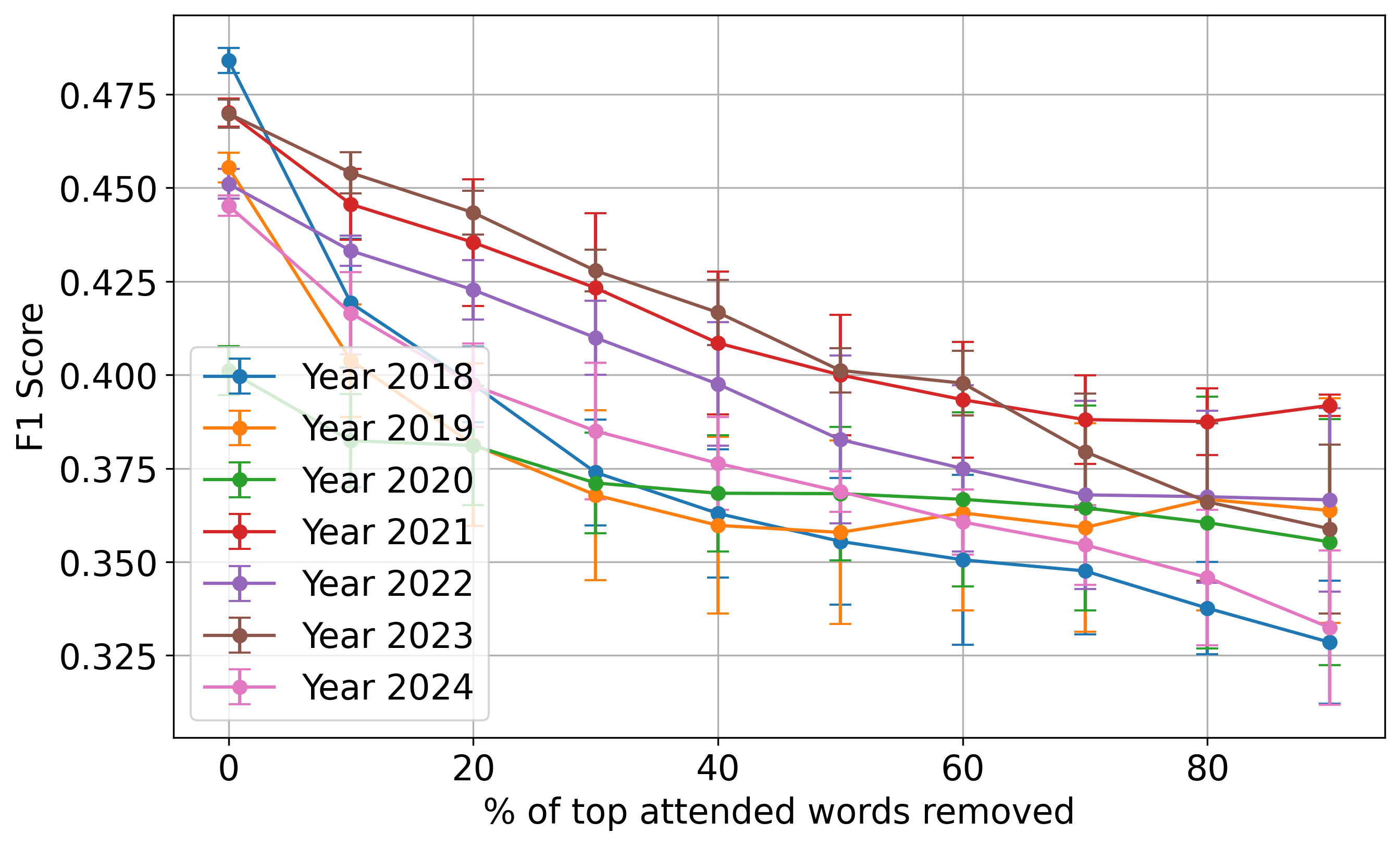}
        \caption[size=small]{}
    \end{subfigure}
    \hfill
    \begin{subfigure}[b]{0.32\textwidth}
        \includegraphics[width=\textwidth]{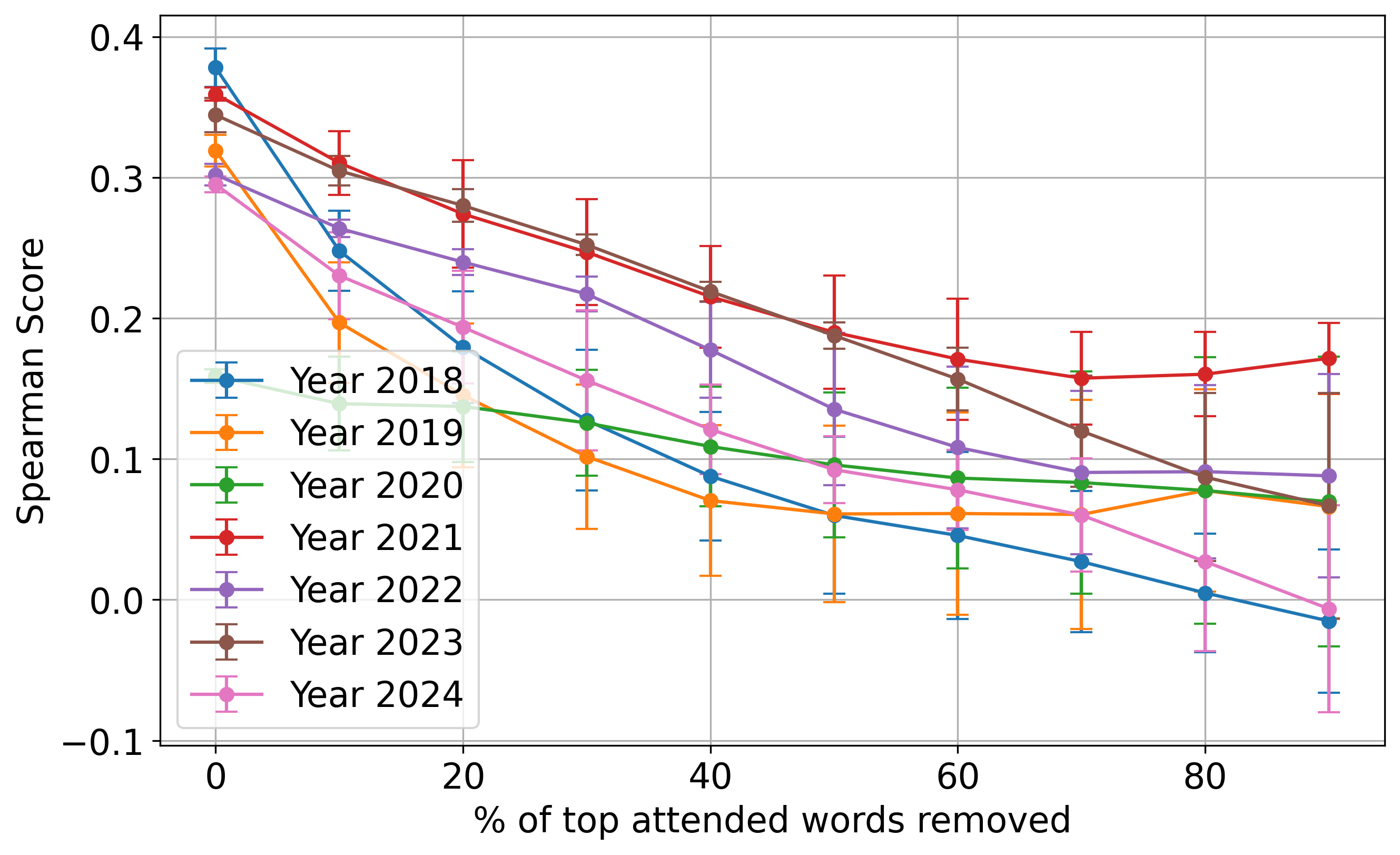}
        \caption[size=small]{}
    \end{subfigure}
    \hfill
    \begin{subfigure}[b]{0.32\textwidth}
        \includegraphics[width=\textwidth]{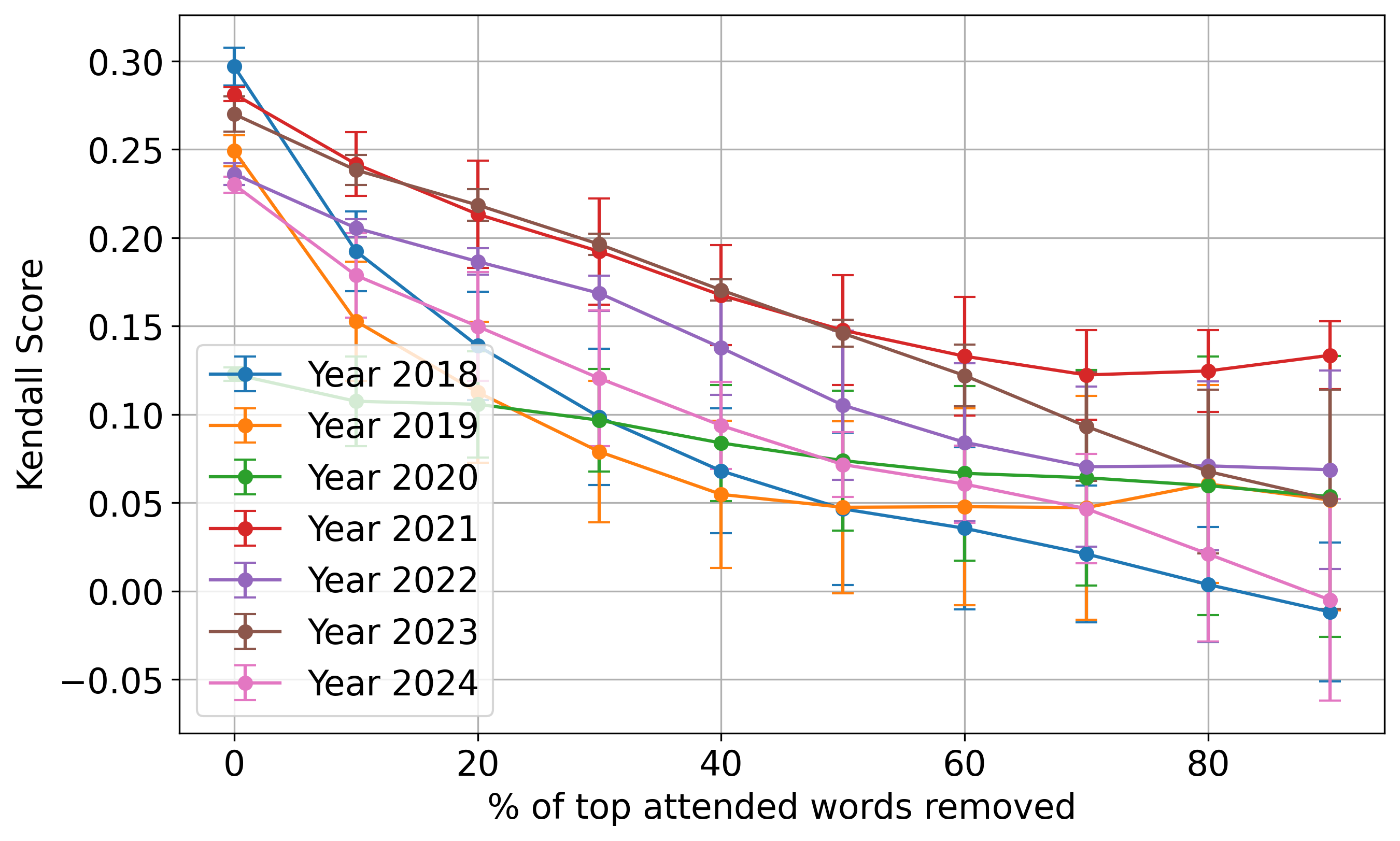}
        \caption[size=small]{}
    \end{subfigure}
    \caption{Performance metrics versus the percentage of top-\(k\%\) attended words removed, evaluated using kurtosis as the risk measure from 2018 to 2024. The plots show the mean and standard deviation (as error bars) across five random seeds for F1 Score (a), Spearman’s Rho (b), and Kendall’s Tau (c).}
    \label{fig:exp_word_kurt}
\end{figure}

\begin{figure}[htb!]
    \centering
    \begin{subfigure}[b]{0.32\textwidth}
        \includegraphics[width=\textwidth]{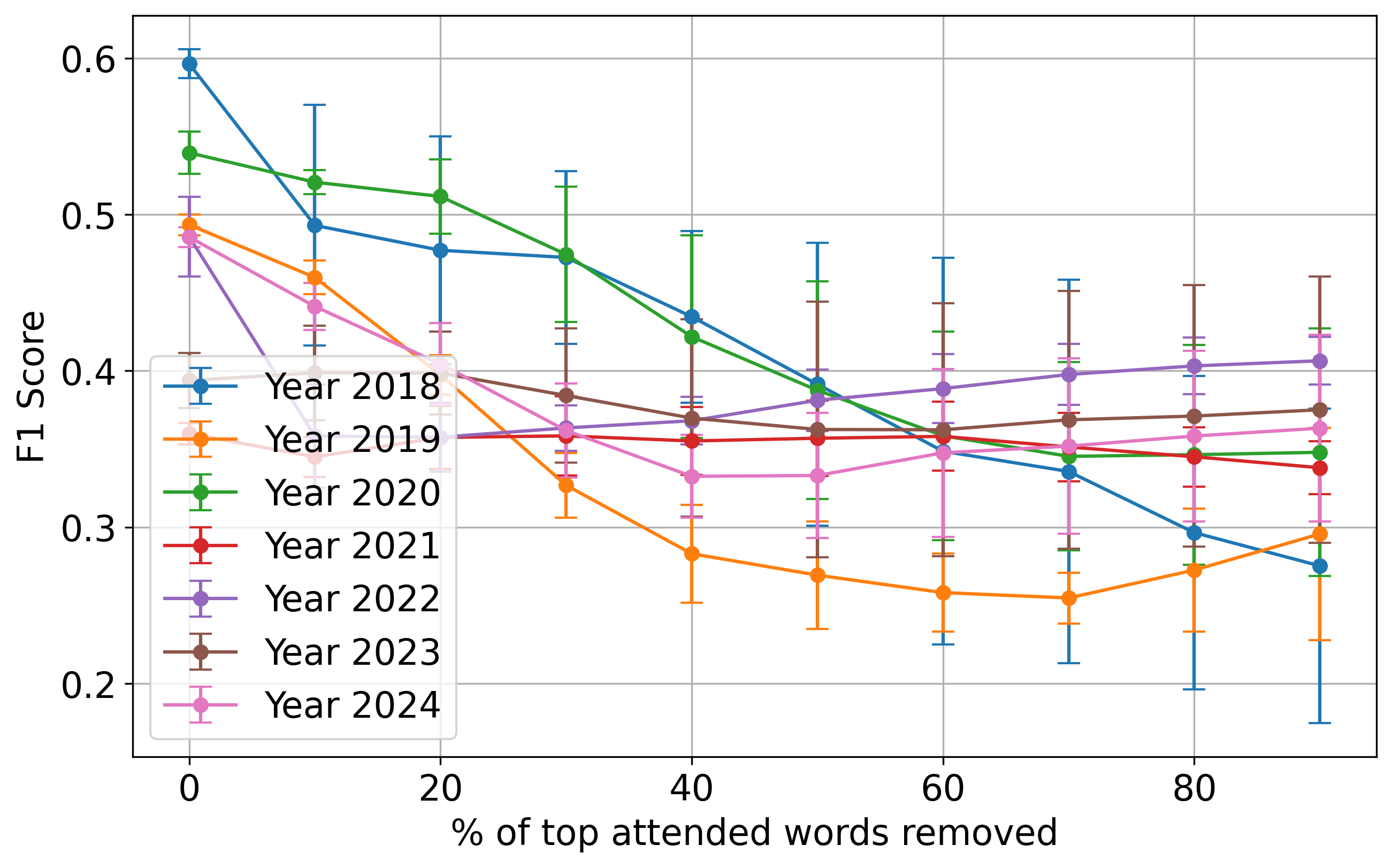}
        \caption[size=small]{}
    \end{subfigure}
    \hfill
    \begin{subfigure}[b]{0.32\textwidth}
        \includegraphics[width=\textwidth]{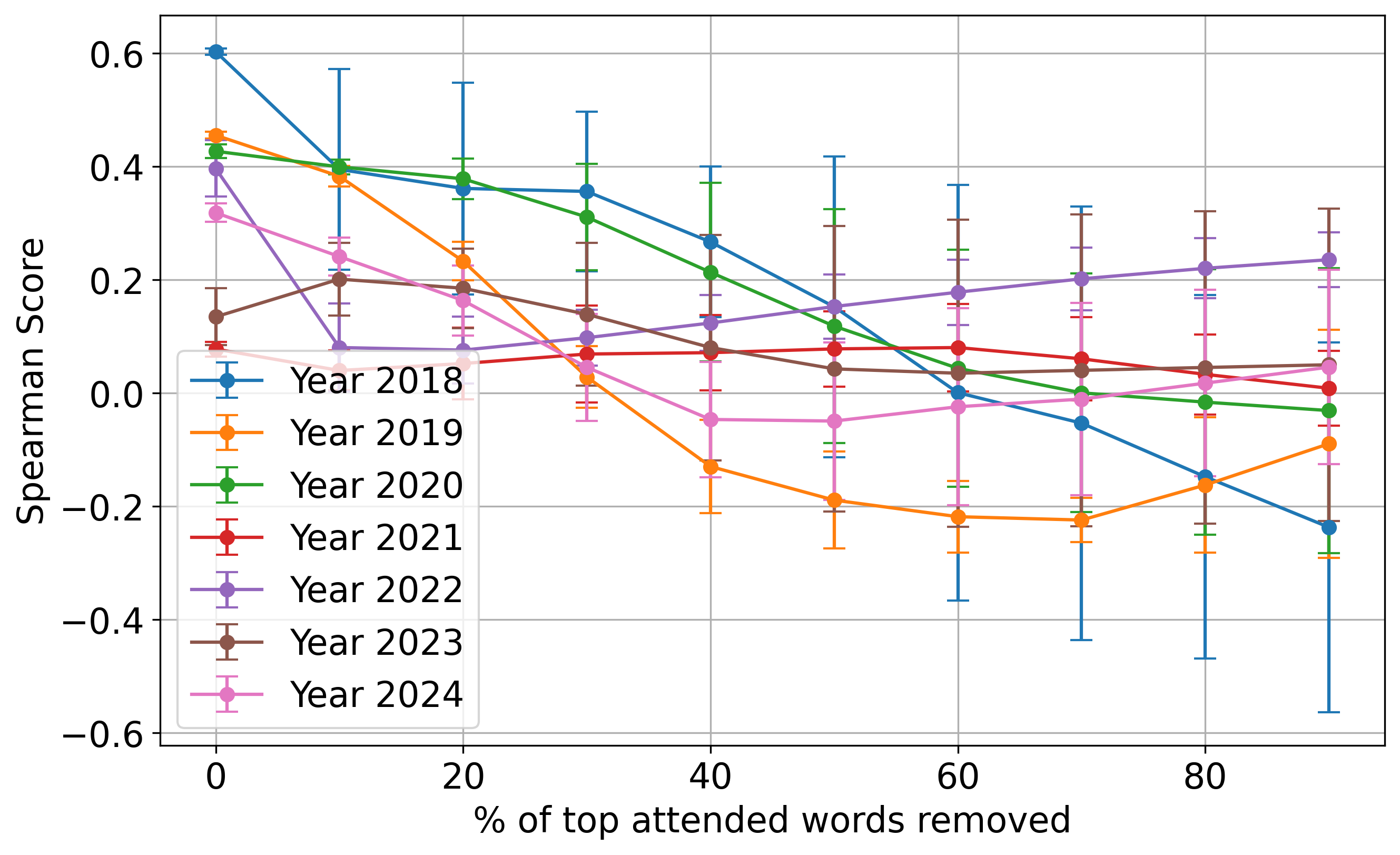}
        \caption[size=small]{}
    \end{subfigure}
    \hfill
    \begin{subfigure}[b]{0.32\textwidth}
        \includegraphics[width=\textwidth]{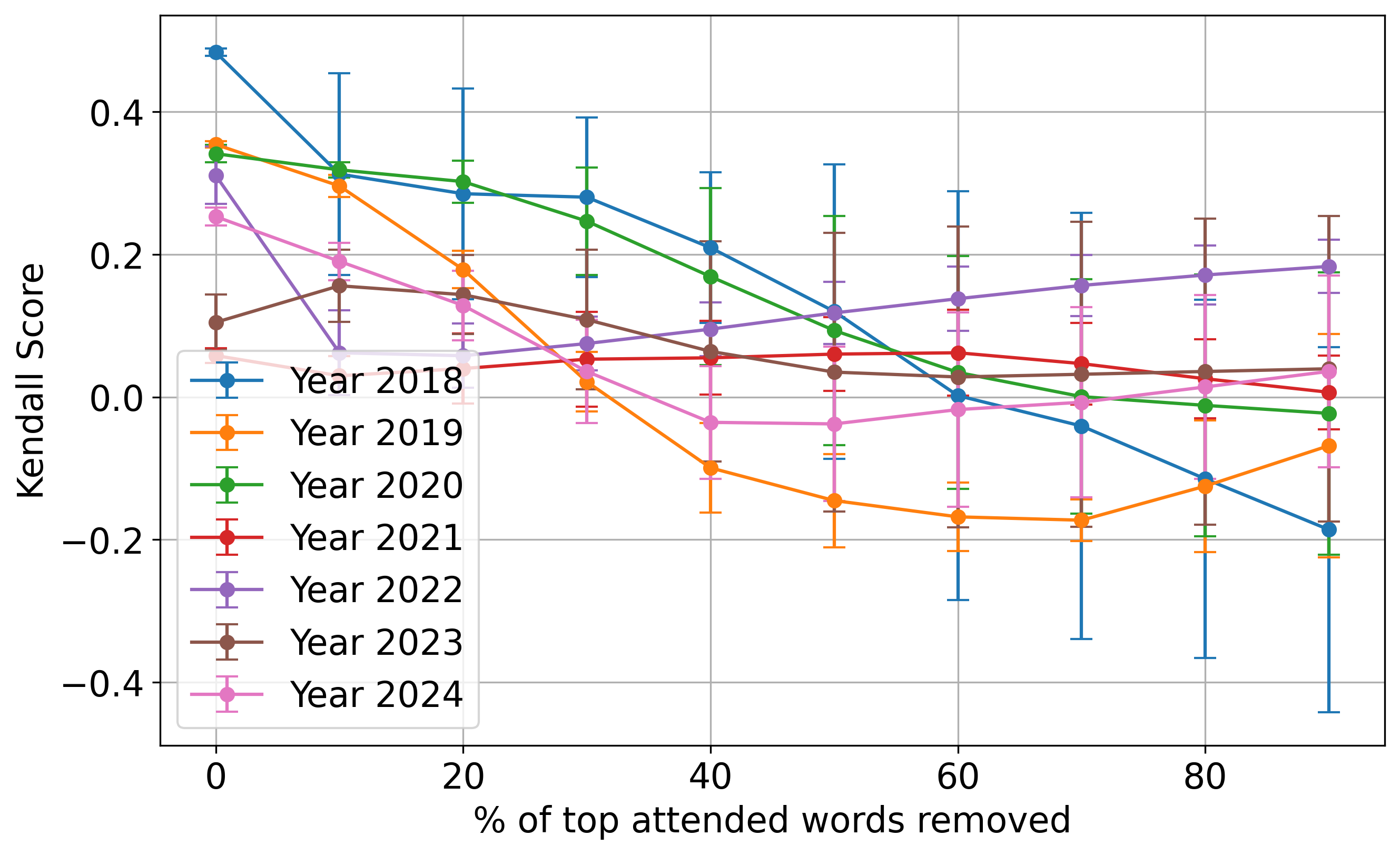}
        \caption[size=small]{}
    \end{subfigure}
    \caption{Performance metrics versus the percentage of top-\(k\%\) attended words removed, evaluated using sortino ratio as the risk measure from 2018 to 2024. The plots show the mean and standard deviation (as error bars) across five random seeds for F1 Score (a), Spearman’s Rho (b), and Kendall’s Tau (c).}
    \label{fig:exp_word_sortino}
\end{figure}
\clearpage

\subsection{Removal of Highly Attended Sentences}

\begin{figure}[htb!]
    \centering
    \begin{subfigure}[b]{0.32\textwidth}
        \includegraphics[width=\textwidth]{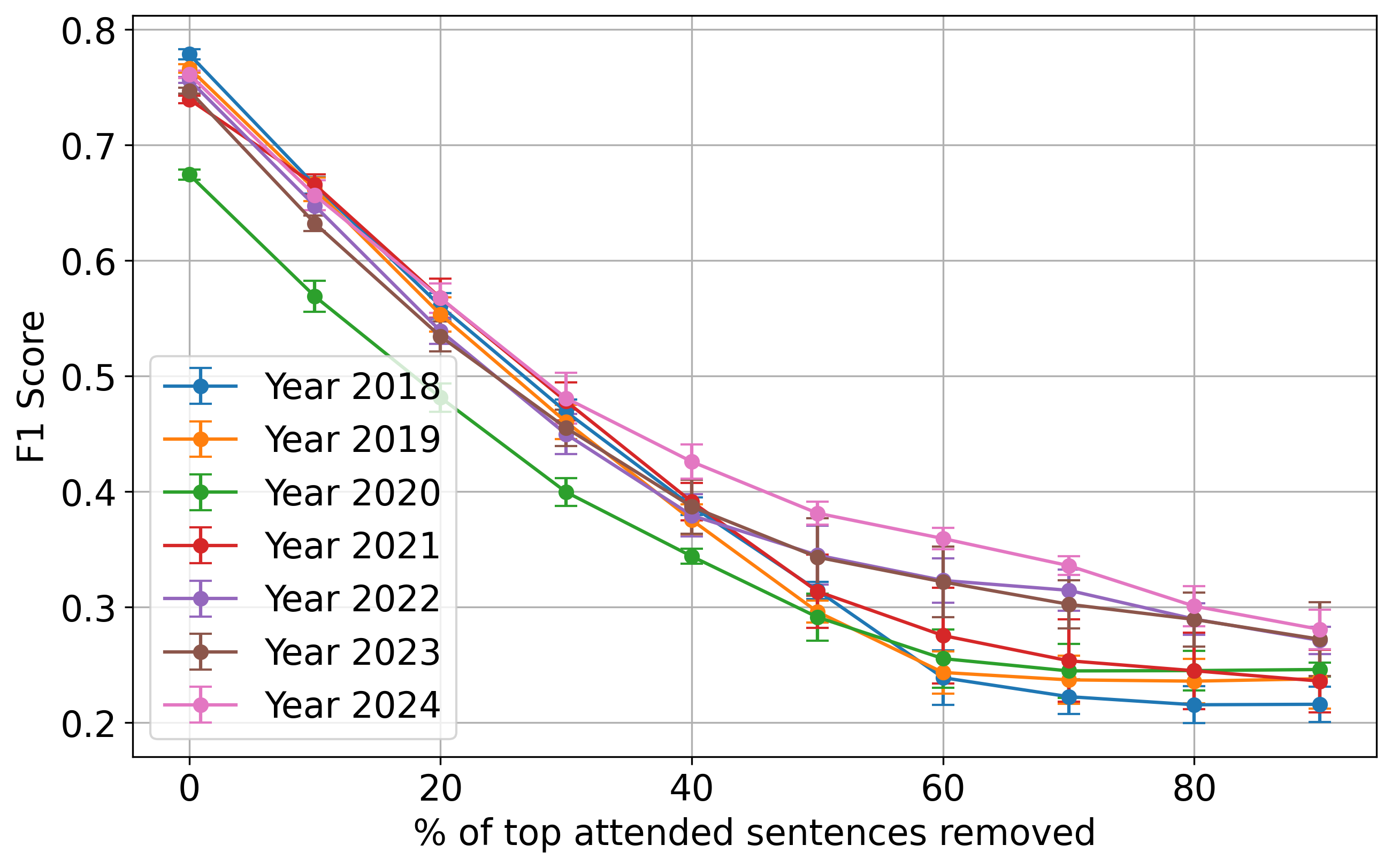}
        \caption[size=small]{}
    \end{subfigure}
    \hfill
    \begin{subfigure}[b]{0.32\textwidth}
        \includegraphics[width=\textwidth]{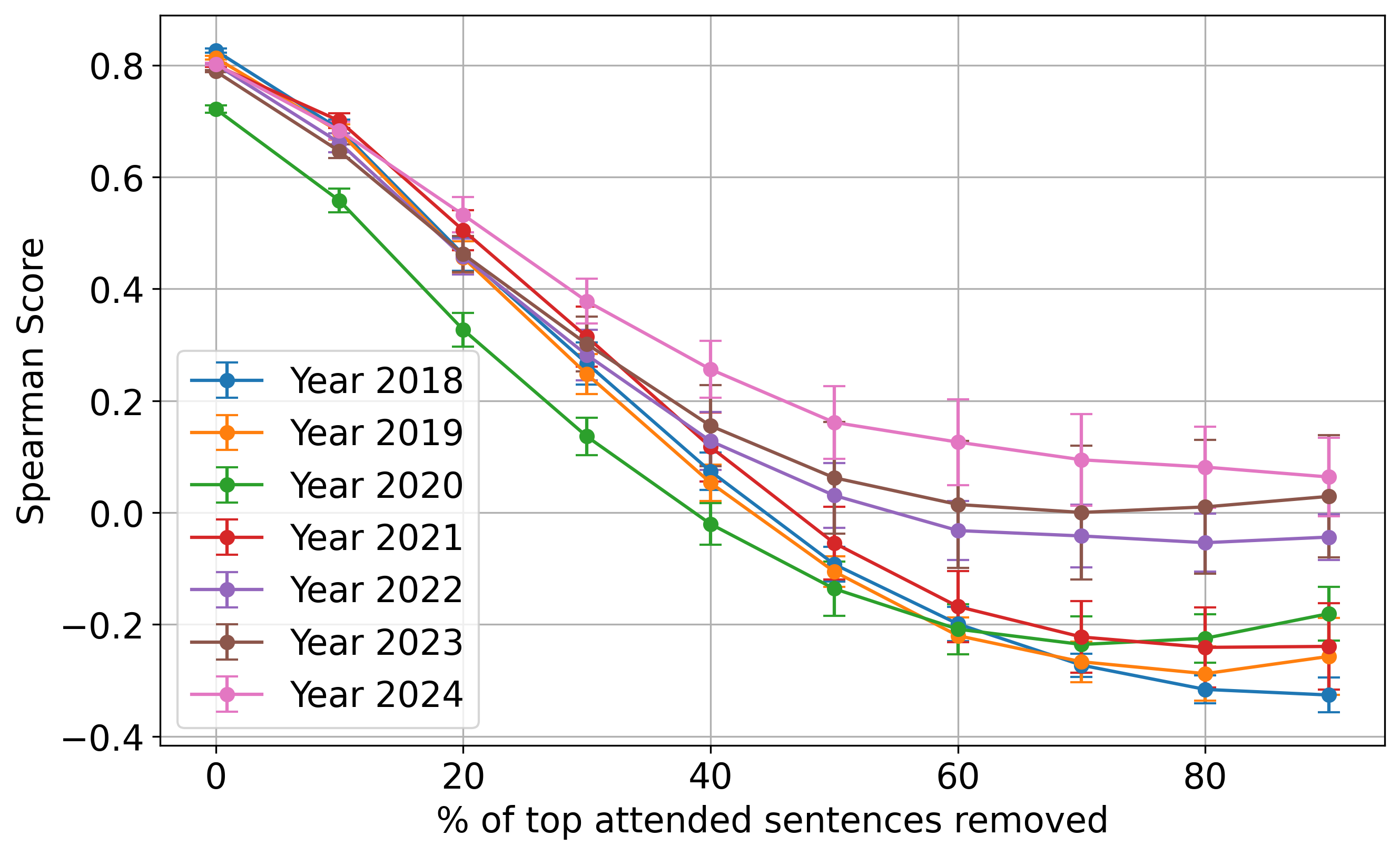}
        \caption[size=small]{}
    \end{subfigure}
    \hfill
    \begin{subfigure}[b]{0.32\textwidth}
        \includegraphics[width=\textwidth]{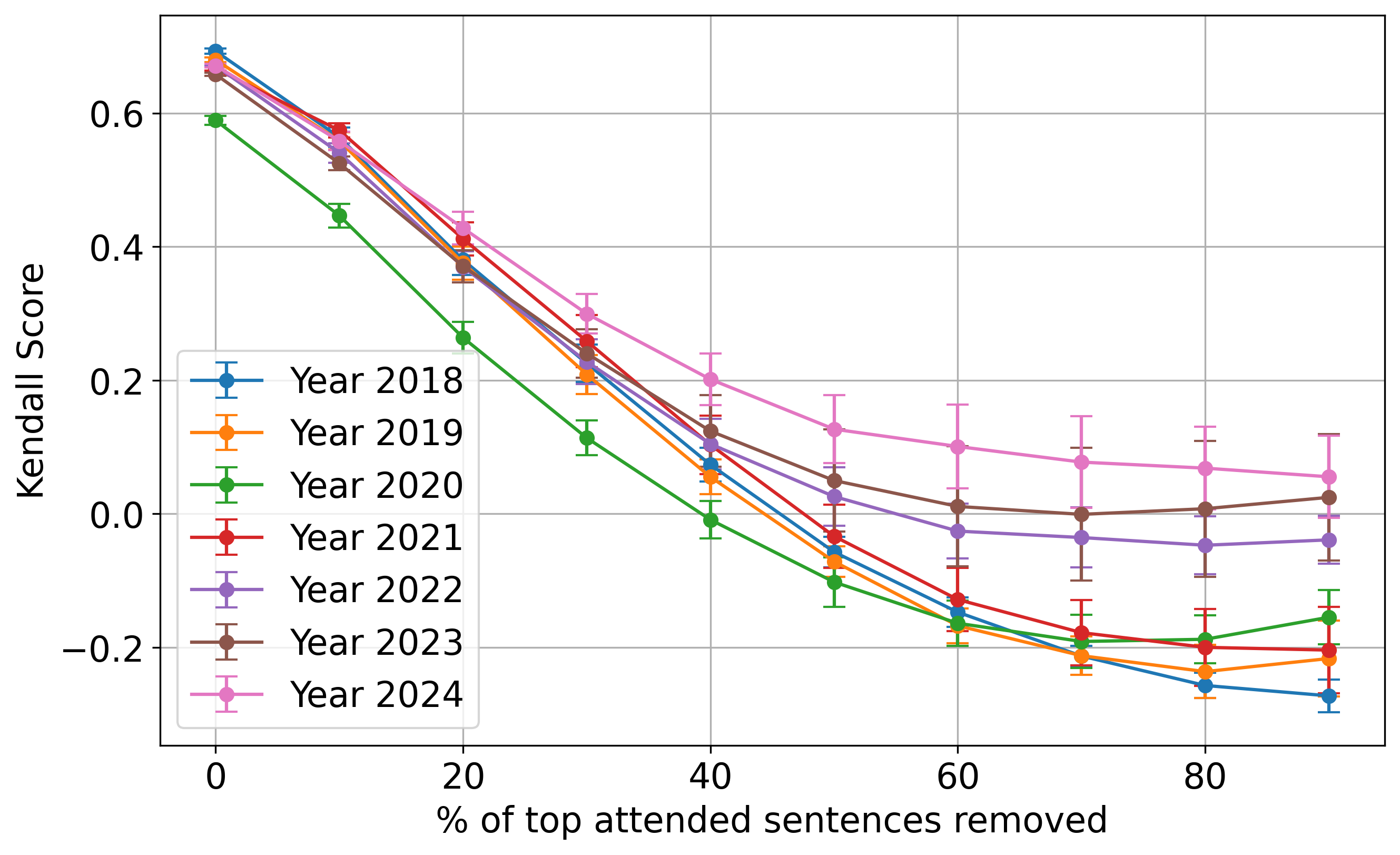}
        \caption[size=small]{}
    \end{subfigure}
    \caption{Performance metrics versus the percentage of top-\(k\%\) attended sentences removed, evaluated using standard deviation as the risk measure from 2018 to 2024. The plots show the mean and standard deviation (as error bars) across five random seeds for F1 Score (a), Spearman’s Rho (b), and Kendall’s Tau (c).}
    \label{fig:exp_sent_std}
\end{figure}

\begin{figure}[htb!]
    \centering
    \begin{subfigure}[b]{0.32\textwidth}
        \includegraphics[width=\textwidth]{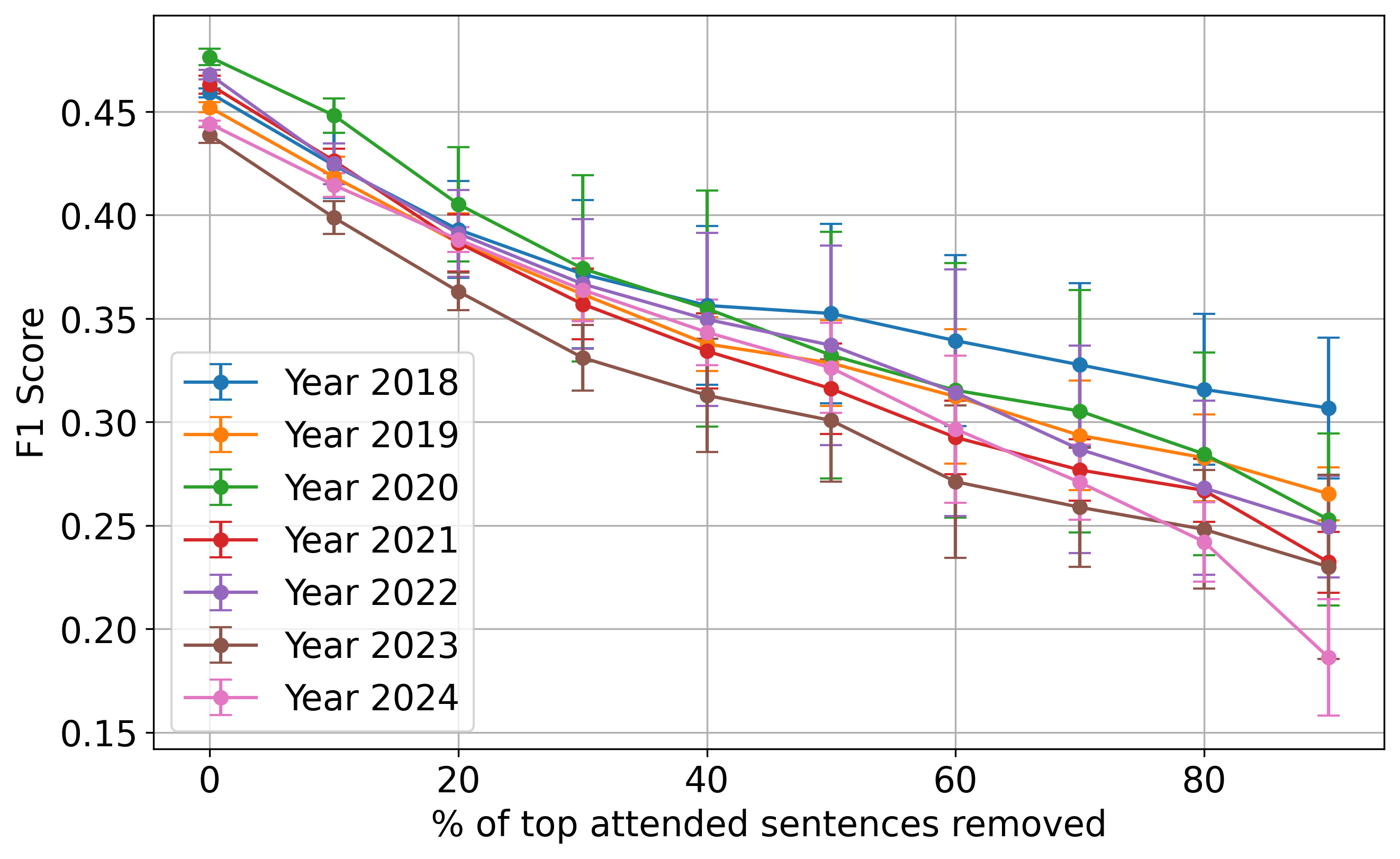}
        \caption[size=small]{}
    \end{subfigure}
    \hfill
    \begin{subfigure}[b]{0.32\textwidth}
        \includegraphics[width=\textwidth]{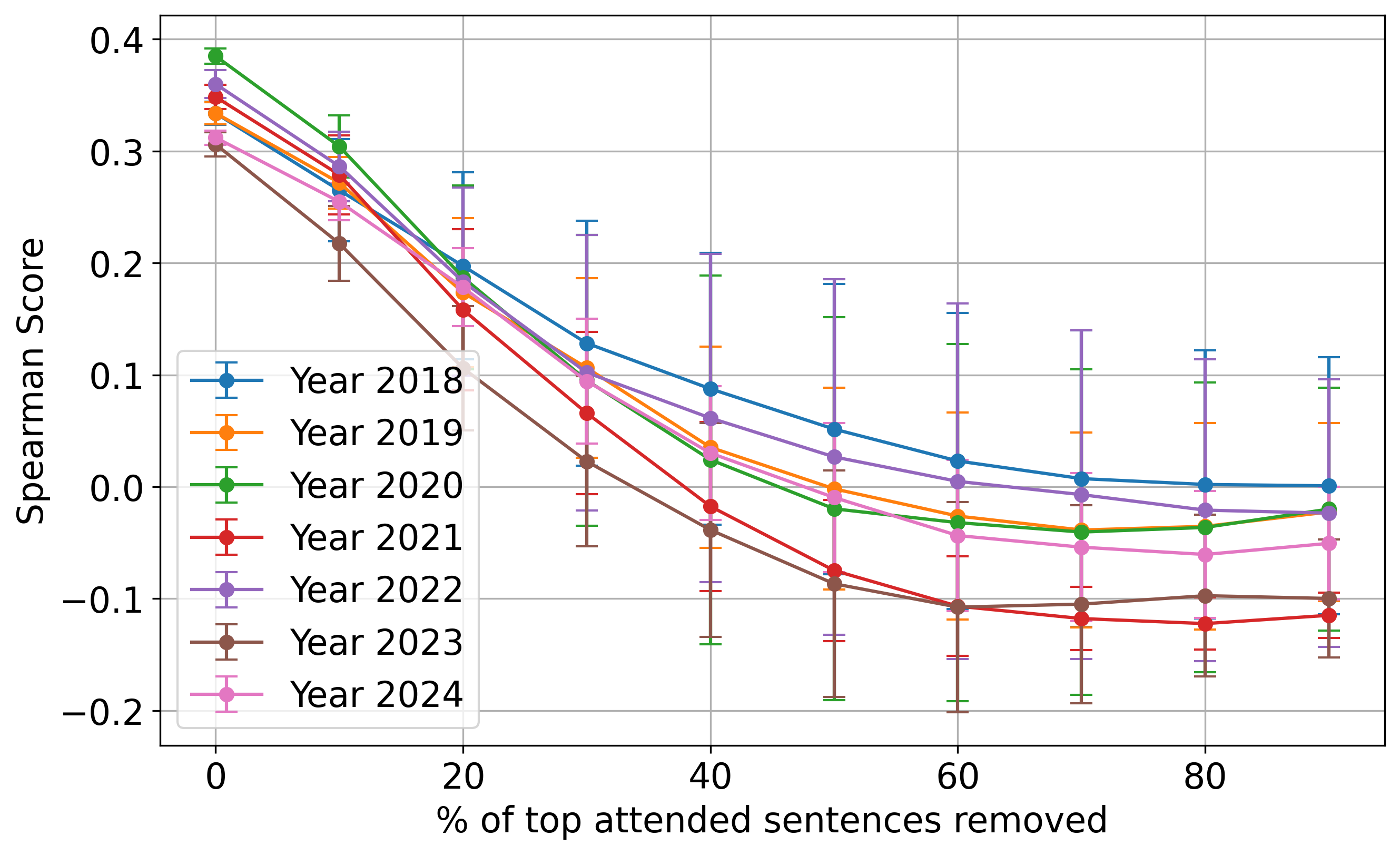}
        \caption[size=small]{}
    \end{subfigure}
    \hfill
    \begin{subfigure}[b]{0.32\textwidth}
        \includegraphics[width=\textwidth]{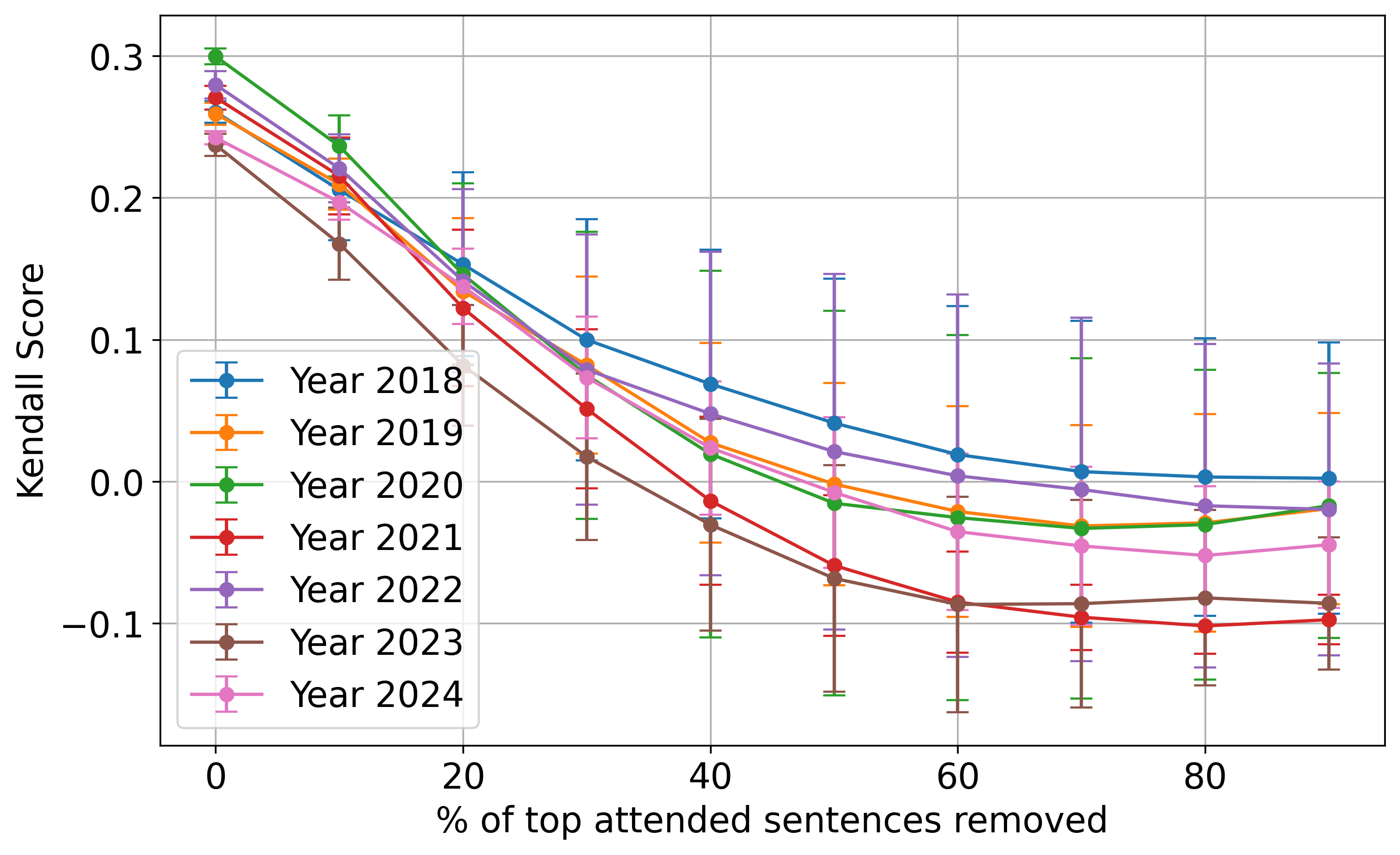}
        \caption[size=small]{}
    \end{subfigure}
    \caption{Performance metrics versus the percentage of top-\(k\%\) attended sentences removed, evaluated using skewness as the risk measure from 2018 to 2024. The plots show the mean and standard deviation (as error bars) across five random seeds for F1 Score (a), Spearman’s Rho (b), and Kendall’s Tau (c).}
    \label{fig:exp_sent_skew}
\end{figure}

\begin{figure}[htb!]
    \centering
    \begin{subfigure}[b]{0.32\textwidth}
        \includegraphics[width=\textwidth]{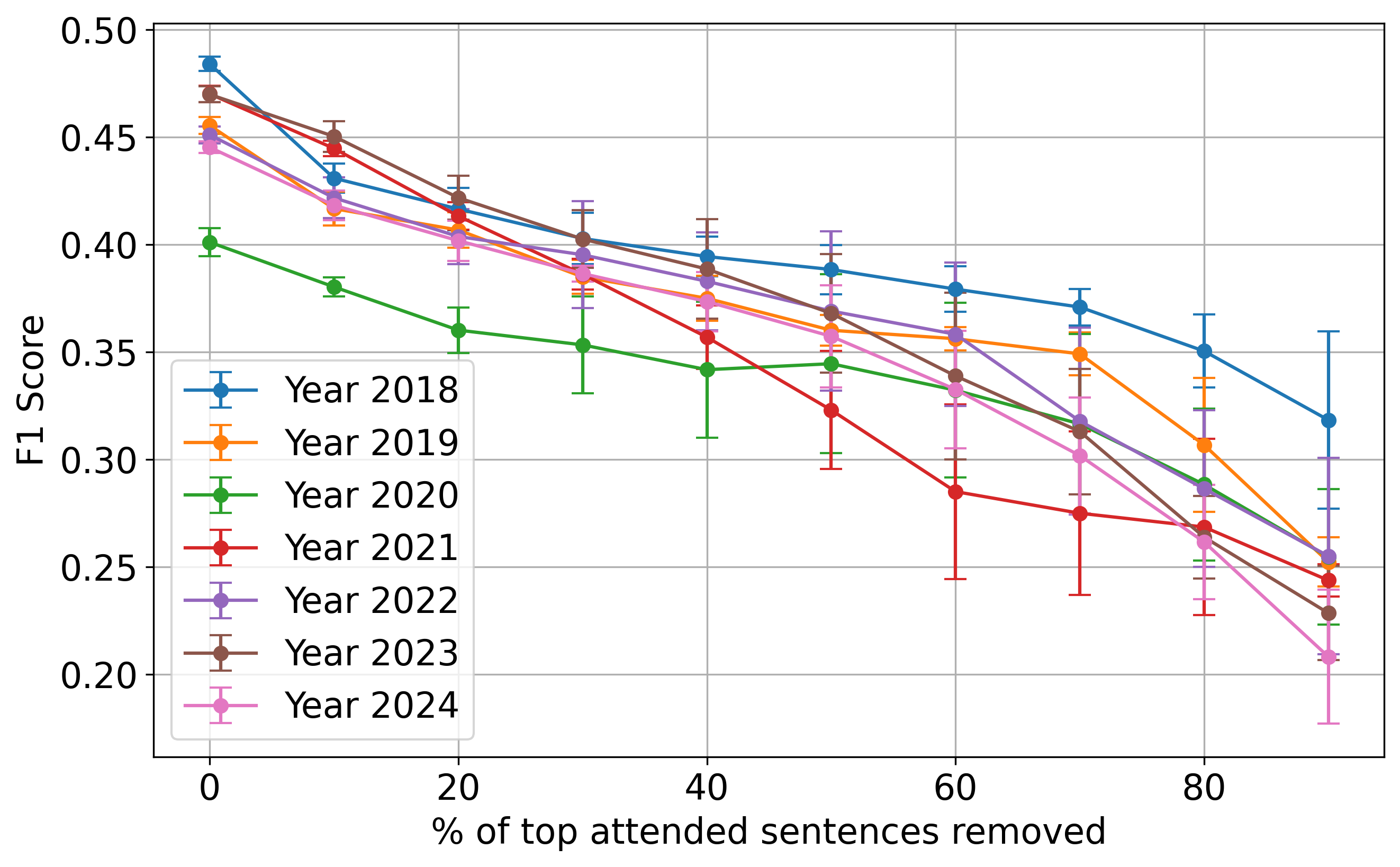}
        \caption[size=small]{}
    \end{subfigure}
    \hfill
    \begin{subfigure}[b]{0.32\textwidth}
        \includegraphics[width=\textwidth]{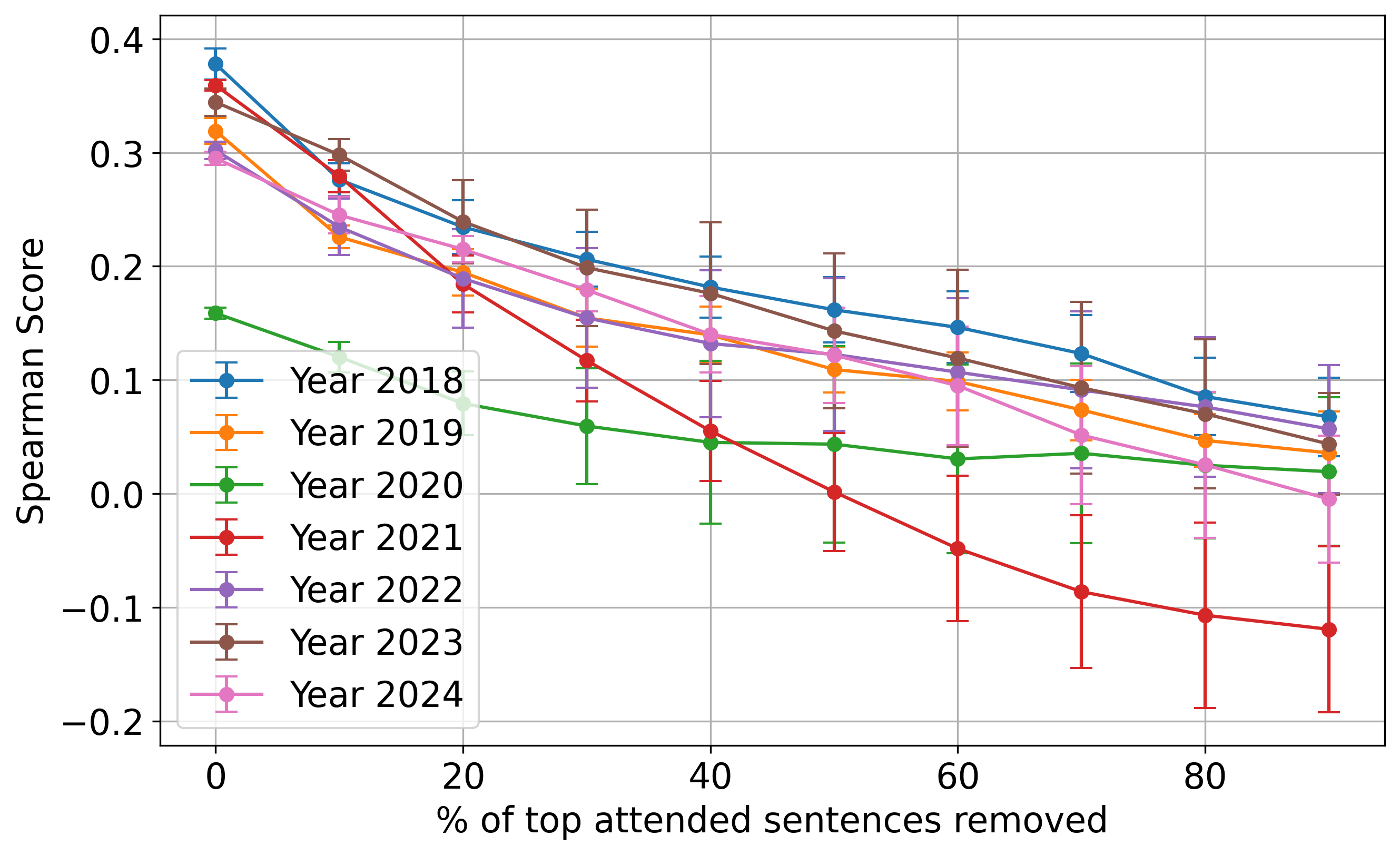}
        \caption[size=small]{}
    \end{subfigure}
    \hfill
    \begin{subfigure}[b]{0.32\textwidth}
        \includegraphics[width=\textwidth]{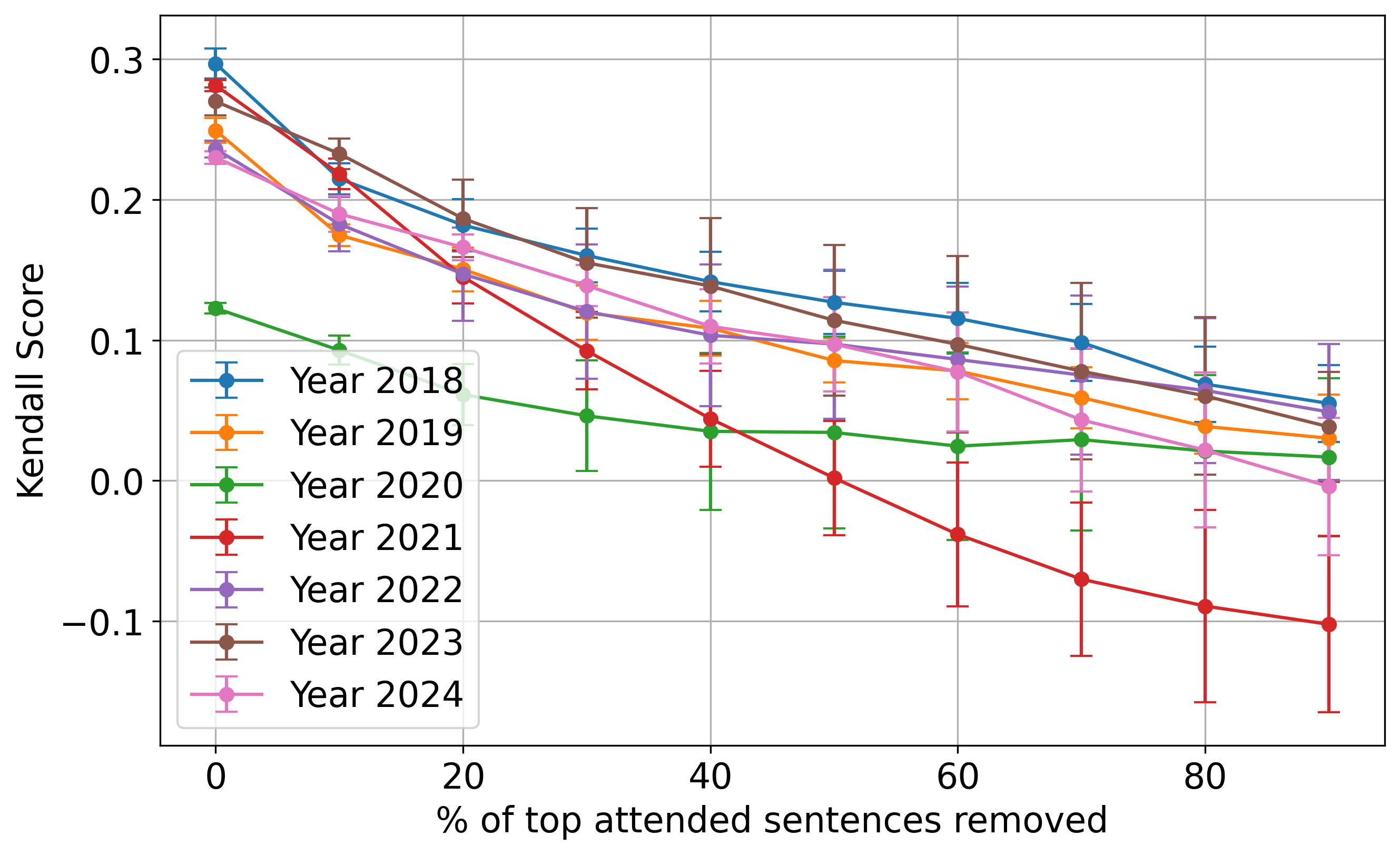}
        \caption[size=small]{}
    \end{subfigure}
    \caption{Performance metrics versus the percentage of top-\(k\%\) attended sentences removed, evaluated using kurtosis as the risk measure from 2018 to 2024. The plots show the mean and standard deviation (as error bars) across five random seeds for F1 Score (a), Spearman’s Rho (b), and Kendall’s Tau (c).}
    \label{fig:exp_sent_kurt}
\end{figure}

\begin{figure}[htb!]
    \centering
    \begin{subfigure}[b]{0.32\textwidth}
        \includegraphics[width=\textwidth]{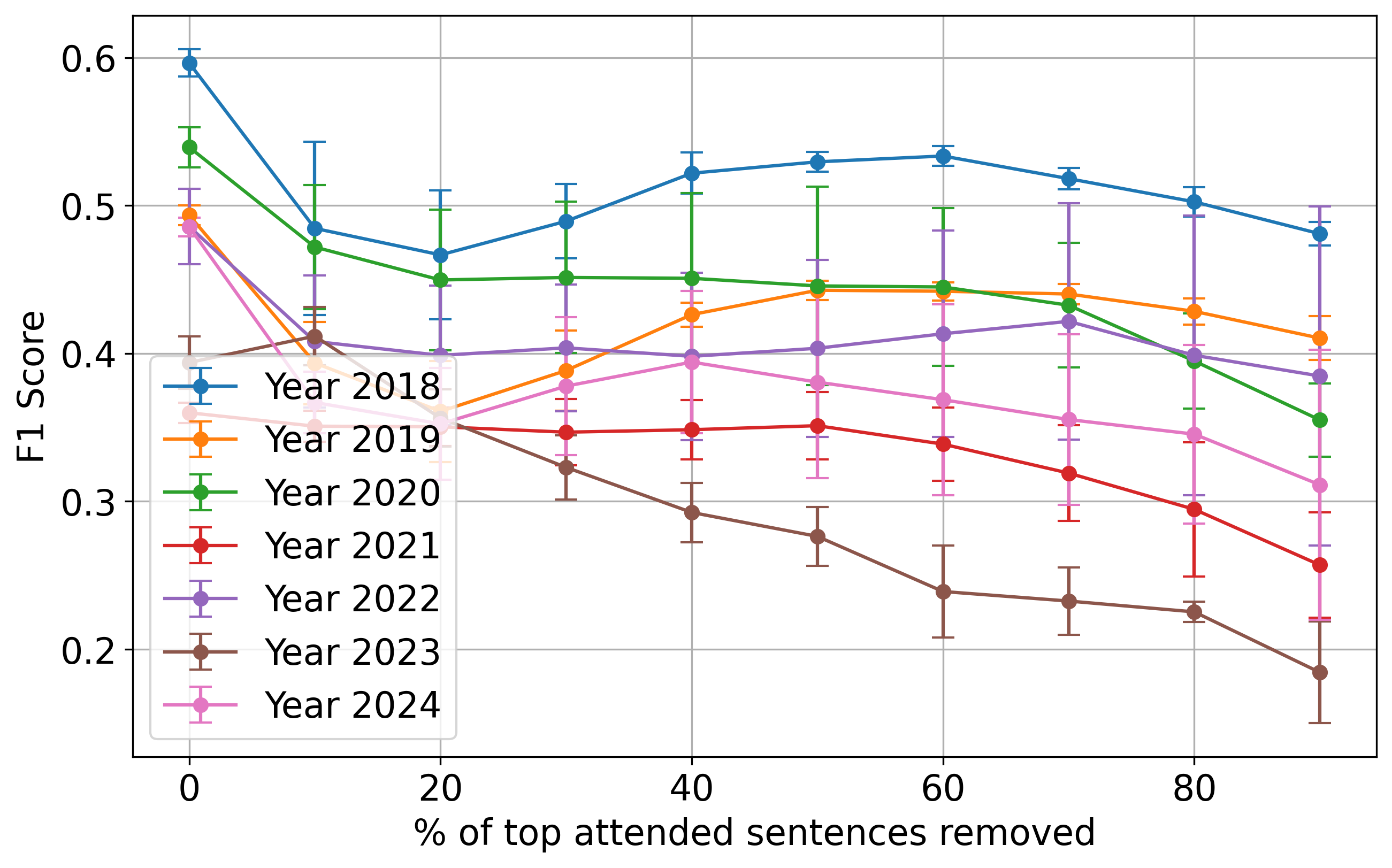}
        \caption[size=small]{}
    \end{subfigure}
    \hfill
    \begin{subfigure}[b]{0.32\textwidth}
        \includegraphics[width=\textwidth]{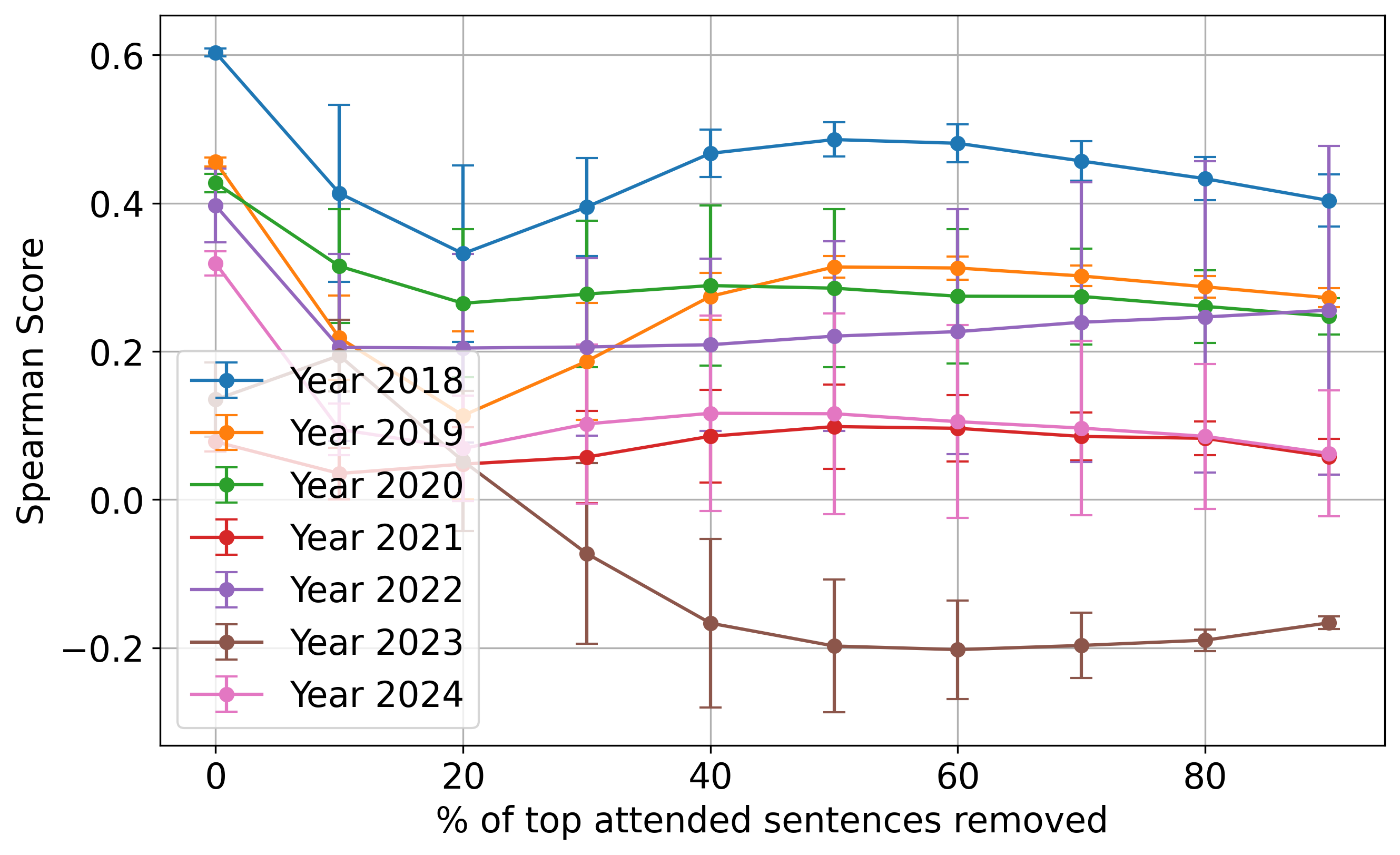}
        \caption[size=small]{}
    \end{subfigure}
    \hfill
    \begin{subfigure}[b]{0.32\textwidth}
        \includegraphics[width=\textwidth]{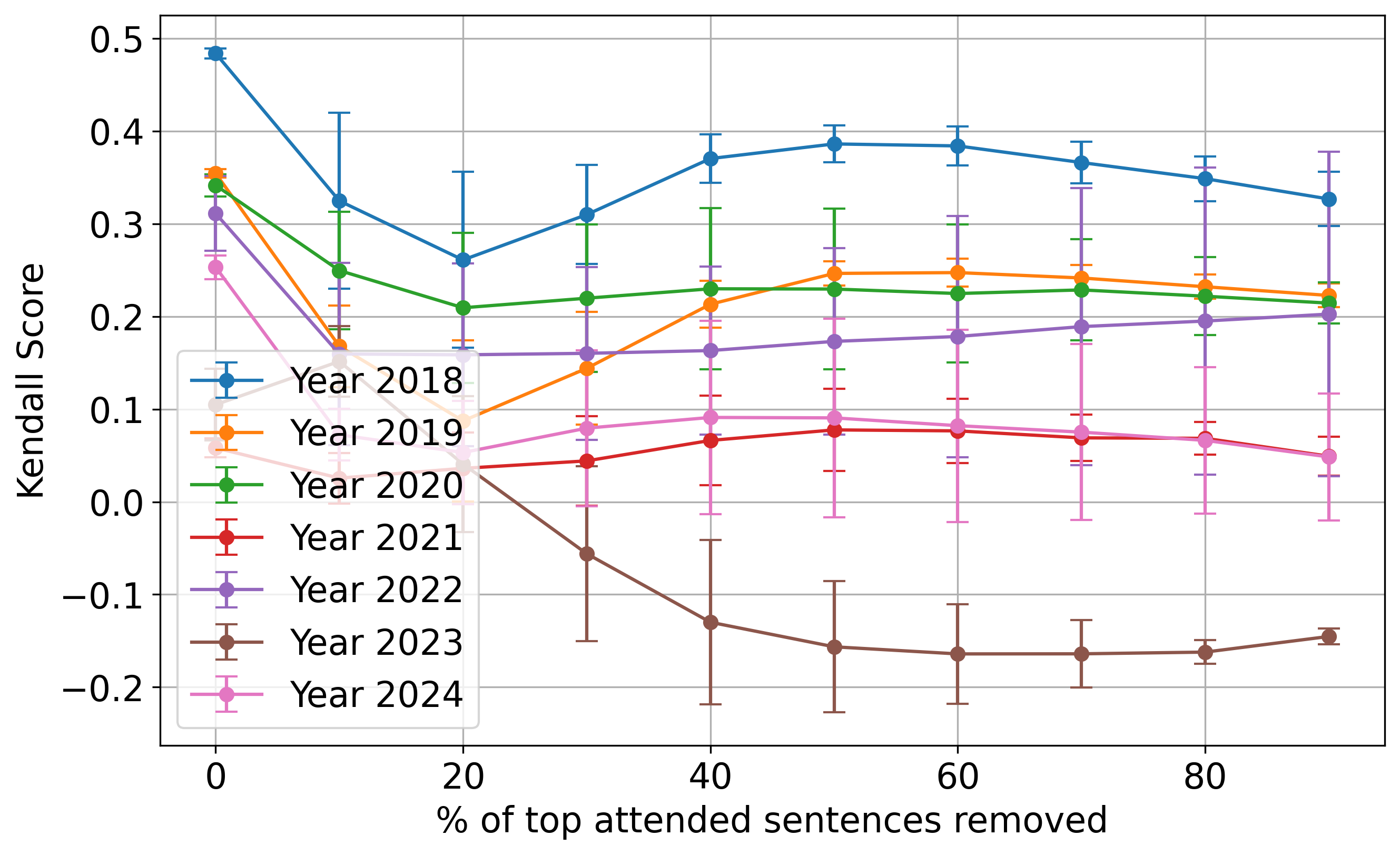}
        \caption[size=small]{}
    \end{subfigure}
    \caption{Performance metrics versus the percentage of top-\(k\%\) attended sentences removed, evaluated using sortino ratio as the risk measure from 2018 to 2024. The plots show the mean and standard deviation (as error bars) across five random seeds for F1 Score (a), Spearman’s Rho (b), and Kendall’s Tau (c).}
    \label{fig:exp_sent_sortino}
\end{figure}

\clearpage
\section{Datasets Statistics for Various Risk Measurement}
\label{appendix:datasets}
\begin{figure}[htb!]
     \FIGURE
     {\includegraphics[width=\textwidth]{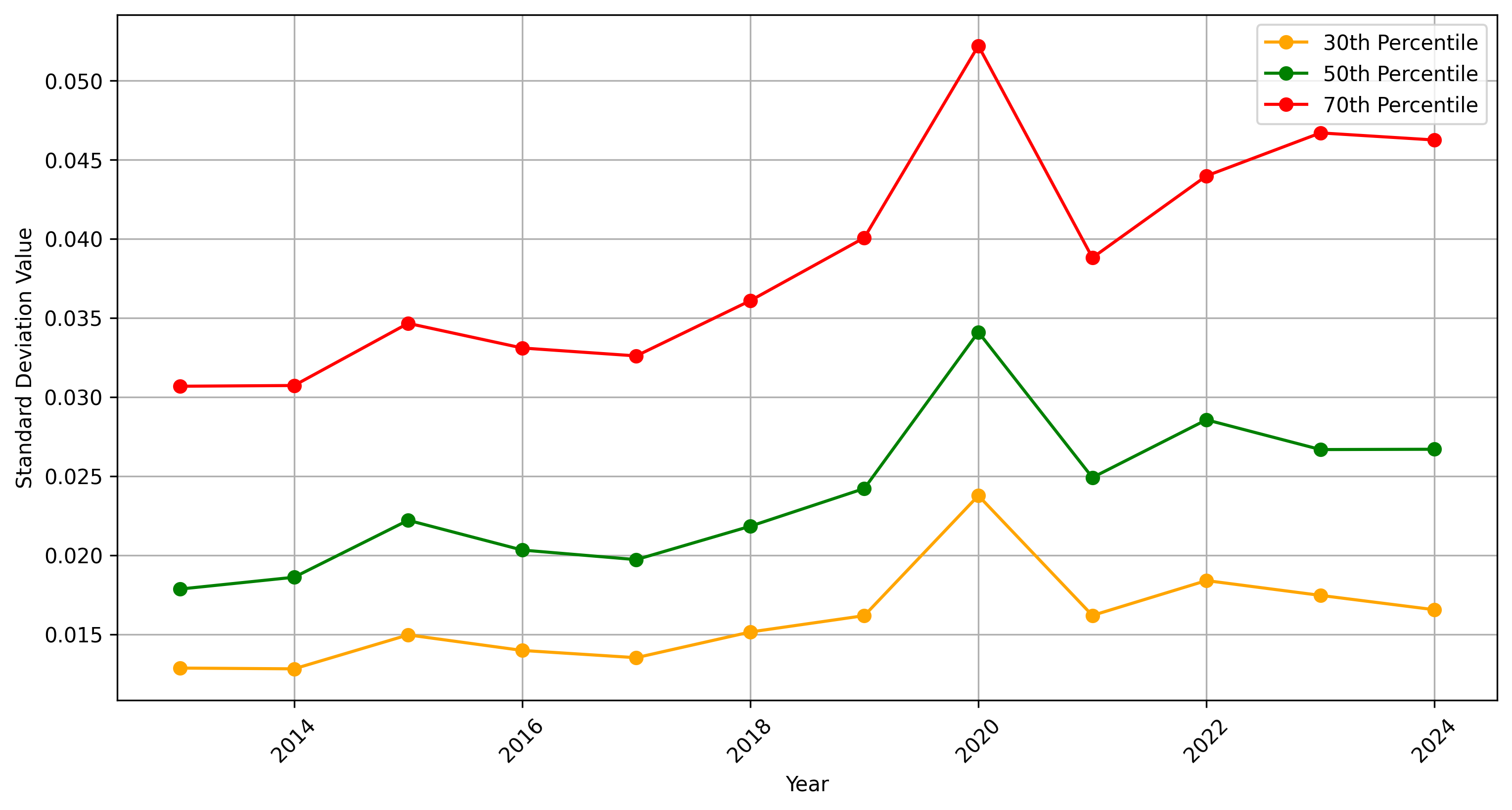}}
     {Standard deviation trends for the 30th, 50th, and 70th percentiles from 2013 to 2024. \label{fig:std_dev}}
     {}
\end{figure}

\begin{figure}[htb!]
     \FIGURE
     {\includegraphics[width=\textwidth]{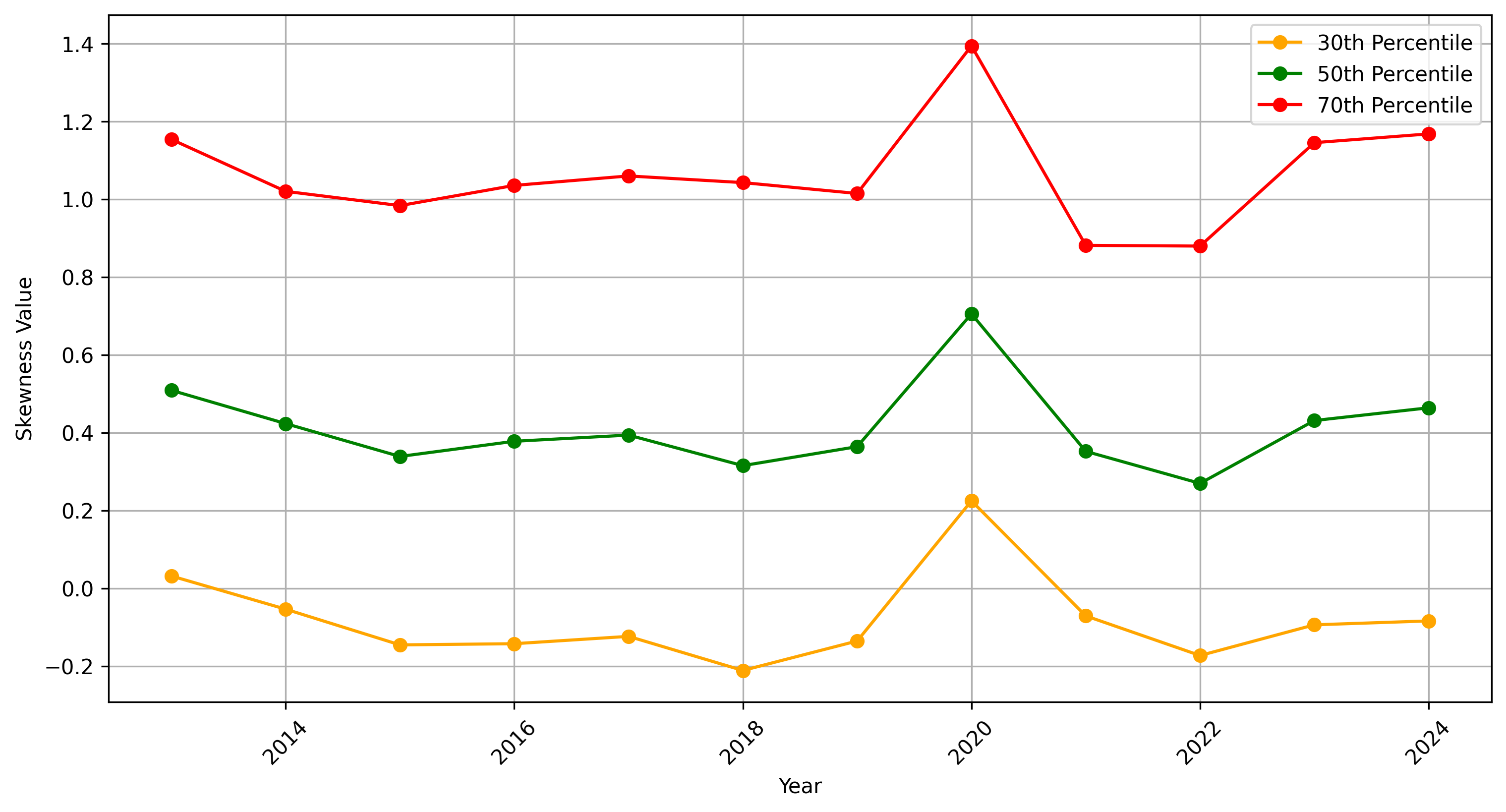}}
     {Skewness trends for the 30th, 50th, and 70th percentiles from 2013 to 2024. \label{fig:skew}}
     {}
\end{figure}

\begin{figure}[htb!]
     \FIGURE
     {\includegraphics[width=\textwidth]{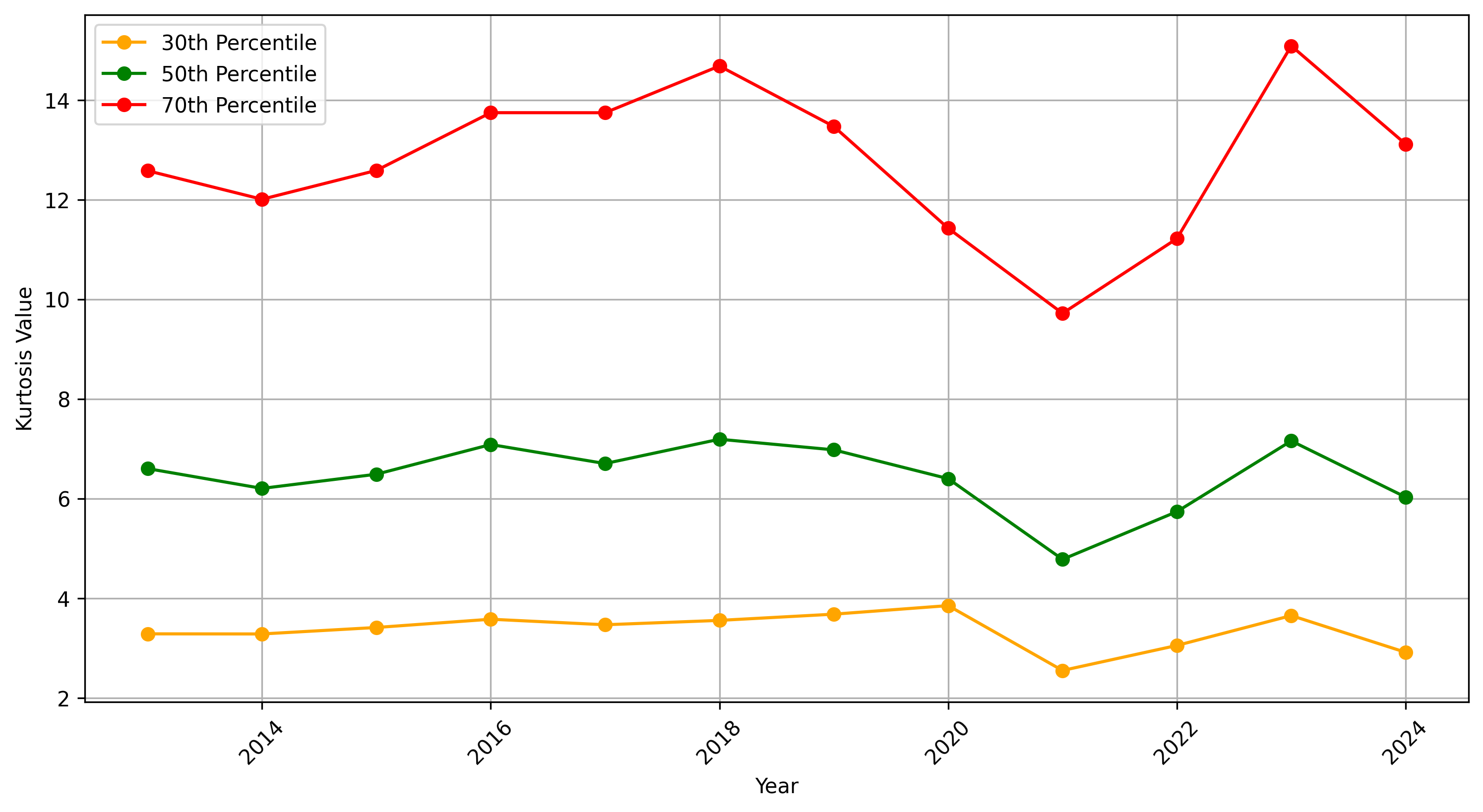}}
     {Kurtosis trends for the 30th, 50th, and 70th percentiles from 2013 to 2024. \label{fig:kurt}}
     {}
\end{figure}

\begin{figure}[htb!]
     \FIGURE
     {\includegraphics[width=\textwidth]{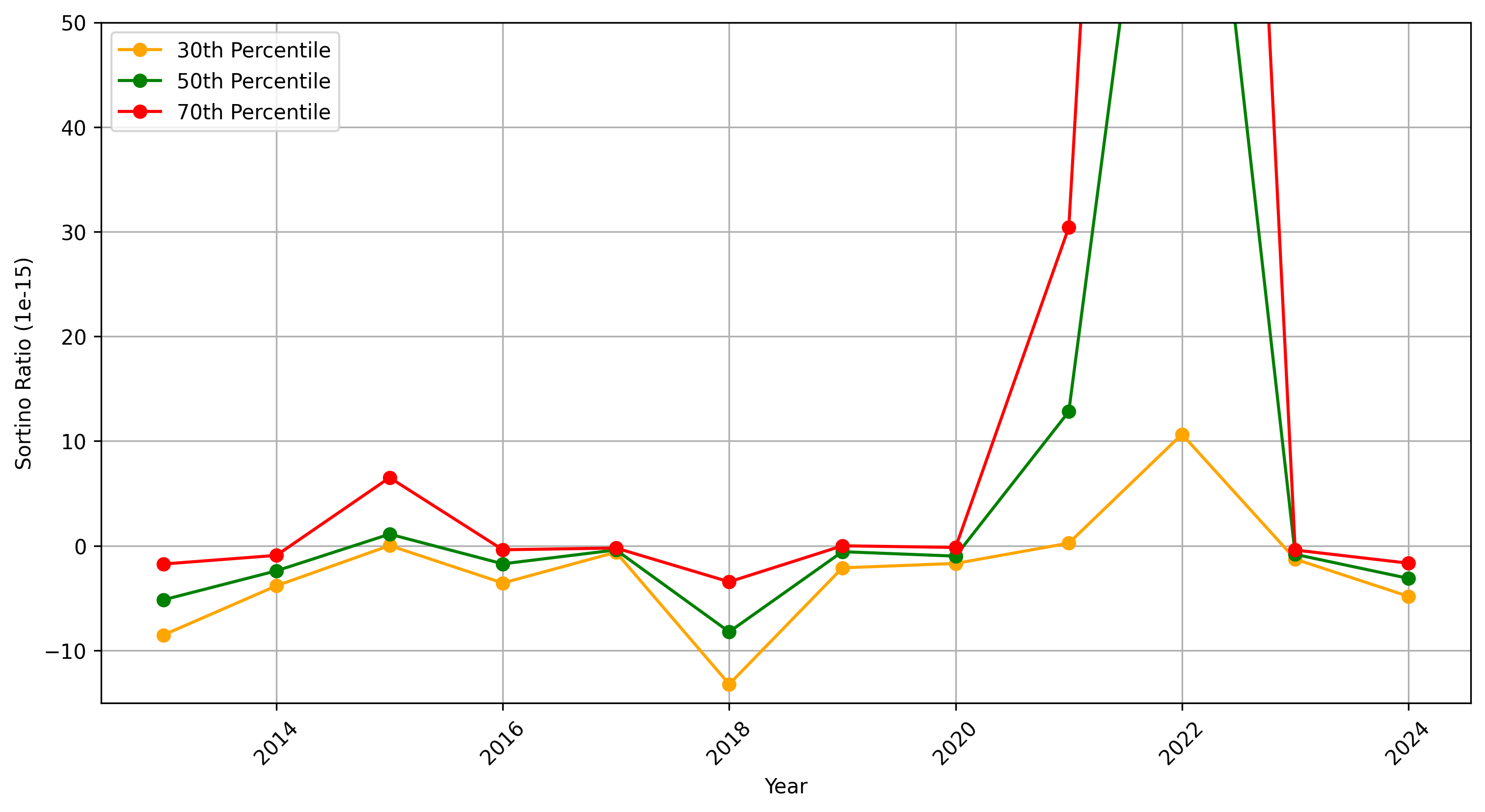}}
     {Sortino ratio trends for the 30th, 50th, and 70th percentiles from 2013 to 2024. \label{fig:sortino}}
     {*Sortino ratio value at Year 2022, 11 for 30th percentile, 95 for 50th Percentile, and 216 for 70th Percentile.}
\end{figure}

\clearpage
\section{Experiments Setup}
Implementation details are available in the accompanying code.
\label{appendix:hyperparam}
\subsection{Hyperparameter Configurations}
\begin{table}[htb!]
\TABLE
{Hyperparameters used for all models \label{tab:hyperparams}}
{\begin{tabular}{@{}l c c c c@{}}
    \hline\up
    Model & Encoder LR & Classifier LR & Epochs & Batch Size \\
    \hline 
    TinyXRA & $1.00 \times 10^{-5}$ & $6.00 \times 10^{-5}$ & 30 & 8 \\
    XRR & $1.00 \times 10^{-5}$ & $6.00 \times 10^{-5}$ & 30 & 8 \\
    TF-IDF & NA & $6.00 \times 10^{-5}$ & 30 & 8 \\
    Llama3.2-1B (SFT) & NA & $5.00 \times 10^{-4}$ & 5 & 1 \\
    Qwen2.5-0.5B (SFT) & NA & $5.00 \times 10^{-4}$ & 5 & 1 \down\\ \hline
\end{tabular}}
{Learning rates (LR) are reported for both encoder and classifier components, where applicable. ``NA" indicates not applicable for the given model. Experiments ran on NVIDIA GeForce RTX 2080 with 11GB of GPU VRAMs}
\end{table}

\subsection{One-shot Prompting Strategy for LLM Evaluation}
\label{appendix:prompt}

\begin{verbatim}
Classify the following text into labels 0, 1, or 2.
Text: "<Support Text 0>"
Label: 0

Text: "<Support Text 1>"
Label: 1

Text: "<Support Text 2>"
Label: 2

Text: "<Test Text>"
Label:
\end{verbatim}
\clearpage
\end{APPENDICES}




\bibliographystyle{informs2014} 
\bibliography{main_docs} 




%
%
%
%

\end{document}